\documentclass[prx, english, twocolumn, longbibliography]{revtex4-1}
\usepackage{graphicx}
\usepackage{amsmath}
\usepackage{amssymb}
\usepackage{amsfonts}
\usepackage{color}
\usepackage{appendix}
\usepackage{braket}
\usepackage{bbm}
\usepackage{subfigure}

\usepackage[colorlinks,urlcolor=blue,citecolor=blue,linkcolor=blue]{hyperref}
\usepackage{lipsum}

\begin{document}
	\title{Degeneracy and defectiveness in non-Hermitian systems with open boundary }
	\author{Yongxu Fu}
	  \email{yancyfoy@mail.ustc.edu.cn}
	\author{Shaolong Wan}
	  \email{slwan@ustc.edu.cn}
	\affiliation{Department of Modern Physics, University of Science and Technology of China, Hefei, 230026, China}
\begin{abstract}
We develop a systematically general theory of one-dimensional~(1D) non-Hermitian systems, elaborating on the energy bands, the band degeneracy, and the defectiveness of eigenstates under open boundary conditions. We analyze the band degeneracy and defectiveness of two typical 1D non-Hermitian models. We obtain the unusual presence and absence of the exceptional points in the generalized non-Hermitian Su-Schrieffer-Heeger model under open boundary conditions. Beyond the general theory, we discover that infernal points exist in 1D non-Hermitian systems, where the energy spectra under open boundary conditions converge on some discrete energy values. We analyze two relevant 1D non-Hermitian models with the existence of infernal points. Moreover, we generalize the infernal points to the infernal knots in four-dimensional systems. The general theory and the infernal points of non-Hermitian systems developed in this paper are also valid in Hermitian systems.
\end{abstract}

\maketitle
\section{Introduction}
Numerous theories on expounding the exotic physics in non-Hermitian systems have been developed in recent years~\cite{bergholtzrev,lee2016,leykam2018,shen2018,kunst2018,yao201801,yao201802,sato2011,kawabata2018,xiong2018,gong2018,longhi2019,yokomizo2019,song2019,imura2019,edvardsson2019,ezawa2019,kawabata2019second,lee2019,okuma2019,su2020,slager2020,jones2020,origin2020,kawabataprx,kawabata20,okugawa2020,kawabatahigher,yang202001,yang202002,knot2021,fu2021,xie2021,longhi2021,li2021}. The most profound achievement is the non-Bloch band theory~\cite{yao201801,yokomizo2019,kawabata20}, which successfully interprets the non-Hermitian skin effect~\cite{yao201801,yao201802,song2019,lee2019,yokomizo2019,origin2020}, a phenomenon of the abundant localized bulk states. In addition, the study of the exceptional points~(EPs), where complex energy bands coalesce, is also an intriguing research field in non-Hermitian systems~\cite{heiss2012,yin2018,alvarez2018,kawabata2019,okugawa2019,bergholtz201901,bergholtz201902,yang2019,li2019,zhang2020,rui2019,xue2020,yokomizo2020,zhong2020,crippa2021,yang2021,denner2021,tidal,mandal2021,delplace2021,ghorashi202101,ghorashi202102,liu2021,bergholtz2021,le2021generalized,wang2021}. Recently, the knot theory has spread to non-Hermitian systems, both on the energy band structure~\cite{jones2020,knot2021} and nodal points~\cite{bergholtz201902,yang2019,li2019,zhang2020,tidal,bergholtz2021}. The exceptional Hopf-link~\cite{bergholtz201902,yang2019,li2019,zhang2020,tidal} and higher-order EPs~\cite{yang2021,mandal2021,delplace2021,ghorashi202101,ghorashi202102,liu2021,bergholtz2021} are attractive courses in non-Hermitian systems.

Although the non-Bloch band theory can depict the bulk energy bands and the localization of the bulk states under open boundary conditions~(OBCs)~\footnote{We emphasize that we take OBC along one specific direction to obtain the open-boundary spectra of non-Hermitian systems in this paper.}, it cannot give the specific forms of the bulk eigenstates in general non-Hermitian systems. In addition, most studies concentrate on the behavior of the EPs in the momentum space, namely, periodic boundary conditions~(PBCs). There are few works focused on the unusual physics of EPs under OBCs~\cite{alvarez2018,rui2019,yokomizo2020}, which is still lacking a general description.    

The EPs are gapless points unique to non-Hermitian systems, which accompany the band degeneracy and defectiveness~(coalescence) of the eigenstates in momentum space~(PBC). However, we need to distinguish the concepts of the band degeneracy and the defectiveness under OBCs. We define the EPs under OBCs as the points where the band degeneracy and defectiveness of the eigenstates both occur. In this paper, we exhaustively develop a systematically general theory of 1D non-Hermitian systems, elaborating on the specific definition of the energy bands, the band degeneracy, and the defectiveness of the eigenstates under OBCs. This general theory contains the energy spectra and the analytic forms of the eigenstates, as well as the degeneracy and defectiveness of them, respectively. In our general theory, we claim that the energy spectrum of non-Hermitian systems under OBCs is comprised of isolated energy bands~(IEBs) and continuous energy bands~(CEBs), and give a specific definition of the band degeneracy under OBCs. The defectiveness at the degenerate points is determined by the dimension of the kernel of the boundary matrix. Beyond the general theory, there possibly exist points in 1D non-Hermitian systems, where the energy spectra under OBCs converge on some discrete energy values. We dub these points the infernal points. Moreover, we can generalize the infernal points to the infernal lines, rings, and knots in higher dimensional non-Hermitian systems. Although we concentrate on non-Hermitian systems in this paper, the general theory and infernal points of non-Hermitian systems are also valid in Hermitian systems.

This paper is organized as follows. In Sec.~\ref{2}, we develop a general theory of energy bands and band degeneracy in 1D non-Hermitian systems, as well as specify the terminology of EPs under OBC. Based on our general theory, we study the zero-energy modes of the generalized non-Hermitian Su-Schrieffer-Heeger~(SSH) model and the non-perturbation of a non-Hermitian two-band model in Sec.~\ref{3}. In Sec.~\ref{4}, we propose the terminology of infernal points in 1D non-Hermitian systems, and generalize them to the infernal knots in higher dimensions. Finally, the conclusion and discussion are given in Sec.~\ref{5}.

\section{The general theory of energy bands and band degeneracy in 1D non-Hermitian systems}
\label{2}
Due to presence of skin effect in 1D non-Hermitian systems, a remarkable difference between complex energy spectra under PBC and those under OBCs, the non-Bloch band theory is developed to obtain the energy spectra under OBCs. However, the non-Bloch band theory cannot expound all possible 1D non-Hermitian systems, and cannot give the eigenstates of the energy bands. In this section, we elaborate on a general theory to analyze the energy bands, as well as the eigenstates, of general 1D non-Hermitian systems under OBCs. Utilizing this theory, we can study the properties of degeneracy and defectiveness of the 1D non-Hermitian systems under OBCs.   
\subsection{The general forms of the eigenstates and the boundary matrix}
A generic tight-binding 1D non-Hermitian Hamiltonian, with $N$ lattice sites, the hopping range $R$, and the internal degrees of freedom per unit cell $q$, is~\footnote{Without loss of generality, we take the same hopping range $R$ in the left and right directions, since the general theory of 1D non-Hermitian systems is not fundamentally affected in the situation with different hopping ranges.} 
\begin{eqnarray}
\hat{H}=\sum_{x=1}^{N}\sum_{n=-R}^{R}\sum_{\mu,\nu=1}^{q}t_{n,\mu\nu}\ket{x,\mu}\bra{x+n,\nu},
\end{eqnarray} 
where $\ket{x,\mu}=\ket{x}\otimes\ket{\mu}\equiv\ket{x}\ket{\mu}$. We assume the trial solution as 
\begin{eqnarray}
\ket{\psi}=\sum_{x=1}^{N}\beta^{x}\ket{x}\ket{u}\equiv\sum_{x=1}^{N}\ket{x}\ket{u_{x}}.
\end{eqnarray}
According to the Schr\"{o}dinger equation $\hat{H}\ket{\psi}=E\ket{\psi}$, we obtain the characteristic equation 
\begin{eqnarray}
\label{charac}
f(E,\beta)\equiv\det{[H(\beta)-E]}=0,
\end{eqnarray}
where $H(\beta)=\sum_{n=-R}^{R}T_{n}\beta^{n}$ and the element of matrix $T_{n}$ is $(T_{n})_{\mu\nu}=t_{n,\mu\nu}$. Motivated by Ref.~\cite{alase2017}, we define the polynomial $P(E,\beta)\equiv\beta^{qR}f(E,\beta)$. By the equation $P(E,\beta)=0$ with a fixed eigenenergy $E$, we can obtain $M$ nonzero solutions of $\beta$ with multiplier $s_{j},j=1,2,\ldots,M$ and $s_{0}$ zero solutions of $\beta$, with $\sum_{j=1}^{M}s_{j}+2s_{0}=2qR$. The eigenstates of the bulk equation are $\ket{\psi_{js}}$, $j=1,2,\ldots,M$, $s=1,2,\dots,s_{j}$ and $\ket{\psi_{s^{\pm}}}$, $s^{\pm}=1,2,\ldots,s_{0}$ for nonzero and zero solutions of $\beta$, respectively (see Appendix~\ref{bulksolu} for details).

We represent the bulk eigenstates of eigenenergy $E$ as the linear superposition of the states in solution space $\left\{\ket{\psi_{js}},\ket{\psi_{s^{-}}},\ket{\psi_{s^{+}}}\right\}$~(Appendix~\ref{bulksolu}), 
\begin{eqnarray}
\psi_{\alpha}=\sum_{j=1}^{M}\sum_{s=1}^{s_{j}}\alpha_{js}\ket{\psi_{js}}+\sum_{s^{-}=1}^{s_{0}}\alpha_{s^{-}}\ket{\psi_{s^{-}}}+\sum_{s^{+}=1}^{s_{0}}\alpha_{s^{+}}\ket{\psi_{s^{+}}}.\nonumber\\
\end{eqnarray}
For simplicity, we denote $J\in\left\{js,s^{-},s^{+}\right\}$, and the above formula becomes
\begin{eqnarray}
\psi_{\alpha}=\sum_{J}\alpha_{J}\ket{\psi_{J}}.
\end{eqnarray}

We apply the OBC and denote the sites of boundary as $b=1,2,\ldots,R,N-R+1,\ldots,N$. The boundary equation is given by~(Appendix~\ref{boundaryeq})
\begin{eqnarray}
[B(E)]_{b\mu,J}\cdot\alpha_{J}=0,
\end{eqnarray}
where $B(E)$ is the boundary matrix, $[B(E)]_{b\mu,J}=\bra{b,\mu}(\hat{H}-E)\ket{\psi_{J}}$. When the zero solutions of $\beta$ are absent, we can obtain another equivalent form of the boundary matrix~(Appendix~\ref{boundaryeq}).

Without loss of generality, we only consider the cases without the zero solutions of $\beta$ for large $N$ in this paper~[Appendix~\ref{boundaryeq}]. We define the sets $P$ and $Q$ as two disjoint subsets of the set of all $js$, such that the number of elements of each subset is $m\equiv qR$. The determinant of the boundary matrix is~(see Appendix~\ref{boundaryeq} for details)
\begin{eqnarray}
\label{deter}
\det{[B(E)]}=\sum_{P,Q}F(\beta_{I\in P},\beta_{J\in Q},E)\prod_{J\in Q}(\beta_{J})^{N},
\end{eqnarray}
where $\beta_{J}=\beta_{j}$ is corresponding to $J=js$. The eigenstates of eigenenergy $E$ with OBCs are the kernel of the boundary matrix $B(E)$. Therefore, there exist eigenstates of eigenenergy $E$ only if $\det{[B(E)]}=0$. In principle, we need to scan all $E\in \mathbb{C}$ and determine whether the determinant of the corresponding boundary matrix vanishes, to obtain the full energy spectra of the systems under OBCs. 
\subsection{The energy bands and band degeneracy}
We number the solutions~(nonzero) of $\beta$ satisfying $|\beta_{1}|\leq\ldots\leq|\beta_{m}|\leq|\beta_{m+1}|\ldots\leq|\beta_{2m}|$, where there are $2m$ nonzero solutions of $\beta$, and analyze the energy bands and degeneracy of 1D non-Hermitian systems under OBCs.  
\subsubsection{Isolated energy bands}
\label{iso}
When $|\beta_{m}|<|\beta_{m+1}|$, the leading-order term of Eq.~(\ref{deter}) with respect to $\prod_{J\in Q}(\beta_{J})^{N}$ is only one, namely,
\begin{eqnarray}
F(\beta_{I\in P_{0}},\beta_{J\in Q_{0}},E_{0})\prod_{J\in Q_{0}}(\beta_{J})^{N},
\end{eqnarray}
where $P_{0}=\left\{1,2,\ldots,m\right\}$, $Q_{0}=\left\{m+1,m+2,\ldots,2m\right\}$, and $E_{0}$ is the corresponding eigenenergy. Since the energy spectra are isolated in this case, we dub them the IEBs. 

If $N$ is infinite, we discuss as follows situations according to the magnitude of $|\prod_{J\in Q}\beta_{J}|$. (i) $|\prod_{J\in Q_{0}}\beta_{J}|<1$, $\det{[B(E_{0})]}=0$ holds and $E_{0}$ is exactly an IEB. (ii) $|\prod_{J\in Q_{i}}\beta_{J}|\geq1$ for some orders, $\det{[B(E_{0})]}=0$ holds only when all of $F(\beta_{I\in P_{i}},\beta_{J\in Q_{i}},E_{0})$ corresponding to $|\prod_{J\in Q_{i}}\beta_{J}|\geq1$ vanish. Hence, if $F(\beta_{I\in P_{i}},\beta_{J\in Q_{i}},E_{0})=0$, $E_{0}$ is exactly an IEB; If $F(\beta_{I\in P_{i}},\beta_{J\in Q_{i}},E_{0})\neq0$, $E_{0}$ is not an eigenenergy.

If $N$ is large but finite, we define 
\begin{eqnarray}
F^{(n)}_{0}=\frac{d^{n}}{dE^{n}}F(\beta_{I\in P_{0}},\beta_{J\in Q_{0}},E)\bigg|_{E_{0}},
\end{eqnarray}
and discuss as follows situations according to $F^{(0)}_{0}$. (i) $F^{(0)}_{0}\neq0$, we expand $F(\beta_{I\in P_{0}},\beta_{J\in Q_{0}},E)$ around $E_{0}$~\footnote{We take the first three-order terms in the expansion of $\det{[B(E)]}$, to discuss the asymptotic behavior of the IEBs. If one of the first three-order terms vanishes, we need to consider the fourth-order term, and it is the same as the subsequent orders.},
\begin{eqnarray}
\label{leading}
F(\beta_{I\in P_{0}},\beta_{J\in Q_{0}},E)\sim F^{(0)}_{0}+F^{(1)}_{0}\Delta E+\frac{1}{2}F^{(2)}_{0}\Delta E^{2},\nonumber\\
\end{eqnarray}
where $\Delta E=E-E_{0}$ is small enough such that we can expand $\det{[B(E)]}$ up to the third-order term. Actually, if $\Delta E$ is not very small, the eigenenergies are away from $E_{0}$, obviously. We obtain finite $\Delta E$ independent of $N$ by $F(\beta_{I\in P_{0}},\beta_{J\in Q_{0}},E)=0$, and $E_{0}$ is not an eigenenergy. (ii) $F^{(0)}_{0}=0$, we need to consider the second-order term of $\det{[B(E)]}$ respect to $\prod_{J\in Q}(\beta_{J})^{N}$, 
\begin{eqnarray}
F(\beta_{I\in P_{1}},\beta_{J\in Q_{1}},E_{0})\prod_{J\in Q_{1}}(\beta_{J})^{N},
\end{eqnarray}
where $P_{1}=\left\{1,2,\ldots,m-1,m+1\right\}$, and $Q_{1}=\left\{m,m+2,\ldots,2m\right\}$. We define
\begin{eqnarray}
F^{(n)}_{1}=\frac{d^{n}}{dE^{n}}F(\beta_{I\in P_{1}},\beta_{J\in Q_{1}},E)\bigg|_{E_{0}}.
\end{eqnarray}
We require $\Delta E$ to be small enough, such that $\det{[B(E)]}$ can be expanded around $E_{0}$ up to the third-order term for large $N$ as
\begin{eqnarray}
\label{expand}
&&\det{[B(E)]}\sim(F^{(1)}_{0}\Delta E+\frac{1}{2}F^{(2)}_{0}\Delta E^{2})\prod_{J\in Q_{0}}(\beta_{J})^{N}\nonumber\\
&&\qquad\qquad\quad+F^{(0)}_{1}\prod_{J\in Q_{1}}(\beta_{J})^{N}.
\end{eqnarray}
If $F^{(1)}_{0}\neq0$, we obtain nonzero $\Delta E$, and $E_{0}$ is not an eigenenergy. Note that, $\Delta E$ may not be small enough to make the expansion Eq.~(\ref{expand}) inappropriate, but we still obtain nonzero $\Delta E$ by $\det{[B(E)]}=0$ with proper expansion unless all $F(\beta_{I\in P_{i}},\beta_{J\in Q_{i}},E_{0})$ of $\det{[B(E)]}$ vanish. If $F^{(1)}_{0}=0$, we obtain
\begin{eqnarray}
\label{exp}
\Delta E\sim\pm\sqrt{-\frac{2F^{(0)}_{1}}{F^{(2)}_{0}}\bigg(\frac{\prod_{J\in Q_{1}}\beta_{J}}{\prod_{J\in Q_{0}}\beta_{J}}\bigg)^{N}}.
\end{eqnarray}
Since $|\frac{\prod_{J\in Q_{1}}\beta_{J}}{\prod_{J\in Q_{0}}\beta_{J}}|<1$, $\Delta E$ exponentially decays to zero as $N$ increasing. When $N$ is infinite, namely, in the thermodynamics limit, $E_{0}$ is an IEB, where the bands $E_{0}+\Delta E$ with finite $N$ decay to exponentially. We assume $|\beta_{1}|\leq\ldots\leq|\beta_{m-1}|<|\beta_{m}|$ and $|\beta_{m+1}|<|\beta_{m+2}|\leq\ldots\leq|\beta_{2m}|$, such that the number of the second-order term of $\det{[B(E)]}$ respect to $\prod_{J\in Q}(\beta_{J})^{N}$ is only one. If $|\beta_{1}|\leq\ldots\leq|\beta_{m-1}|=|\beta_{m}|$ and (or) $|\beta_{m+1}|=|\beta_{m+2}|\leq\ldots\leq|\beta_{2m}|$, there are more than one second-order terms respect to $\prod_{J\in Q}(\beta_{J})^{N}$, such that the term under the square root of Eq.~(\ref{exp}) is the addition of more than one exponentially displaced terms, but the exponentially decaying behavior of $\Delta E$ still holds as $N$ increases.

\subsubsection{Continuous energy bands}
\label{con}
When $|\beta_{m}|=|\beta_{m+1}|$, there are two leading-order terms of Eq.~(\ref{deter}) with respect to $\prod_{J\in Q}(\beta_{J})^{N}$,
\begin{eqnarray}
\sum_{i=0,1}F(\beta_{I\in P_{i}},\beta_{J\in Q_{i}},E)\prod_{J\in Q_{i}}(\beta_{J})^{N},
\end{eqnarray}
where $P_{i}$ and $Q_{i}$, $i=0,1$, are defined in Sec.~\ref{iso}. In the thermodynamics limit, we obtain the generalized Brillouin zone~(GBZ) and the non-Bloch Hamiltonian $H(\beta)$ over the GBZ~\cite{yao201801,yokomizo2019,yang202002}.
In general, different energy bands $E^{\mu}(\beta)$, which are obtained by solving the characteristic Eq.~(\ref{charac}), correspond to different sub-auxiliary generalized Brillouin zones~(sub-aGBZs)~[Appendix \ref{gbz}]. We denote the sub-aGBZ of each energy band $E^{\mu}(\beta)$ as $\beta_{(p,p+1)}^{\mu}(\theta)$, where $\theta\in[0,2\pi]$ and $p=1,2,\ldots,2m-1$. The sub-GBZs correspond to the sub-aGBZs with $p=m$, which we specially denote as $\beta_{GBZ}^{\mu}(\theta)$~\footnote{In this paper, we concentrate on the general theory of 1D non-Hermitian systems without conjugated time-reversal symmetry~(TRS$^{\dagger}$). The continuous bands condition of systems in symplectic class~(with TRS$^{\dagger}$) is given by $|\beta_{m-1}|=|\beta_{m}|$ or $|\beta_{m+1}|=|\beta_{m+2}|$, where $m$ is an even integer~\cite{kawabata20}. The general theory of 1D non-Hermitian systems is not fundamentally affected in symplectic class.}. Substituting these sub-aGBZs into each expression $E^{\mu}(\beta)$ of the energy band, we obtain the sub-auxiliary energy bands~(sub-AEBs), denoted as $E^{\mu}\big(\beta_{(p,p+1)}^{\mu}(\theta)\big)$. The physical sub-continuous energy bands~(sub-CEBs), $E^{\mu}\big(\beta_{GBZ}^{\mu}(\theta)\big)$, correspond to the sub-AEBs with $p=m$~(Appendix \ref{gbz}). 

If we consider the cases with finite $N$, we need to follow the discussion in Sec.~\ref{iso}. However, the leading-order term respect to $\prod_{J\in Q}(\beta_{J})^{N}$ of $\det{[B(E)]}$ is the sum of the two terms with $\left\{P_{0},Q_{0}\right\}$ and $\left\{P_{1},Q_{1}\right\}$, and the second-order term respect to $\prod_{J\in Q}(\beta_{J})^{N}$ is the term with $\left\{P_{2},Q_{2}\right\}$, where $P_{2}$ and $Q_{2}$ are dependent on concrete systems. It is tedious to deal with the cases with finite $N$. In this paper, we only concentrate on the CEBs in the thermodynamics limit, when $|\beta_{m}|=|\beta_{m+1}|$.  

\subsubsection{Band degeneracy}
We claim that the energy spectrum of non-Hermitian systems under OBCs is constituted by IEBs and CEBs. Based on the above, we study the band degeneracy and the defectiveness of the eigenstates of non-Hermitian systems under OBCs.

According to the theory of the IEBs in Sec.~\ref{iso}, we expound the band degeneracy of IEBs. We refer to the degeneracy of two IEBs at energy value $E_{0}$, only if they exponentially displace from $E_{0}$ with finite $N$, and are degenerate at $E_{0}$ in the thermodynamics limit. Notice that, in most systems, the IEBs correspond to the topological modes, resulting in a topological phase.  

Utilizing the theory of the GBZs in Sec.~\ref{con}, we study the band degeneracy of AEBs. Two sub-AEBs are degenerate at energy value $E_{0}$, only if $E^{\mu}\big(\beta_{(p,p+1)}^{\mu}\big)=E^{\nu}\big(\beta_{(p',p'+1)}^{\nu}\big)=E_{0}$ for some points on the two corresponding sub-aGBZs $\beta_{(p,p+1)}^{\mu}$ and $\beta_{(p',p'+1)}^{\nu}$. We concentrate on the case with $\mu\neq\nu$ and $p=p'=m$, namely, the physical degeneracy between different CEBs, and discuss the other non-physical cases in Appendix~\ref{appband}. Assume the two CEBs are degenerate at energy value $E_{0}$. Since the values of $\beta$ on sub-GBZs correspond to the solutions of equation $f(E_{0},\beta)=0$ with $|\beta_{m}|=|\beta_{m+1}|$, two sub-CEBs are degenerate at one energy value, only when the two sub-GBZs of these two sub-CEBs are degenerate at some points. However, the inverse of this statement is not always true~(Appendix~\ref{appband}). Noteworthily, the GBZs in non-Hermitian systems, are the generalization of the momentum space, $k\in[0,2\pi]$, in Hermitian systems~(the GBZ on unit circle). Consequently, the degeneracy of the sub-CEBs is the generalization of the band degeneracy, $E^{\mu}(k)=E^{\nu}(k)=E_{0}$, in Hermitian systems.

After finding the degenerate energies, we need to determine whether the eigenstates of the degenerate energies are defective, namely, the presence or absence of EPs under OBCs. Assuming there are $d$ energy bands degenerate at $E_{0}$, we need to calculate the boundary matrix $B(E_{0})$~(Appendix~\ref{boundaryeq}). If the dimension of the kernel of $B(E_{0})$, $dim[B(E_{0})]$~(the number of the eigenstates respect to $E_{0}$), is less than $d$, the energy $E_{0}$ is an EP, and the defective degree of this point is $d-dim[B(E_{0})]$.

\begin{figure*}
	\centering
	\subfigure[]{\includegraphics[width=0.45\textwidth]{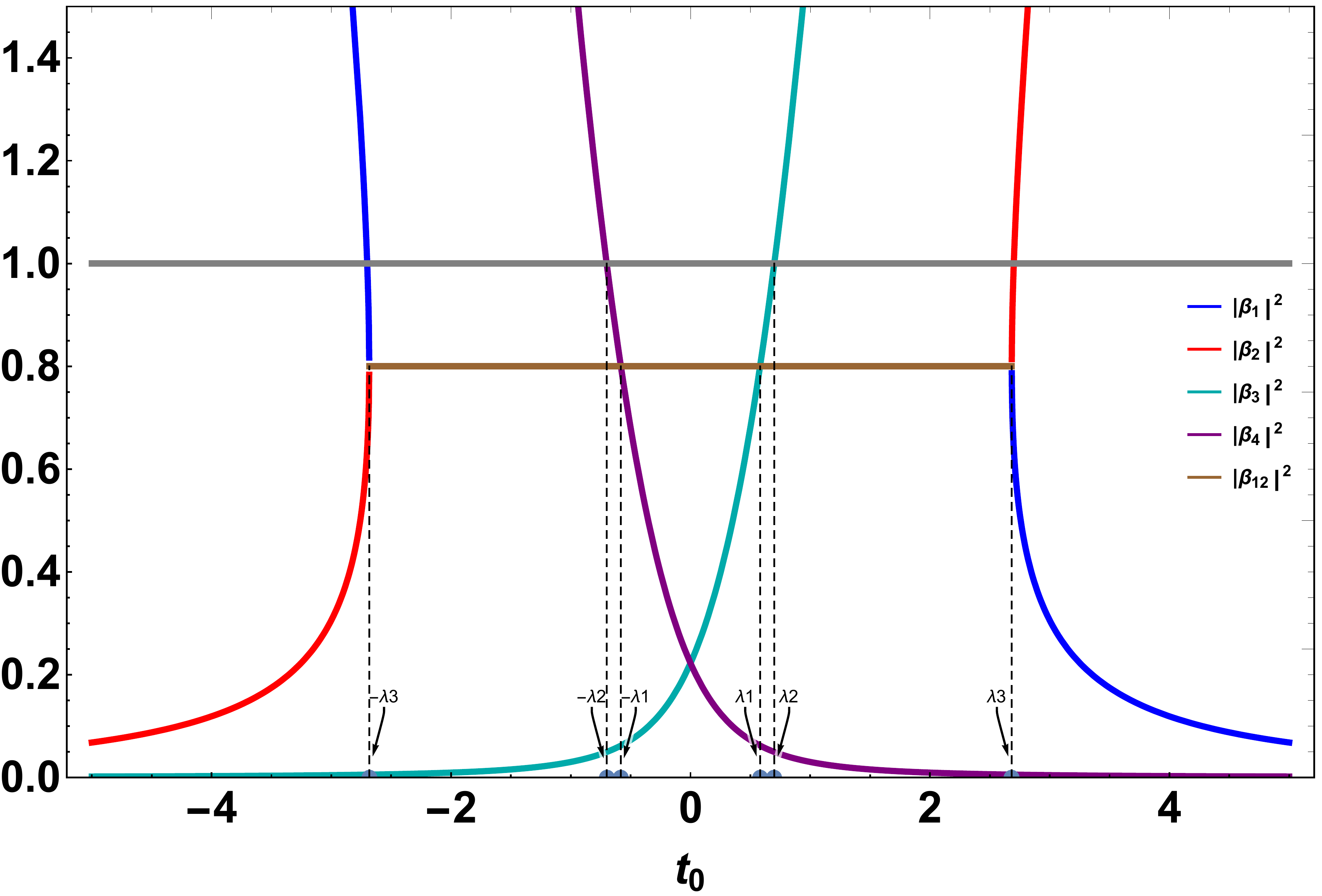}}\\
	\subfigure[]{\includegraphics[width=0.30\textwidth]{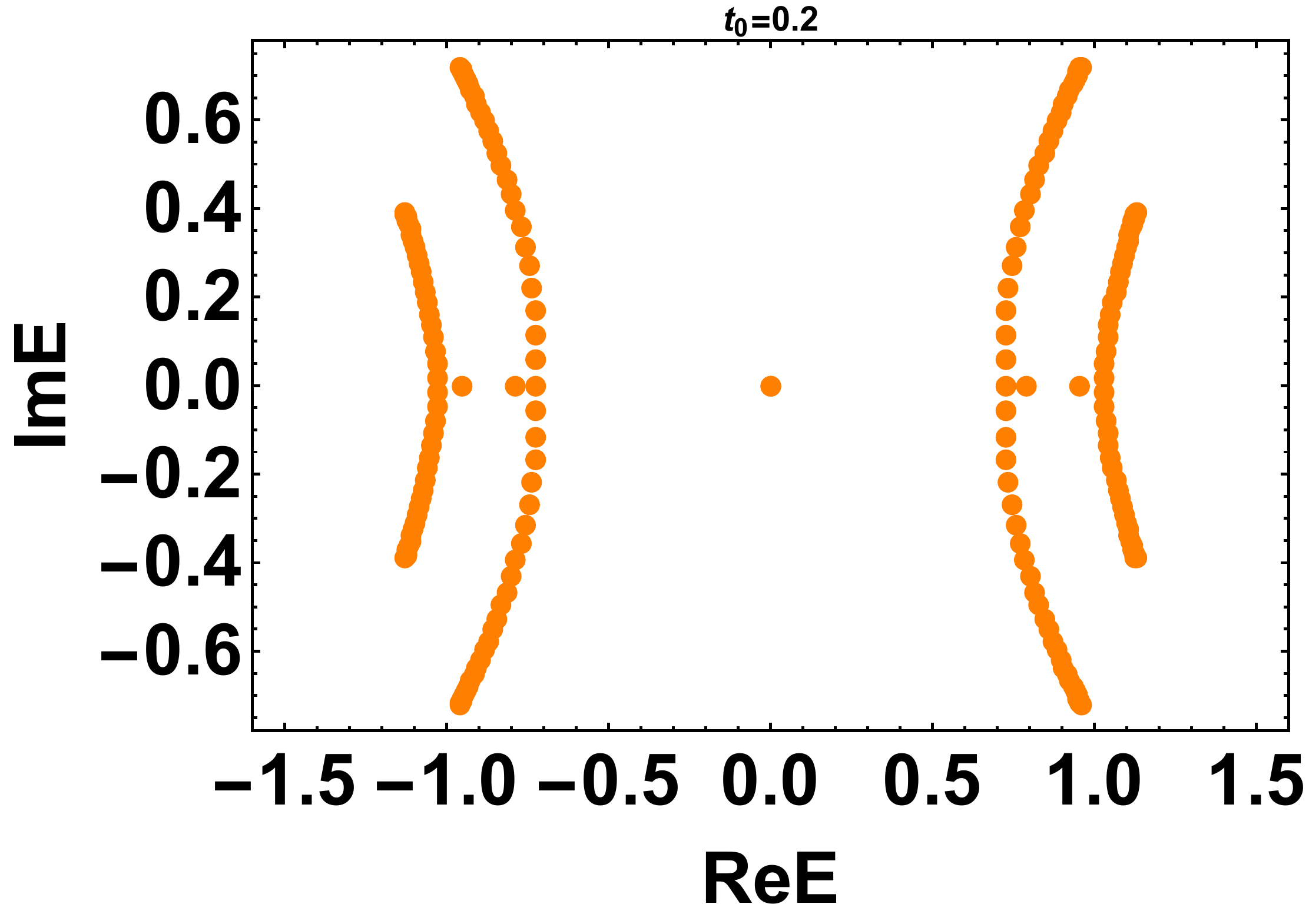}}
	\subfigure[]{\includegraphics[width=0.30\textwidth]{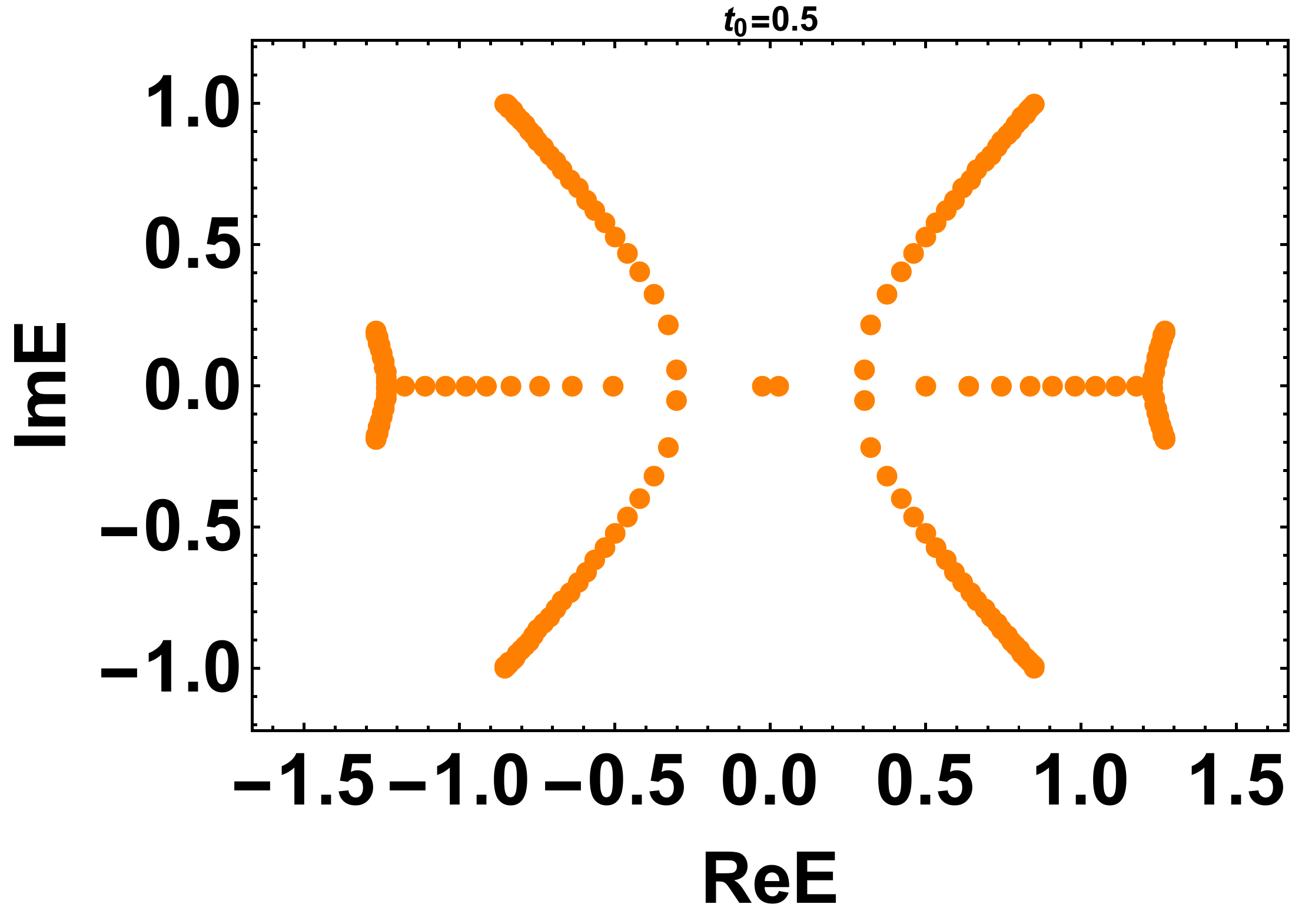}}
	\subfigure[]{\includegraphics[width=0.30\textwidth]{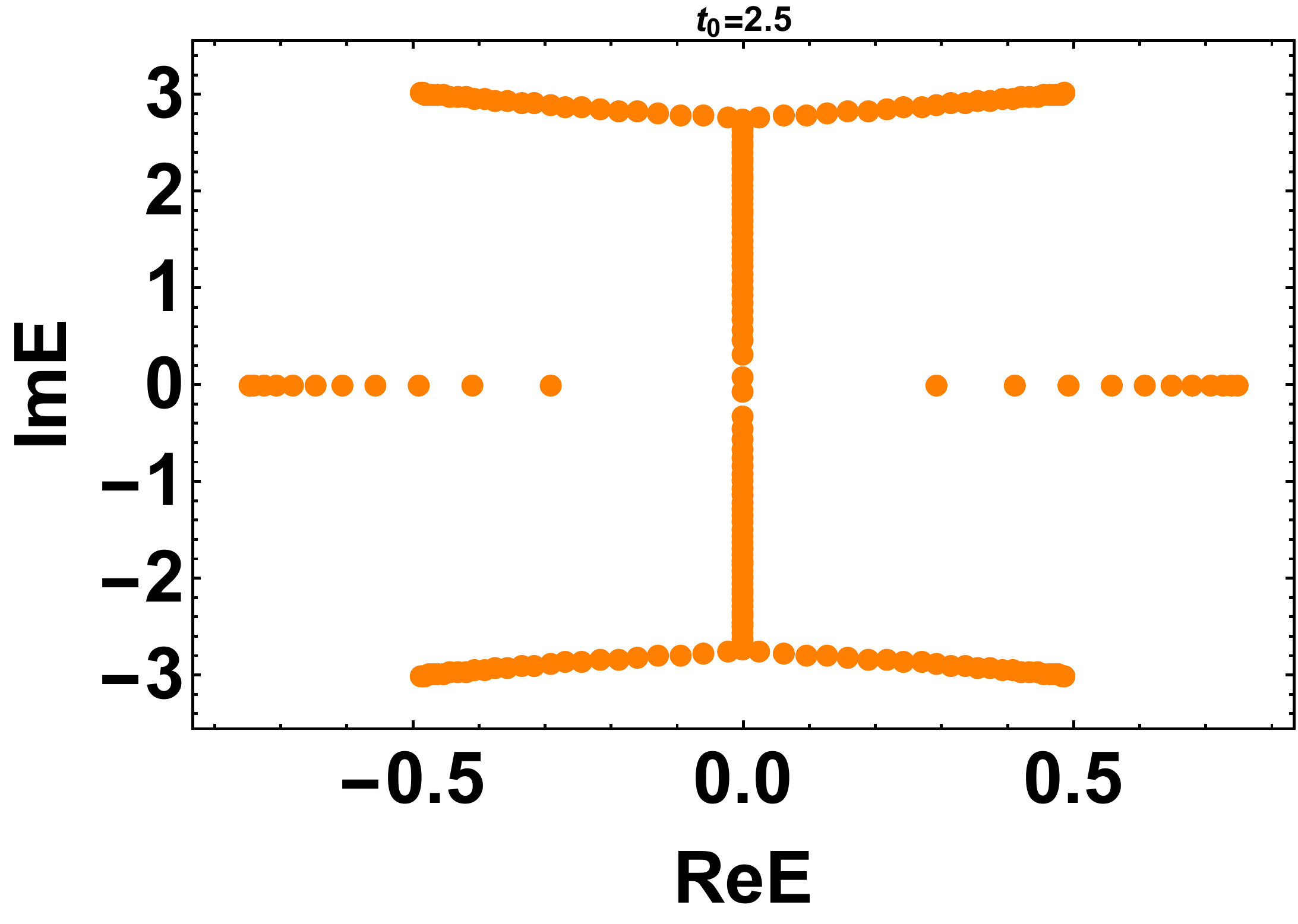}}\\
	\subfigure[]{\includegraphics[width=0.31\textwidth]{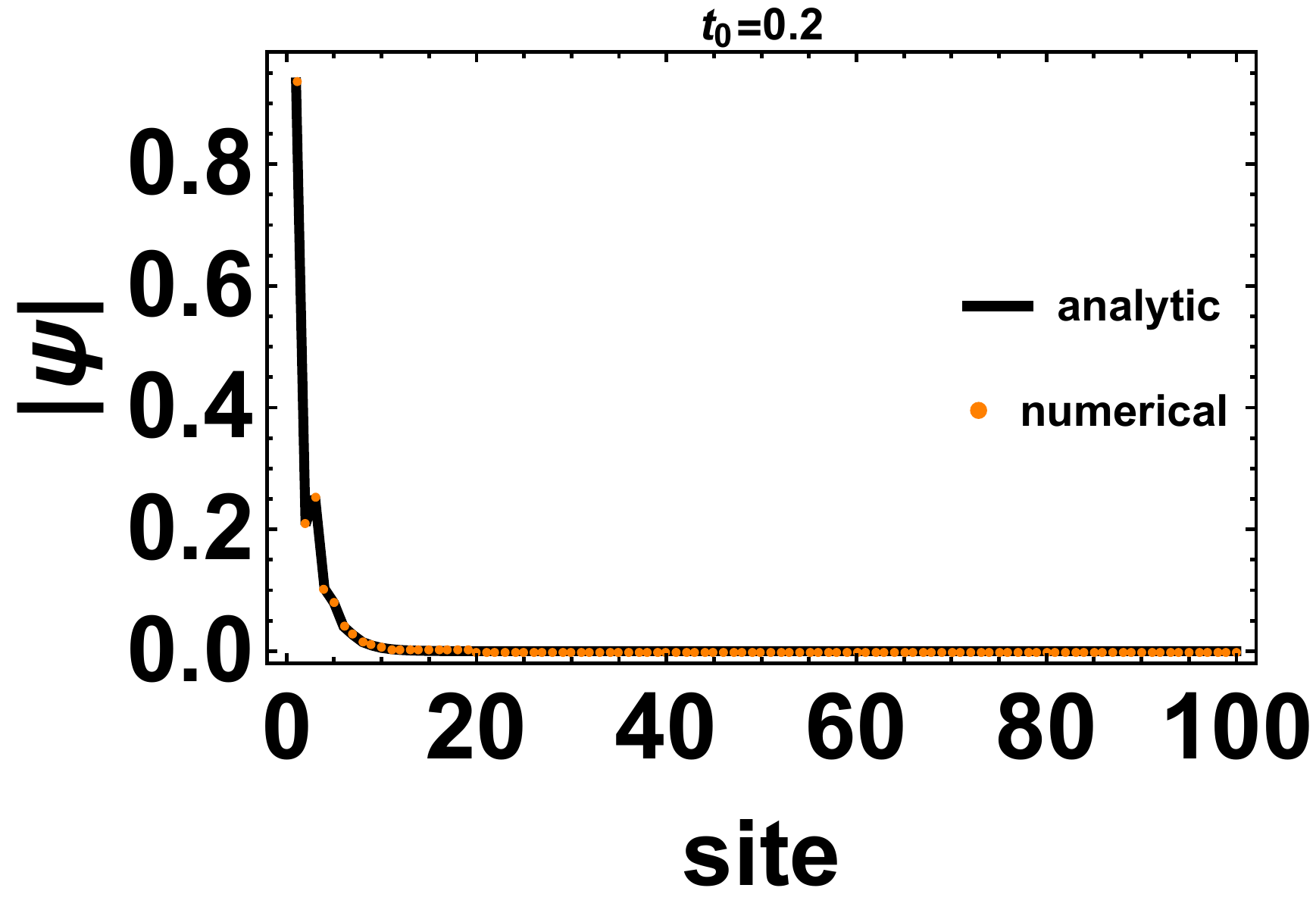}}
	\subfigure[]{\includegraphics[width=0.31\textwidth]{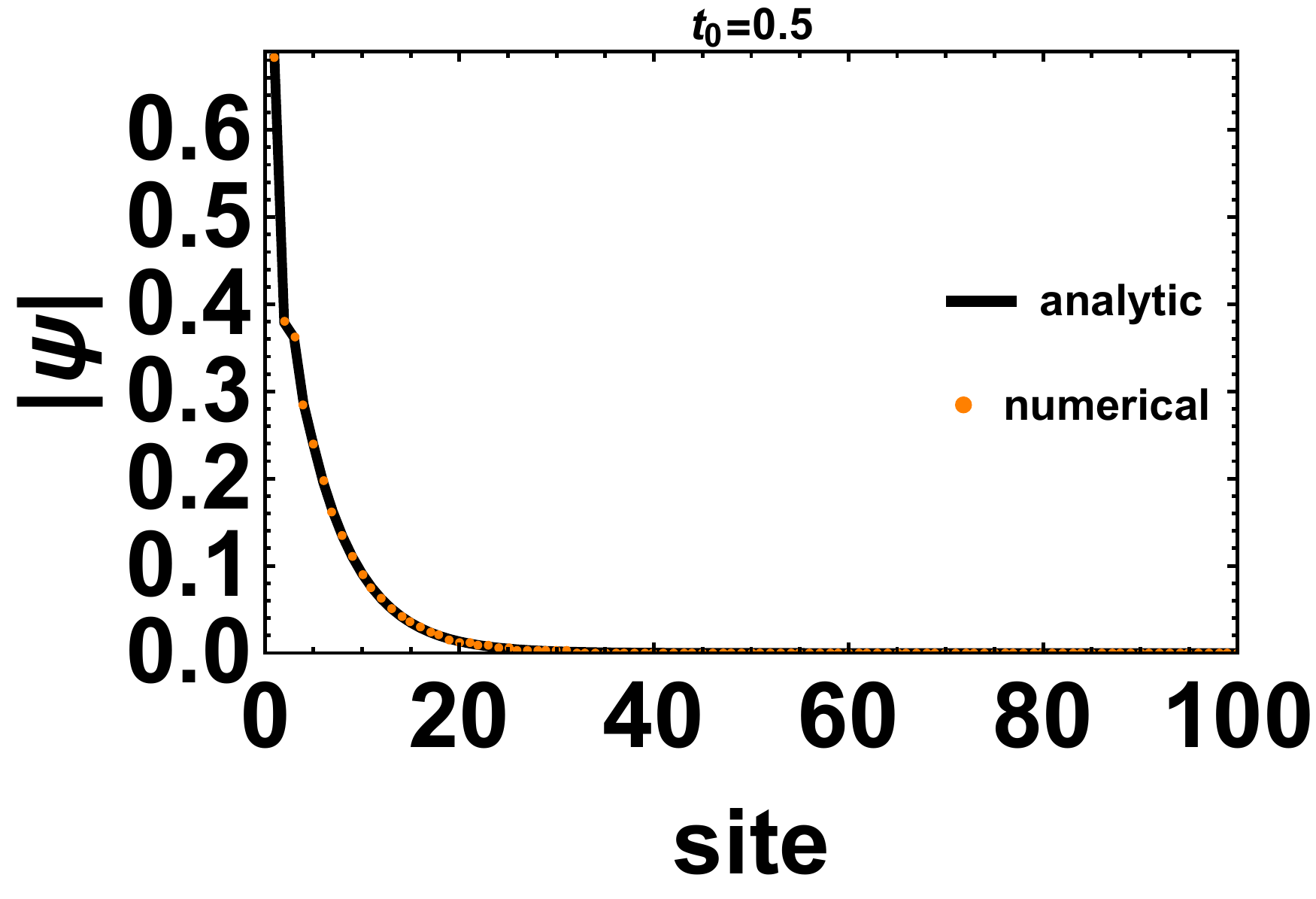}}
	\subfigure[]{\includegraphics[width=0.31\textwidth]{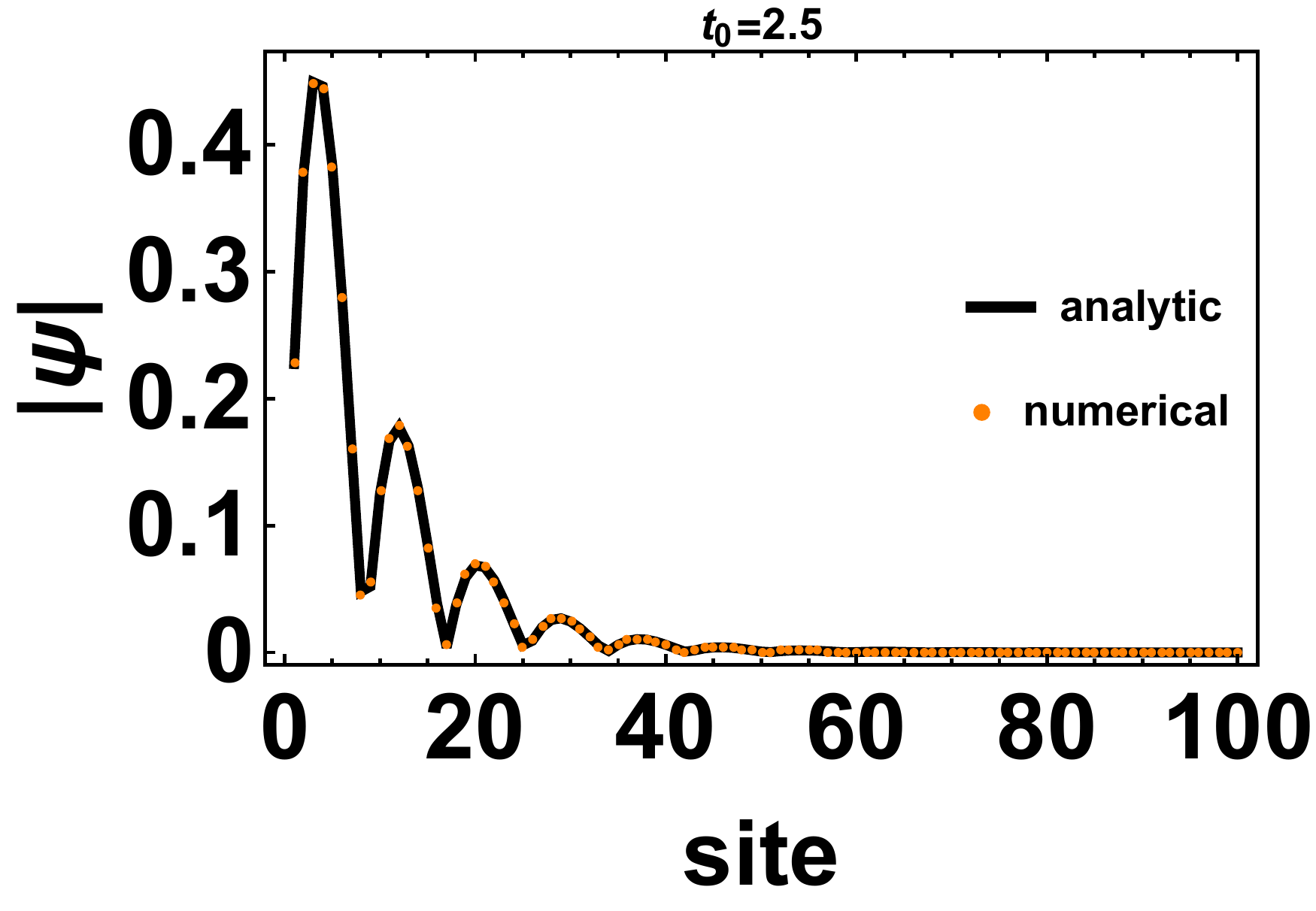}}
	\caption{(a) The square of the norms of the four solutions in Eq.~(\ref{solus}), where $\lambda_{1}\approx0.581378$, $\lambda_{2}\approx0.7$, and $\lambda_{3}\approx2.68328$. When $-\lambda_{3}\leq t_{0}\leq\lambda_{3}$, $|\beta_{1}|^{2}=|\beta_{2}|^{2}=|\beta_{12}|^{2}=0.8$~(brown line). The complex energy spectra of the model in Eq.~(\ref{gssh}) with $t_{0}=0.2$, $t_{0}=0.5$, and $t_{0}=2.5$, are shown in (b), (c), and (d), respectively. The distribution of the analytic eigenstate~(black line) and the numerical eigenstate~(orange dots) of zero energy with $t_{0}=0.2$, $t_{0}=0.5$, and $t_{0}=2.5$, are shown in (e), (f), and (g), respectively. The number of the lattice sites is 100.}
	\label{fig1}
\end{figure*}

\section{Typical 1D non-Hermitian models with band degeneracy}
\label{3}
We apply our general theory to typical 1D non-Hermitian models under OBCs, which possess band degeneracy of IEBs or CEBs. We consider the general 1D two-band non-Hermitian model, whose non-Bloch Hamiltonian over the GBZ is
\begin{eqnarray}
H(\beta)=\begin{bmatrix}
h_{1}(\beta)&h_{+}(\beta)\\
h_{-}(\beta)&h_{2}(\beta)
\end{bmatrix}.
\end{eqnarray}
The two energy bands of this model are
\begin{eqnarray}
E_{\pm}(\beta)=\frac{1}{2}\Big[h_{p}(\beta)\pm\sqrt{h_{m}(\beta)^{2}+4h_{+}(\beta)h_{-}(\beta)}\Big],
\end{eqnarray}
where $h_{p}(\beta)=h_{1}(\beta)+h_{2}(\beta)$ and $h_{m}(\beta)=h_{1}(\beta)-h_{2}(\beta)$. If $h_{p}(\beta)$ is independent of $\beta$, the characteristic equations of the two energy bands, $f(E_{+},\beta)=0$ and $f(E_{-},\beta)=0$, are the same. Consequently, the two sub-GBZs of the two sub-CEBs are the same, namely, degenerate sub-GBZs. In this section, we study two-band models with and without degenerate sub-GBZs, respectively.   

\subsection{The degenerate zero-energy modes of the generalized non-Hermitian SSH model}
Consider the Hamiltonian of the generalized SSH model in real space,
\begin{eqnarray}
\label{gssh}
\hat{H}_{c}=\sum_{x}\Big(T_{0}\ket{x}\bra{x}+T_{1}\ket{x}\bra{x+1}+T_{-1}\ket{x+1}\bra{x}\Big),\nonumber\\
\end{eqnarray}
where $T_{0}=\begin{bmatrix}
0&-t_{0}\\
t_{0}&0
\end{bmatrix}$, 
$T_{1}=\begin{bmatrix}
0&t_{1}^{+}\\
t_{1}^{-}&0
\end{bmatrix}$,
$T_{-1}=\begin{bmatrix}
0&t_{-1}^{+}\\
t_{-1}^{-}&0
\end{bmatrix}$, and the number of lattice sites is $N$. 

Due to the chiral symmetry, the two energy bands of this model are only possibly degenerate at zero energy. By the characteristic equation respect to zero energy of this model, we obtain the four solutions of $\beta$,
\begin{eqnarray}
\label{solus}
&&\beta_{1,2}=\frac{1}{2t_{1}^{+}}\big(t_{0}\pm\sqrt{t_{0}^{2}-4t_{1}^{+}t_{-1}^{+}}\big),\nonumber\\
&&\beta_{3,4}=\frac{1}{2t_{1}^{-}}\big(-t_{0}\pm\sqrt{t_{0}^{2}-4t_{1}^{-}t_{-1}^{-}}\big).
\end{eqnarray}  
The non-Bloch Hamiltonian over the GBZ is
\begin{eqnarray}
\label{nonbloch}
H_{c}(\beta)=\begin{bmatrix}
0&h_{+}(\beta)\\
h_{-}(\beta)&0
\end{bmatrix},
\end{eqnarray}
where $h_{+}(\beta)=-t_{0}+t_{1}^{+}\beta+t_{-1}^{+}\beta^{-1}$ and $h_{-}(\beta)=t_{0}+t_{1}^{-}\beta+t_{-1}^{-}\beta^{-1}$. Since $h_{p}(\beta)=0$, the two sub-GBZs of the two CEBs are degenerate; in other words, there is only one GBZ for this model. The boundary matrix respect to zero energy of this model is
\begin{eqnarray}
\label{boundssh}
B(0)=\begin{bmatrix}
t_{-1}^{+}&t_{-1}^{+}&0&0\\
0&0&t_{-1}^{-}&t_{-1}^{-}\\
t_{1}^{+}\beta_{1}^{N}&t_{1}^{+}\beta_{2}^{N}&0&0\\
0&0&t_{1}^{-}\beta_{3}^{N}&t_{1}^{-}\beta_{4}^{N}
\end{bmatrix}.
\end{eqnarray}

We set the parameters as $t_{1}^{+}=1.5, t_{1}^{-}=0.9,t_{-1}^{+}=1.2,t_{-1}^{-}=-0.2$. We plot the square of the norms of the four solutions in Fig.~\ref{fig1}(a). Notice that, $|\beta_{1}|^{2}=|\beta_{2}|^{2}=|\beta_{12}|^{2}=0.8$~[brown line in Fig.~\ref{fig1}(a)] when $t_{0}^{2}\leq4t_{1}^{+}t_{-1}^{+}$, namely, $-\lambda_{3}\leq t_{0}\leq\lambda_{3}$ with $\lambda_{3}=\sqrt{4t_{1}^{+}t_{-1}^{+}}\approx2.68328$. The $t_{0}$ coordinate of the intersection of curves $|\beta_{12}|^{2}$ and $|\beta_{3}|^{2}$~($|\beta_{4}|^{2}$) can be calculated by solving $|\beta_{3}|^{2}=0.8$~($|\beta_{4}|^{2}=0.8$), resulting in $\lambda_{1}\approx0.581378$~($-\lambda_{1}$). The $t_{0}$ coordinate of the intersection of curves $|\beta|^{2}=1$ and $|\beta_{3}|^{2}$~($|\beta_{4}|^{2}$) is $\lambda_{2}\approx0.7$~($-\lambda_{2}$). We number these solutions by their norms. In the thermodynamics limit, when the norms of the middle of the two solutions are not equal, the two IEBs degenerate at zero energy, namely, the topological edge modes, are present~(absent), resulting in a topological~(trivial) phase; When they are equal, the two CEBs are degenerate at zero energy, resulting in a semimetal phase.

When $0\leq t_{0}<\lambda_{1}$, that is $|\beta_{4}|^{2}\leq|\beta_{3}|^{2}<|\beta_{1}|^{2}=|\beta_{2}|^{2}<1$, the degenerate zero energies are IEBs, resulting in a topological phase. In the thermodynamics limit, we obtain two eigenstates $\alpha_{1}=(1,-1,0,0)^{T}$ and $\alpha_{2}=(0,0,1,-1)^{T}$ by the boundary matrix~(Appendix \ref{detail1}). When $-\lambda_{1}< t_{0}\leq 0$, that is $|\beta_{3}|^{2}\leq|\beta_{4}|^{2}<|\beta_{1}|^{2}=|\beta_{2}|^{2}<1$, we obtain the same result. Noteworthily, by Eq.~(\ref{exp}) in Sec.~\ref{iso}, the two zero-energy topological modes displace exponentially from zero energy for finite system size $N$, and they are degenerate at zero energy in the thermodynamics limit~(Appendix \ref{detail1}).

When $\lambda_{1}\leq t_{0}\leq\lambda_{3}$, that is $|\beta_{4}|^{2}<|\beta_{1}|^{2}=|\beta_{2}|^{2}\leq|\beta_{3}|^{2}$, the degenerate zero energies are CEBs, resulting in a semimetal phase. In the thermodynamics limit, the CEBs correspond to the GBZ, where $\beta_{1}$ and $\beta_{2}$~(and $\beta_{3}$ if $t_{0}=\lambda_{1}$) lie. Although the non-Bloch Hamiltonian Eq.~(\ref{nonbloch}) is defective in the interval $\lambda_{1}<t_{0}\leq\lambda_{3}$~\cite{yokomizo2020}, the defectiveness of eigenstates is determined by the boundary matrix. Hence, whether a degenerate energy of CEBs is an EP under OBCs, is not determined by the non-Bloch Hamiltonian but the kernel of the boundary matrix. In the interval $\lambda_{1}\leq t_{0}<\lambda_{2}$, $|\beta_{4}|^{2}<|\beta_{1}|^{2}=|\beta_{2}|^{2}\leq|\beta_{3}|^{2}<1$, and we obtain two zero-energy eigenstates by the boundary matrix; thus, EPs do not exist~(Appendix \ref{detail1}). However, in the interval $\lambda_{2}\leq t_{0}\leq\lambda_{3}$, $|\beta_{4}|^{2}<|\beta_{1}|^{2}=|\beta_{2}|^{2}<1\leq|\beta_{3}|^{2}$, and we obtain only one eigenstates by the boundary matrix, thus the points at this interval are EPs~(Appendix \ref{detail1}). When $-\lambda_{3}\leq t_{0}\leq-\lambda_{1}$, $|\beta_{3}|^{2}<|\beta_{1}|^{2}=|\beta_{2}|^{2}\leq|\beta_{4}|^{2}$, and we obtain the same results as above. We plot the complex energy spectra with $t_{0}=0.2$, $t_{0}=0.5$, and $t_{0}=2.5$ in Figs.~\ref{fig1}(b)-\ref{fig1}(d), respectively, where the emergence of imprecise zero-energy modes near the origin is attributed to the finite system size $N$ in the numerical calculations and they tend to zero energy with $N\rightarrow\infty$. We also plot the distribution of the analytic eigenstate and the numerical eigenstate of zero energy with $t_{0}=0.2$, $t_{0}=0.5$, and $t_{0}=2.5$ in Figs.~\ref{fig1}(e)-\ref{fig1}(g), respectively, and the analytic- and the numerical-results are consistent with each other~(Appendix~\ref{detail1}). When $t_{0}>\lambda_{3}$ and $t_{0}<-\lambda_{3}$, the degenerate zero-energy modes are absent, due to the absence of the exponentially displaced IEBs $\Delta E$~[Eq.~(\ref{exp})] with finite $N$ and zero-energy CEBs, resulting in a trivial insulator phase. 

\begin{figure*}
	\centering
	\subfigure[]{\includegraphics[width=0.3\textwidth]{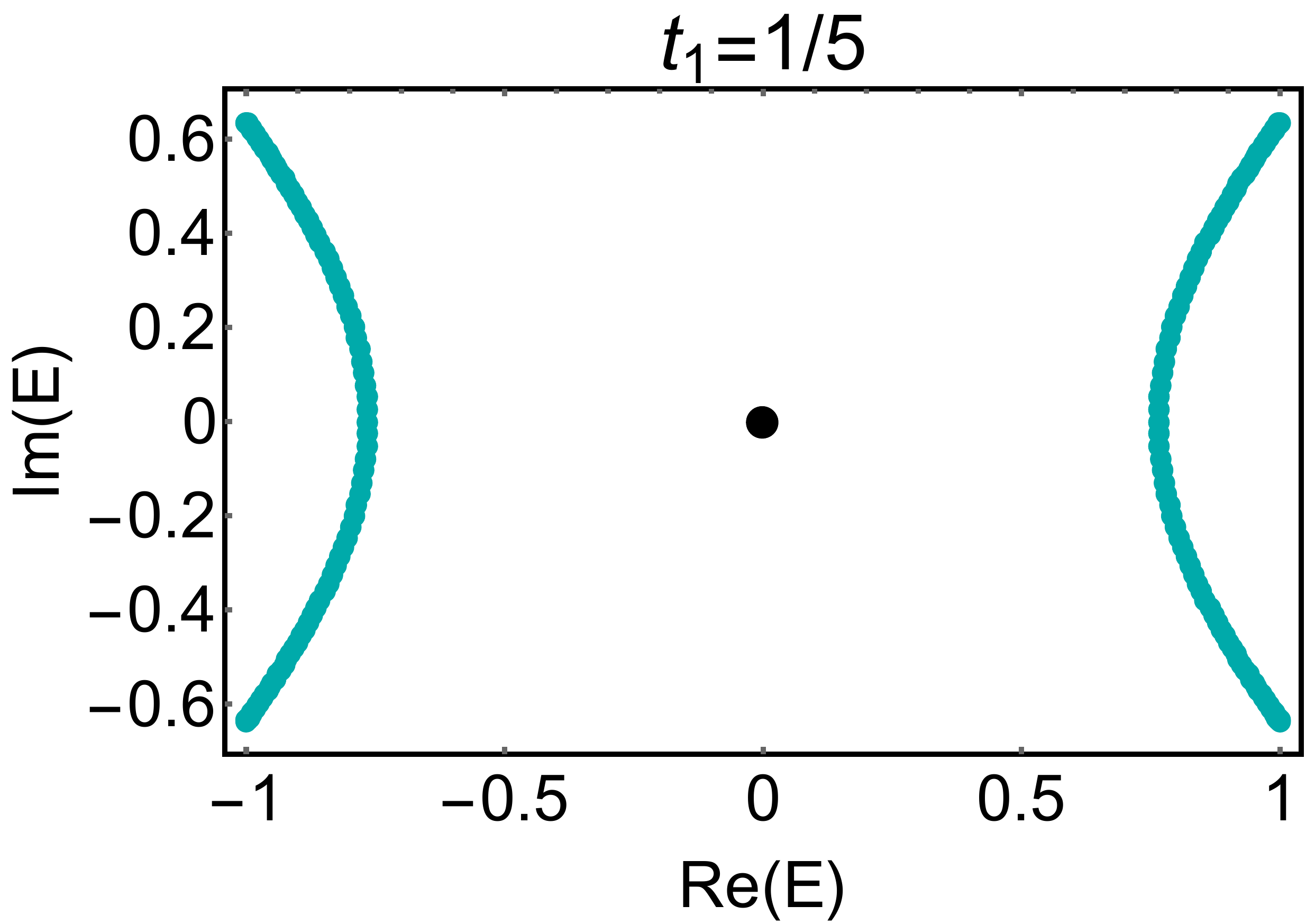}}
	\subfigure[]{\includegraphics[width=0.3\textwidth]{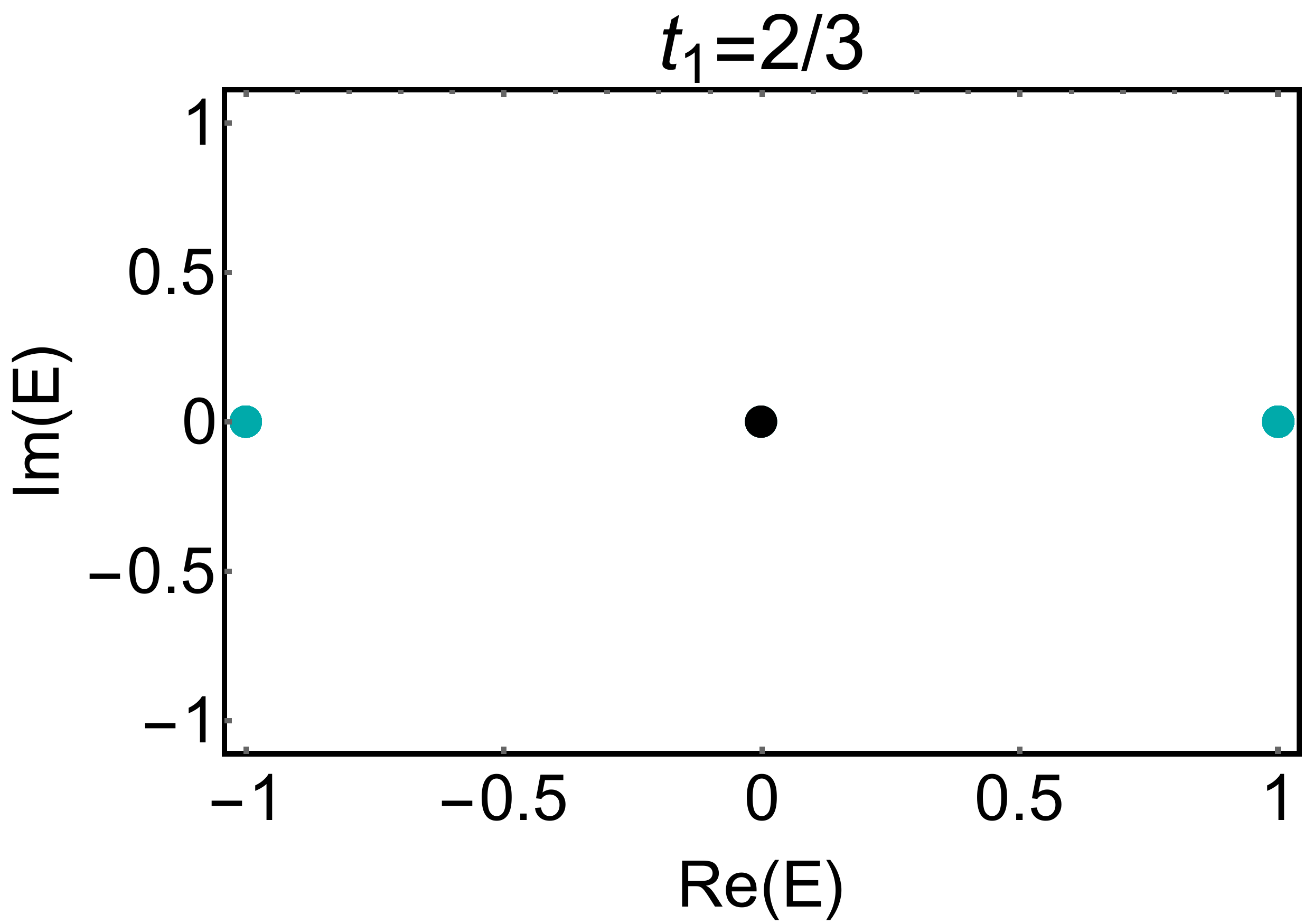}}
	\subfigure[]{\includegraphics[width=0.3\textwidth]{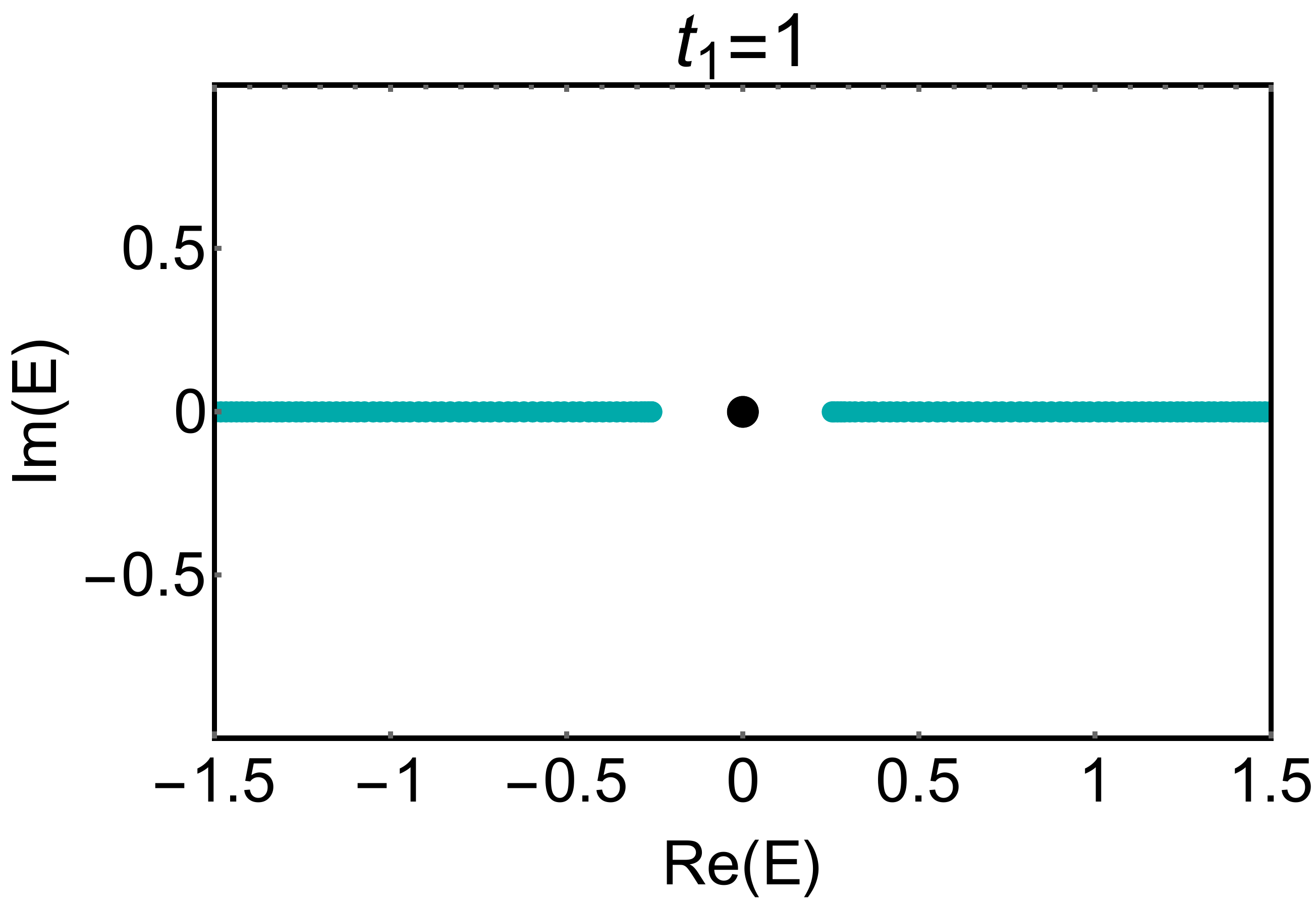}}
	\caption{The complex energy spectra of the non-Hermitian SSH model~[Eq.~(\ref{ssh})], with parameters $\gamma=2/3$ and $t_{2}=1$ under OBCs. The two CEBs~(cyan lines) and IEBs~(black dot) with $t_{1}=1/5$ and $t_{1}=1$, are shown in (a) and (c), respectively. The two infernal points~(cyan dots) and IEBs~(black dot) with $t_{1}=2/3$ are shown in (b) .} 
	\label{fig2}
\end{figure*}

\subsection{The non-perturbation of the two-band model with non-degenerate sub-GBZs}
Consider other two-band model with non-degenerate sub-GBZs~\cite{yang202002}. The non-Bloch Hamiltonian of this model reads 
\begin{eqnarray}
\label{twoband}
H_{T}(\beta)=\begin{bmatrix}
h_{1}(\beta)&c\\
c&h_{2}(\beta)
\end{bmatrix},
\end{eqnarray}
where $h_{1}(\beta)=a_{0}+a_{1}\beta+a_{-1}\beta^{-1}$ and $h_{2}(\beta)=b_{0}+b_{1}\beta+b_{-1}\beta^{-1}$. This model decouples into two separate single-band models with $c=0$, but it is non-perturbative with a tiny $c$~\cite{yang202002}.

We interpret the non-perturbation based on the theory of band degeneracy. When $c=0$, the CEBs of the two separate single-band models, $h_{1}(\beta)$ and $h_{2}(\beta)$, correspond to the circular GBZs, $C_{GBZ}^{1}$ and $C_{GBZ}^{2}$, with radius $r_{1}=\sqrt{|\frac{a_{-1}}{a_{1}}|}$ and $r_{2}=\sqrt{|\frac{b_{-1}}{b_{1}}|}$, respectively. Although the two GBZs are separate as long as $r_{1}\neq r_{2}$, a part of the two CEBs is superposed, in the region satisfying $h_{1}(\beta_{1})=h_{2}(\beta_{2})$ with $\beta_{1}\in C_{GBZ}^{1}$ and $\beta_{2}\in C_{GBZ}^{2}$. The theory of band degeneracy seems to be invalid in this case. However, the two GBZs are obtained by choosing the solutions of $h_{1}(\beta)-E=0$ and $h_{2}(\beta)-E=0$, satisfying equal norms separately. The characteristic equations of the two separate models  $h_{1}(\beta)$ and $h_{2}(\beta)$ are irrelevant; hence the band degeneracy of them makes no sense. 

When $c\neq0$, the two separate models are coupled. The two CEBs are
\begin{eqnarray}
E_{\pm}(\beta)=\frac{1}{2}\big[h_{p}(\beta)\pm\sqrt{c^{2}+h_{m}(\beta)^{2}}\big],
\end{eqnarray} 
where $h_{p}(\beta)=h_{1}(\beta)+h_{2}(\beta)$ and $h_{m}(\beta)=h_{1}(\beta)-h_{2}(\beta)$. Since $h_{p}(\beta)$ is dependent on $\beta$, the two CEBs correspond to different sub-GBZs, $\beta_{GBZ}^{1}$ and $\beta_{GBZ}^{2}$, respectively. If $c$ is tiny enough, the two CEBs tend to be degenerate at one point $E_{0}$, of which the corresponding point on the GBZs is the intersection point of the two sub-GBZs. We number the solutions of the characteristic equation $\det{[H_{T}(\beta)-E_{0}]}=0$ as $|\beta_{1}^{T}|\leq|\beta_{2}^{T}|\leq|\beta_{3}^{T}|\leq|\beta_{4}^{T}|$. Consequently, we obtain $|\beta_{2}^{T}|=|\beta_{3}^{T}|$. In addition, the band degeneracy is sensitive to the magnitude of $c$, that is, the degenerate point disappears if $c$ is not tiny enough.

\section{Infernal points and infernal knots in non-Hermitian systems}
\label{4}
\subsection{Infernal points in 1D non-Hermitian systems}
The band degeneracy of the CEBs is based on the theory of GBZs~(non-Bloch band theory), which depicts the continuous part of the energy spectra. However, there possibly exist the points in the parameter space of a 1D non-Hermitian system, where the CEBs under OBCs converge on some discrete energy values. Since the energy spectra at these points are not continuous anymore, the theory of GBZs, as well as the non-Bloch band theory, is invalid. We define the points, where the number of degenerate bands scales with the large enough system size $N$ under OBCs, as the infernal points, and we can calculate the energy values at infernal points by the formal theory constructed in Appendix~\ref{infernal}.

Consider the well-known non-Hermitian SSH model~\cite{lee2016,kunst2018,yao201801,imura2019}, whose Hamiltonian in real space reads
\begin{eqnarray}
\label{ssh}
&&\hat{H}_{nssh}=\sum_{x=1}^{N}\big[(t_{1}+\gamma)\ket{x,A}\bra{x,B}+(t_{1}-\gamma)\ket{x,B}\bra{x,A}\big]\nonumber\\
&&\qquad\quad+\sum_{x=1}^{N-1}t_{2}\big[\ket{x,B}\bra{x+1,A}+\ket{x+1,A}\bra{x,B}\big],
\end{eqnarray}
where $N$ is the number of lattice sites, and $A,B$ denote the sub-lattices. The skin effect and topological invariant have been studied in previous works~\cite{lee2016,kunst2018,yao201801,imura2019}. The points $t_{1}=\pm\gamma$ are the infernal points, where the two CEBs~[cyan lines in Fig.~\ref{fig2}(a) and Fig.~\ref{fig2}(c)] in other parameter points converge on two discrete energy values $E=\pm t_{2}$[cyan dots in Fig.~\ref{fig2}(b)], respectively. We can obtain two eigenstates, $\psi_{\pm t_{2}}$, with respect to the two energies by directly solving the Schr\"{o}dinger equation under OBCs~(Appendix \ref{infernal}). In addition, the two IEBs~(topological zero modes) are degenerate at the zero energy~(black dot in Fig.~\ref{fig2}) with only one edge state~(Appendix \ref{infernal}), which is an EP respect to $E=0$. 

In the region $t_{1}\neq\pm\gamma$, the two CEBs are generated by running around the GBZ~(two degenerate sub-GBZs), that is, the number of energy values on the two CEBs is corresponding to the number of $\beta$ values on GBZ. When $N$ is large enough and the theory of GBZs is valid, there are $inf=\frac{1}{2}(N-2)$ energy values on each two CEBs and two IEBs degenerate at zero energy, totally $2N$ energy values under OBC. Hence, when $t_{1}$ evolves from $t_{1}\neq\pm\gamma$ to $t_{1}=\pm\gamma$ continuously, the $inf$ energy values on each CEBs converge on the discrete two $E=\pm t_{2}$, with one eigenstate for each energy value, which is why we call the points $t_{1}=\pm\gamma$ the infernal points. If we transform the matrix form at $t_{1}=\pm\gamma$ of this model under OBCs to the Jordan block form, the algebra multiplier and geometric multiplier are $2N$ and $3$, respectively. When $N$ is very large, the Jordan block form of this model is an extremely defective matrix.

We consider another 1D four-band model, whose Hamiltonian in real space reads
\begin{eqnarray}
\label{fourband}
\hat{H}_{F}=\sum_{x}\Big(M_{0}\ket{x}\bra{x}+T_{+}\ket{x}\bra{x+1}+T_{-}\ket{x+1}\bra{x}\Big),\nonumber\\
\end{eqnarray}
where $M_{0}=t\tau_{0}\sigma_{x}+i\tau_{0}\sigma_{y}$, $T_{+}=\frac{1}{2}(\tau_{x}\sigma_{0}-i\tau_{y}\sigma_{z})$, $T_{-}=\frac{1}{2}(\tau_{x}\sigma_{0}+i\tau_{y}\sigma_{z})$, and the number of lattice sites is $N$.
The infernal points emerge at $t=\pm1$, with the discrete energies $E=\pm1$ and four IEBs degenerate at zero energy~(Appendix \ref{infernal}). However, there are four edge states with respect to $E=0$, which is not an EP, and $N$ eigenstates at each of the energy $E=\pm1$~(ppendix \ref{infernal}). The Jordan block form of the matrix at $t=\pm1$ of the four-band model is also an extremely defective matrix, of which the algebra multiplier and geometric multiplier are $4N$ and $2N+4$, respectively. 

We emphasize that the mismatch between algebraic multiplier and geometric multiplier at infernal points is dependent on concrete non-Hermitian models. For the non-Hermitian SSH model and the four-band model given above, the algebraic multiplier and geometric multiplier at infernal points under OBCs are mismatched scaling with the system size. However, there exist models of which the algebraic multiplier and geometric multiplier at infernal points under OBCs are equal, i.e., a diagonalizable Hamiltonian under OBCs. For example, consider a 1D non-Hermitian model, whose Hamiltonian in real space is 
\begin{eqnarray}
\hat{H}=\sum_{x}\Big(t_{x}^{0}\ket{x}\bra{x}+t^{+}_{x}\ket{x}\bra{x+1}+t^{-}_{x}\ket{x+1}\bra{x}\Big),\nonumber\\
\end{eqnarray}
with $t_{x}^{0}=t_{1}\sigma_{x}+i\gamma\sigma_{y}$, $t^{+}_{x}=\frac{1}{2}(\sigma_{x}-i\sigma_{z})$, $t^{-}_{x}=\frac{1}{2}(\sigma_{x}+i\sigma_{z})$, and $N$ lattice sites. The infernal point of this model is $t_{1}=0$, where there are each $N-1$ bands degenerate at energy value $\sqrt{1-\gamma^{2}}$ and $-\sqrt{1-\gamma^{2}}$, but the Hamiltonian at this point under OBCs is diagonalizable. The infernal points can also apply to Hermitian systems, but the algebraic multiplier and geometric multiplier at the infernal points under OBCs are always equal due to Hermiticity. Note that the infernal points in Hermitian systems are also the flat bands, where the Bloch band theory is valid. We define the infernal points, where the algebraic multiplier and geometric multiplier are mismatched~(matched), as nontrivial~(trivial) infernal points. The nontrivial infernal points are unique to non-Hermitian systems.
\subsection{Infernal knots} 
In the last decade, topological nodal-knot semimetals in Hermitian systems~\cite{knot1,knot2,knot3,knot4,knot5,knot6,knot7,knot8,bergholtzhyper} and non-Hermitian systems~\cite{bergholtz201902,yang2019,li2019,zhang2020,tidal,mandal2021,delplace2021,liu2021,bergholtz2021,ghorashi202101,ghorashi202102}, have been studied in many previous works. The knotted structure of the non-Hermitian bands has also been proposed recently~\cite{jones2020,knot2021}. Based on our research on infernal points, we generalize it to infernal knots in four-dimensional non-Hermitian systems under OBCs. 

A nodal torus knot in momentum space can be constructed as follows~\cite{knot5,bergholtzhyper}. Define a complex polynomial $\mathcal{K}=\xi^{p}+\zeta^{q}$ lying on $S^{3}$, where $\xi$, $\zeta$ are complex variables, satisfying $|\xi|^{2}+|\zeta|^{2}=1$, and $p$, $q$ are coprime integers. Map $(\xi,\zeta)$ to the three-dimensional Brillouin zone, namely, $\xi(k_{x},k_{y},k_{z})$ and $\zeta(k_{x},k_{y},k_{z})$. The combination of the zeros of $\mathcal{K}(k_{x},k_{y},k_{z})$ is a $(p,q)$-torus knot lying on the Brillouin zone. If p and q are not coprime, the zeros of $\mathcal{K}(k_{x},k_{y},k_{z})$ represent torus links with $n = GCD(p,q)$ components of $(\frac{p}{n},\frac{q}{n})$-torus knots.

\begin{figure}
	\centering
	\subfigure[]{\includegraphics[width=0.16\textwidth]{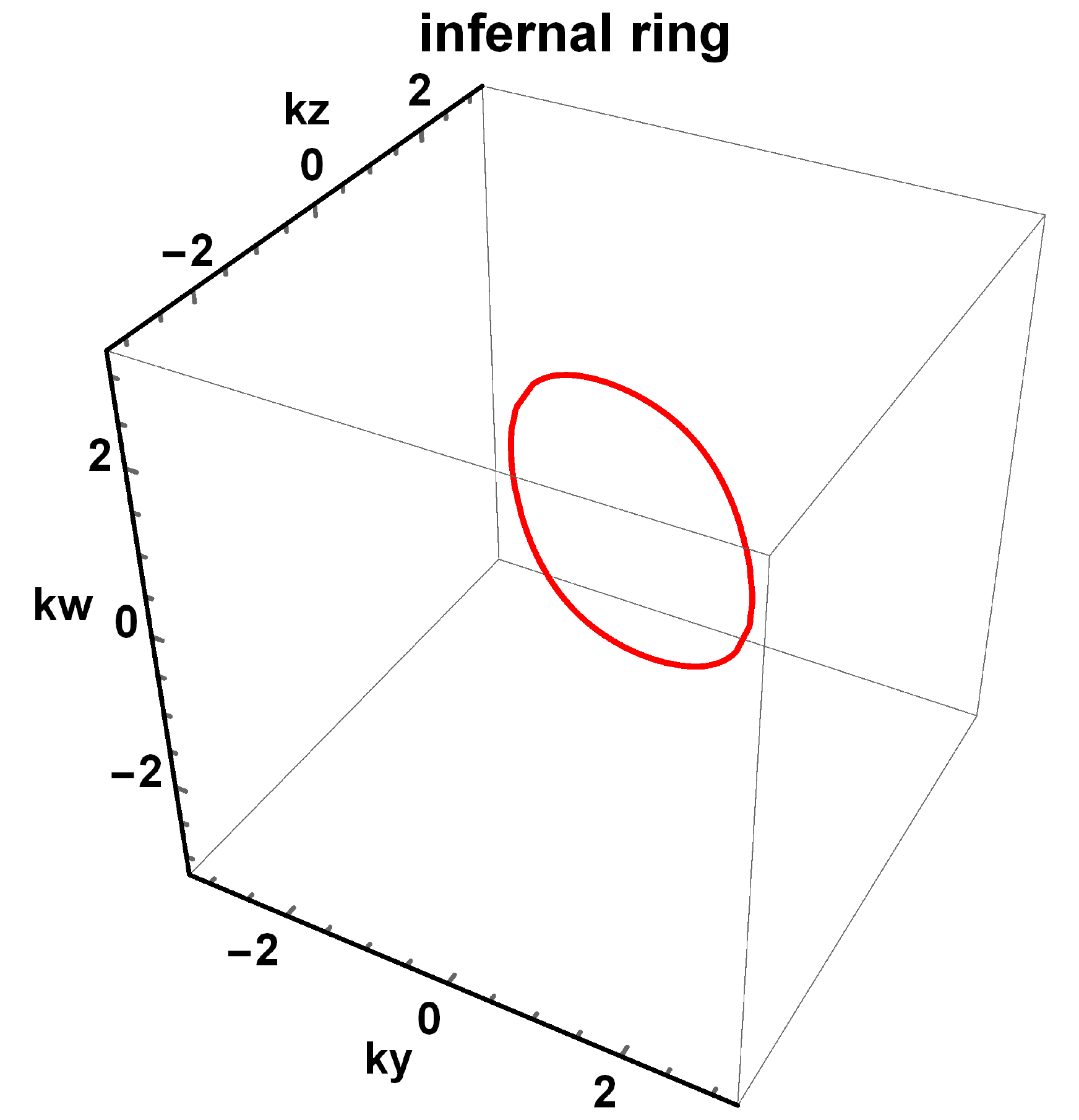}}
	\subfigure[]{\includegraphics[width=0.15\textwidth]{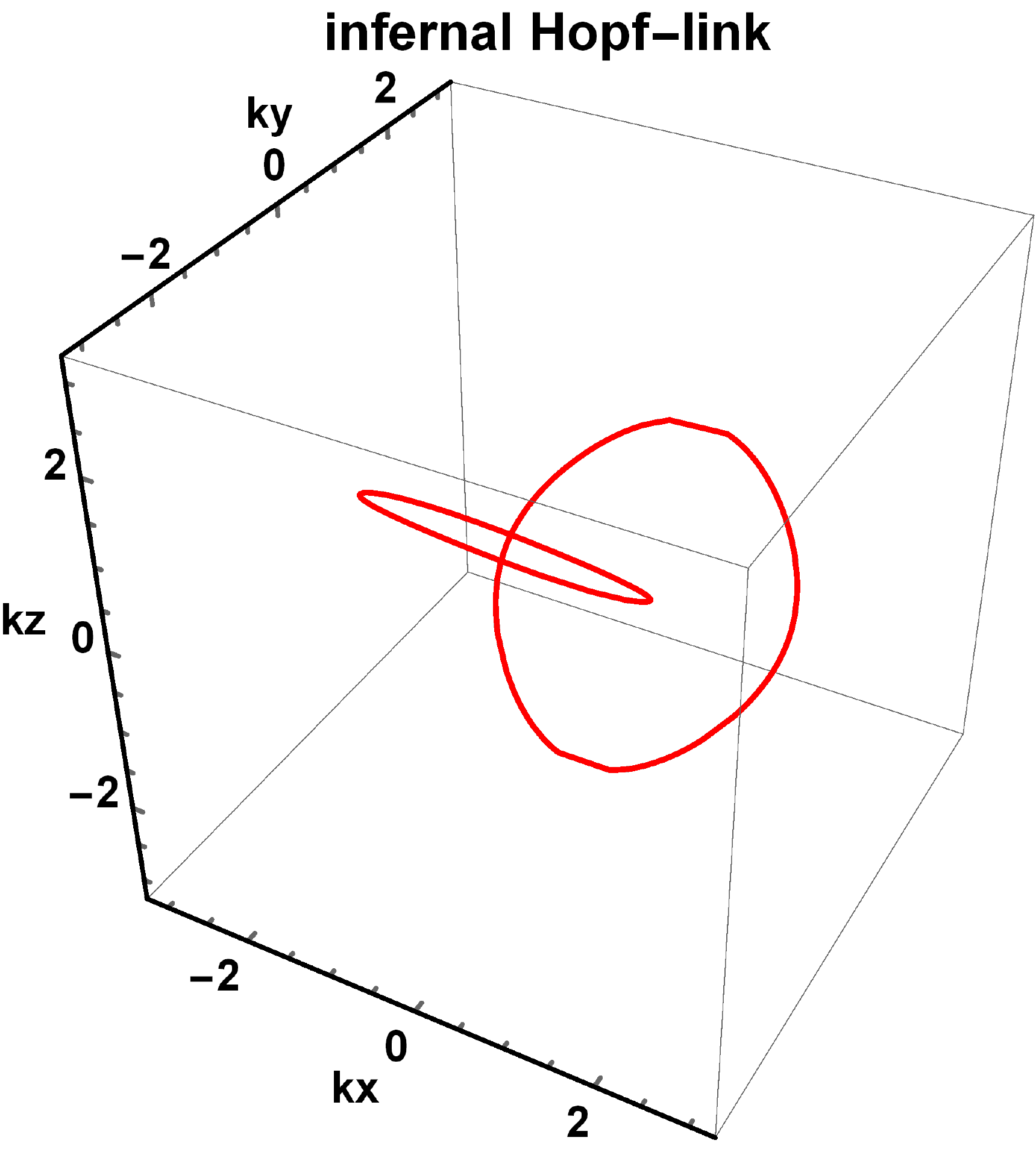}}
	\subfigure[]{\includegraphics[width=0.16\textwidth]{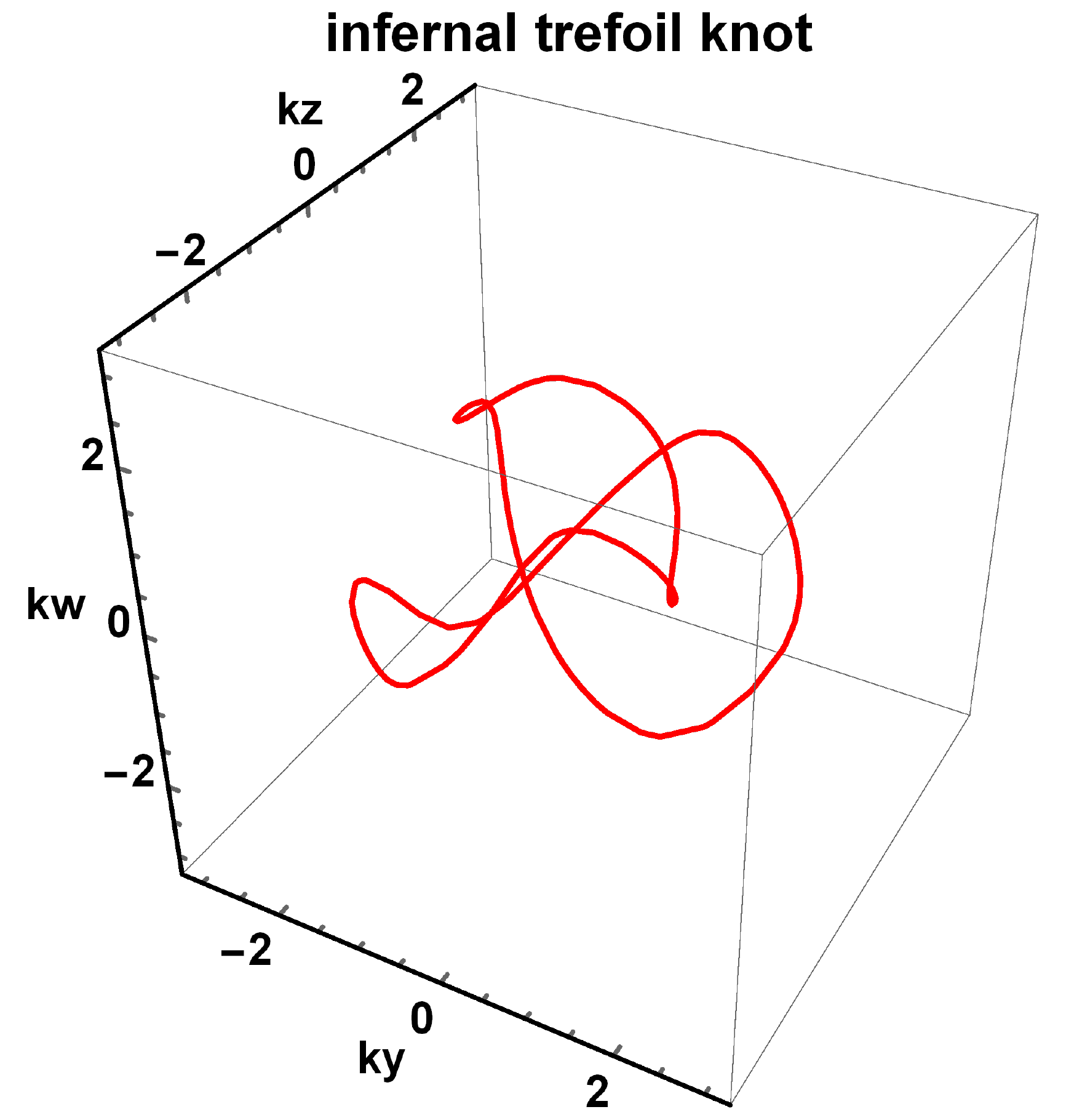}}
	\caption{The infernal ring, Hopf-link, and trefoil knot are shown in (a), (b), and (c), respectively.}
	\label{fig3}
\end{figure}

We generalize the infernal points~($t_{1}=\pm\gamma$) of the non-Hermitian SSH model~[Eq.~(\ref{ssh})] to the infernal knots in four-dimensional systems. If we replace $\gamma$ by $\mathcal{K}(k_{y},k_{z},k_{w})\pm t_{1}$ directly, the infernal points $t_{1}=\pm\gamma$ are replaced by $\mathcal{K}(k_{y},k_{z},k_{w})=0$, thus we obtain infernal torus knot or links in the Brillouin zone. For example, let us take~\cite{knot5}
\begin{eqnarray}
&&\xi(k_{y},k_{z},k_{w})=N_{1}+iN_{2},\nonumber\\
&&\zeta(k_{y},k_{z},k_{w})=N_{3}+iN_{4},
\end{eqnarray}
with
\begin{eqnarray}
&&N_{1}=\sin{k_{y}},N_{2}=\sin{k_{z}},N_{3}=\sin{k_{w}},\nonumber\\
&&N_{4}=\cos{k_{y}}+\cos{k_{z}}+\cos{k_{w}}-2.
\end{eqnarray} 
Define $n_{i}=N_{i}/N$ with $N=\sqrt{\sum_{i=1}^{4}N_{i}^{2}}$. Then, $\vec{n} =(n_{1},n_{2},n_{3},n_{4})$ maps the compactified Brillouin zone to $S^{3}$, and the imagine of $\vec{N} =(N_{1},N_{2},N_{3},N_{4})$ is topologically equivalent to $S^{3}$. Now, the zeros of $\mathcal{K}(k_{y},k_{z},k_{w})=\xi(k_{y},k_{z},k_{w})^{p}+\zeta(k_{y},k_{z},k_{w})^{q}$ represent the infernal torus knots and links with coprime and non-coprime $(p,q)$, respectively. For instance, $(p,q)=(1,1)$, $(2,2)$, and $(2,3)$ represent the infernal ring, Hopf-link, and trefoil knot, respectively~(shown in Fig.~\ref{fig3}). Moreover, we can also generalize the infernal points to two- and three-dimensional systems under OBCs, forming infernal points, lines, or rings.
\section{Discussion and conclusion}
\label{5}
In this paper, we develop a systematically general theory of 1D non-Hermitian systems, elaborating on the energy bands, degeneracy and defectiveness under OBCs. We find that the energy spectrum of non-Hermitian systems is constituted by IEBs and CEBs, and discuss the degeneracy and defectiveness of them in detail. The IEBs correspond to the topological edge modes, while the CEBs correspond to the bulk bands, which are obtained by the GBZs in non-Bloch band theory. As usual, the band degeneracy and defectiveness of the eigenstates do not emerge simultaneously under OBCs, in other words, the existence of EPs is determined by whether the degeneracy and defectiveness both occur. The defectiveness at a degenerate point is determined by the kernel of the boundary matrix. We apply our general theory to two typical 1D non-Hermitian models, and analyze the degeneracy and defectiveness of them. Beyond the general theory, there exist infernal points in some 1D non-Hermitian systems, where the energy spectra under OBCs converge on some discrete energy values. We construct a formal theory to calculate the energy values at the infernal points. We study the infernal points, as well as their eigenstates of two relevant 1D non-Hermitian models analytically. Moreover, we generalize the infernal points to the infernal knots in four-dimensional non-Hermitian systems. Actually, we can generalize the infernal points to any dimensional systems, resulting in infernal points, lines, rings, or surfaces, etc.

\section*{Acknowledgment}
This work was supported by NSFC Grant No.11275180.

\appendix
\begin{widetext}
\section{The bulk eigenstates and bulk equation of general 1D non-Hermitian systems}
\label{bulksolu}
By the characteristic equation $P(E,\beta)=0$ with a fixed eigenenergy $E$ in the main text, we can obtain $M$ nonzero solutions of $\beta$ with multiplier $s_{j},j=1,2,\ldots,M$ and $s_{0}$ zero solutions of $\beta$, and $\sum_{j=1}^{M}s_{j}+2s_{0}=2qR$. The eigenstates for nonzero solutions $\beta_{j}$~\cite{alase2017} are
\begin{eqnarray}
\label{nozerosolu}
\psi_{js}=\sum_{x=1}^{N}\ket{x}\ket{u_{x,js}}=\sum_{x=1}^{N}\ket{x}\sum_{v=1}^{s_{j}}\frac{x^{v-1}}{(v-1)!}\beta_{j}^{x-v+1}\ket{u_{jsv}},\nonumber\\
\end{eqnarray}
where $\ket{u_{jsv}}$ is the component of vector $\ket{u_{js}}=(\ket{u_{js1}},\ket{u_{js2}},\ldots,\ket{u_{jss_{j}}})^{T}$ and $H_{s_{j}}(\beta_{j})\ket{u_{js}}=E\ket{u_{js}}$. The matrix $H_{s_{j}}(\beta_{j})$ is given by 
\begin{eqnarray}
H_{s_{j}}(\beta_{j})=
\begin{bmatrix}
H^{(0)}(\beta_{j})&H^{(1)}(\beta_{j})&\frac{1}{2}H^{(2)}(\beta_{j})&\ldots&\frac{1}{(s_{j}-1)!}H^{(s_{j}-1)}(\beta_{j})\\
0&\ddots&\ddots&\ddots&\vdots\\
\vdots&\ddots&\ddots&\ddots&\frac{1}{2}H^{(2)}(\beta_{j})\\
\vdots&\ddots&\ddots&\ddots&H^{(1)}(\beta_{j})\\
0&\cdots&\cdots&0&H^{(0)}(\beta_{j})
\end{bmatrix},
\end{eqnarray}
where $H^{(v)}(\beta)=\frac{d^{v}}{d\beta^{v}}H(\beta)$. The eigenstates for zero solutions~\cite{alase2017} are
\begin{eqnarray}
\label{zerosolu}
&&\ket{\psi_{s^{-}}}=\sum_{x=1}^{s_{0}}\ket{x}\ket{u_{x,s^{-}}},\nonumber\\
&&\ket{\psi_{s^{+}}}=\sum_{x=1}^{s_{0}}\ket{N-s_{0}+x}\ket{u_{x,s^{+}}},\nonumber\\
&&s^{\pm}=1,2,\ldots,s_{0},
\end{eqnarray}
where $\ket{u_{x,s^{-}}}$ and $\ket{u_{x,s^{+}}}$ are the components of vectors $\ket{u_{s^{-}}}=(\ket{u_{1,s^{-}}},\ket{u_{2,s^{-}}}\ldots,\ket{u_{s_{0},s^{-}}})^{T}$ and $\ket{u_{s^{+}}}=(\ket{u_{1,s^{+}}},\ket{u_{2,s^{+}}}\ldots,\ket{u_{s_{0},s^{+}}})^{T}$, respectively. These two vectors are obtained by equations
\begin{eqnarray}
&&K^{-}_{s_{0}}(E)\ket{u_{s^{-}}}=0,\nonumber\\
&&K^{+}_{s_{0}}(E)\ket{u_{s^{+}}}=0,
\end{eqnarray}
where $K^{-}_{s_{0}}(E)=\tilde{K}^{-}_{s_{0}}(E)$, $K^{+}_{s_{0}}(E)=\left[\tilde{K}^{+}_{s_{0}}(E)\right]^{T}$,
\begin{eqnarray}
\tilde{K}^{\pm}_{s_{0}}(E)=\begin{bmatrix}
K^{(0)}_{\pm}(E,0)&K^{(1)}_{\pm}(E,0)&\frac{1}{2}K^{(2)}_{\pm}(E,0)&\cdots&\frac{1}{(s_{0}-1)!}K^{(s_{0}-1)}_{\pm}(E,0)\\
0&\ddots&\ddots&\ddots&\vdots\\
\vdots&\ddots&\ddots&\ddots&\frac{1}{2}K^{(2)}_{\pm}(E,0)\\
\vdots&\ddots&\ddots&\ddots&K^{(1)}_{\pm}(E,0)\\
0&\cdots&\cdots&0&K^{(0)}_{\pm}(E,0)
\end{bmatrix},
\end{eqnarray}
and $K_{\pm}(E,\beta)=\beta^{R}\left[H(\beta^{\mp1})-E\right]$, with $T$ being the transpose of block elements of the block matrix. In general, the numbers of the eigenstates $\ket{\psi_{s^{\pm}}}$ are $s_{0}^{\pm}$, respectively, which are not always equal and depend on concrete non-Hermitian systems. Accordingly, the two matrices $K^{\pm}_{s_{0}}$ are replaced by $K^{\pm}_{s_{0}^{\pm}}$, respectively. Without loss of generality, we only concentrate on the cases with $s_{0}^{+}=s_{0}^{-}=s_{0}$ in this paper.

We obtain the solution space $\left\{\ket{\psi_{js}},\ket{\psi_{s^{-}}},\ket{\psi_{s^{+}}}\right\}$ of eigenenergy $E$. We represent the bulk eigenstate of eigenenergy $E$ as the linear superposition of the states in solution space, 
\begin{eqnarray}
\psi_{\alpha}=\sum_{j=1}^{M}\sum_{s=1}^{s_{j}}\alpha_{js}\ket{\psi_{js}}+\sum_{s^{-}=1}^{s_{0}}\alpha_{s^{-}}\ket{\psi_{s^{-}}}+\sum_{s^{+}=1}^{s_{0}}\alpha_{s^{+}}\ket{\psi_{s^{+}}}.\nonumber\\
\end{eqnarray}
For simplicity, we denote $J\in\left\{js,s^{-},s^{+}\right\}$ and the above formula reads
\begin{eqnarray}
\psi_{\alpha}=\sum_{J}\alpha_{J}\ket{\psi_{J}}.
\end{eqnarray}
Substituting Eqs.~(\ref{nozerosolu}) and (\ref{zerosolu}) into Schr\"{o}dinger equation $\hat{H}\ket{\psi_{\alpha}}=E\ket{\psi_{\alpha}}$, we obtain the bulk equation
\begin{eqnarray}
\label{bulkeq2}
\sum_{J}\sum_{n=-R}^{R}T_{n}\ket{u_{x+n,J}}\alpha_{J}=E\sum_{J}\ket{u_{x,J}}\alpha_{J},
\end{eqnarray}
namely,
\begin{eqnarray}
\label{bulkeq3}
\sum_{n=-R}^{R}T_{n}\ket{u_{x+n,J}}=E\ket{u_{x,J}}
\end{eqnarray}
for arbitrary $\alpha_{J}$. When the multiple solutions and zero solutions of $\beta$ are absent, the bulk equation is reduced to
\begin{eqnarray}
H(\beta_{J})\ket{u_{J}}=E\ket{u_{J}},
\end{eqnarray}
where $H(\beta_{J})=\sum_{n=-R}^{R}T_{n}\beta^{n}_{J}$ is the non-Bloch Hamiltonian~\cite{yao201801,yokomizo2019}.
\section{The boundary equations and boundary matrix of general 1D non-Hermitian systems}
\label{boundaryeq}
We apply the OBCs and denote the sites of the boundary as $b=1,2,\ldots,R,N-R+1,\ldots,N$. Acting $\bra{b,\mu},\mu=1,2,\ldots,q$ from the left on the bulk equation Eq.~(\ref{bulkeq2}), we obtain the boundary equation
\begin{eqnarray}
\label{bouneq}
\sum_{J}\left[\sum_{n}T_{n}\braket{\mu|u_{b+n,J}}-E\braket{\mu|u_{b,J}}\right]\alpha_{J}=0,
\end{eqnarray}
and the boundary matrix $B(E)$,
\begin{eqnarray}
\label{boundarymatrix1}
&&[B(E)]_{b\mu,J}=\left[\sum_{n}T_{n}\ket{u_{b+n,J}}-E\ket{u_{b,J}}\right]_{\mu}\nonumber\\
&&\qquad\qquad\quad\equiv\bra{b,\mu}\hat{H}-E\ket{\psi_{J}},
\end{eqnarray}
where the summations are $\sum_{n=-b+1}^{R}$ and $\sum_{n=-R}^{N-b}$ at left- and right boundaries, respectively.

When the zero solutions of $\beta$ are absent, we subtract the boundary Eq.~(\ref{bouneq}) from the bulk Eq.~(\ref{bulkeq2}), and obtain the second form of the boundary equation
\begin{eqnarray}
&&\sum_{j=1}^{M}\sum_{s=1}^{s_{j}}\sum_{n=x_{L}}^{R}T_{-n}\ket{u_{x_{L}-n,js}}\alpha_{js}=0,\nonumber\\
&&\sum_{j=1}^{M}\sum_{s=1}^{s_{j}}\sum_{n=x_{R}}^{R}T_{n}\ket{u_{N-x_{R}+n+1,js}}\alpha_{js}=0,
\end{eqnarray}
at the left- and right boundaries, respectively, where $x_{L},x_{R}=1,2,\ldots,R$.

Utilizing Eqs.~(\ref{nozerosolu}), (\ref{zerosolu}) and (\ref{boundarymatrix1}), we define
\begin{eqnarray}
&&f^{js}_{b,\mu}(\beta_{j},E):=\bra{b,\mu}\hat{H}-E\ket{\psi_{js}},\, b=1,\ldots,R,\nonumber\\
&&g^{js}_{b,\mu}(\beta_{j},E):=\bra{b,\mu}\hat{H}-E\ket{\psi_{js}}\beta^{-N}_{j},\, b=N-R+1,\ldots,N,\nonumber\\
&&f^{-s^{-}}_{b,\mu}(E):=\bra{b,\mu}\hat{H}-E\ket{\psi_{s^{-}}}, \,b=1,\ldots,R,\nonumber\\
&&g^{-s^{-}}_{b,\mu}(E):=\bra{b,\mu}\hat{H}-E\ket{\psi_{s^{-}}}, \,b=N-R+1,\ldots,N,\nonumber\\
&&f^{+s^{+}}_{b,\mu}(E):=\bra{b,\mu}\hat{H}-E\ket{\psi_{s^{+}}}, \,b=1,\ldots,R,\nonumber\\
&&g^{+s^{+}}_{b,\mu}(E):=\bra{b,\mu}\hat{H}-E\ket{\psi_{s^{+}}}, \,b=N-R+1,\ldots,N.
\end{eqnarray}
The boundary matrix $B(E)$ reads~(omit the variables for simplicity)
\begin{eqnarray}
\label{boundarymatrix2}
B(E)=
\begin{bmatrix}
f^{11}_{1,1}&\cdots&f^{Ms_{M}}_{1,1}&f^{-1}_{1,1}&\cdots&f^{-s_{0}}_{1,1}&f^{+1}_{1,1}&\cdots&f^{+s_{0}}_{1,1}\\
\vdots&\vdots&\vdots&\vdots&\vdots&\vdots&\vdots&\vdots&\\
f^{11}_{R,q}&\cdots&f^{Ms_{M}}_{R,q}&f^{-1}_{R,q}&\cdots&f^{-s_{0}}_{R,q}&f^{+1}_{R,q}&\cdots&f^{+s_{0}}_{R,q}\\
g^{11}_{N-R+1,1}\beta_{1}^{N}&\cdots&g^{Ms_{M}}_{N-R+1,1}\beta_{M}^{N}&g^{-1}_{N-R+1,1}&\cdots&g^{-s_{0}}_{N-R+1,1}&g^{+1}_{N-R+1,1}&\cdots&g^{+s_{0}}_{N-R+1,1}\\
\vdots&\vdots&\vdots&\vdots&\vdots&\vdots&\vdots&\vdots&\\
g^{11}_{N,q}\beta_{1}^{N}&\cdots&g^{Ms_{M}}_{N,q}\beta_{M}^{N}&g^{-1}_{N,q}&\cdots&g^{-s_{0}}_{N,q}&g^{+1}_{N,q}&\cdots&g^{+s_{0}}_{N,q}
\end{bmatrix}.
\end{eqnarray}
When the zero solutions of $\beta$ are absent, we define
\begin{eqnarray}
&&f^{'js}_{b,\mu}(\beta_{j},E):=\sum_{n=b}^{R}[T_{-n}\ket{u_{b-n,js}}]_{\mu}, \,b=1,\ldots,R,\nonumber\\
&&g^{'js}_{b,\mu}(\beta_{j},E):=\sum_{n=b}^{R}[T_{n}\ket{u_{N-b+n+1,js}}]_{\mu}\beta_{j}^{-N}, \,b=1,\ldots,R. \nonumber\\
\end{eqnarray}
The second form of boundary matrix reads~(omit the variables for simplicity)
\begin{eqnarray}
\label{boundarymatrix3}
B^{'}(E)=
\begin{bmatrix}
f^{'11}_{1,1}&\cdots&f^{'1s_{1}}_{1,1}&\cdots&f^{'M1}_{1,1}&\cdots&f^{'Ms_{M}}_{1,1}\\
\vdots&\vdots&\vdots&\vdots&\vdots&\vdots&\vdots\\
f^{'11}_{R,q}&\cdots&f^{'1s_{1}}_{R,q}&\cdots&f^{'M1}_{R,q}&\cdots&f^{'Ms_{M}}_{R,q}\\
g^{'11}_{1,1}\beta_{1}^{N}&\cdots&g^{'1s_{1}}_{1,1}\beta_{1}^{N}&\cdots&g^{'M1}_{1,1}\beta_{M}^{N}&\cdots&g^{'Ms_{M}}_{1,1}\beta_{M}^{N}\\
\vdots&\vdots&\vdots&\vdots&\vdots&\vdots&\vdots\\
g^{'11}_{R,q}\beta_{1}^{N}&\cdots&g^{'1s_{1}}_{R,q}\beta_{1}^{N}&\cdots&g^{'M1}_{R,q}\beta_{M}^{N}&\cdots&g^{'Ms_{M}}_{R,q}\beta_{M}^{N}\\
\end{bmatrix},
\end{eqnarray}
which is equivalent to the boundary matrix $B(E)$ without zero solutions of $\beta$. Noteworthily, the two forms of boundary matrix respect the same boundary equation
\begin{eqnarray}
[B(E)]_{b\mu,J}\cdot\alpha_{J}=[B^{'}(E)]_{b\mu,J}\cdot\alpha_{J}=0.
\end{eqnarray}
When the zero solutions and multiple solutions of $\beta$ are both absent, the boundary matrix is reduced to
\begin{eqnarray}
B(E)=\begin{bmatrix}
f_{1}(\beta_{1},E)&\cdots&f_{1}(\beta_{2qR},E)\\
\vdots&\vdots&\vdots\\
f_{qR}(\beta_{1},E)&\cdots&f_{qR}(\beta_{2qR},E)\\
g_{1}(\beta_{1},E)\beta_{1}^{N}&\cdots&g_{1}(\beta_{2qR},E)\beta_{2qR}^{N}\\
\vdots&\vdots&\vdots\\
g_{qR}(\beta_{1},E)\beta_{1}^{N}&\cdots&g_{qR}(\beta_{2qR},E)\beta_{2qR}^{N}
\end{bmatrix},
\end{eqnarray}
which is the same as the form of boundary matrix in Ref.~\cite{yokomizo2019}.

The eigenstates of eigenenergy $E$ with OBCs are the kernel of the boundary matrix $B(E)$. Therefore, there exist eigenstates of eigenenergy $E$ only if $\det{[B(E)]}=0$. We define the sets $P$ and $Q$ as two disjoint subsets of the set
$\left\{js,s^{-},s^{+}\right\}$, such that the number of elements of each subset is $qR$. The determinant of the boundary matrix Eq.~(\ref{boundarymatrix2}) is
\begin{eqnarray}
\label{der}
\det{[B(E)]}=\sum_{P,Q}F(\beta_{I\in P},\beta_{J\in Q},E)\prod_{J\in Q}(\beta_{J})^{N},
\end{eqnarray}
where $\beta_{J}=\beta_{j}$ and $\beta_{J}=1$ correspond to $J=js$ and $J=s^{\pm}$, respectively. The function $F(\beta_{I\in P},\beta_{J\in Q},E)$ reads
\begin{eqnarray}
&&F(\beta_{I\in P},\beta_{J\in Q},E)=\sum_{\mathcal{P},\mathcal{Q}}(-1)^{sgn(\mathcal{P},\mathcal{Q})}\nonumber\\
&&\times f^{\mathcal{P}(1)}_{1,1}\ldots f^{\mathcal{P}(qR)}_{R,q}g^{\mathcal{Q}(1)}_{N-R+1,1}\ldots g^{\mathcal{Q}(qR)}_{N,q},
\end{eqnarray}
where $\mathcal{P}$ and $\mathcal{Q}$ are the permutations of $P$ and $Q$, respectively, $sgn(\mathcal{P},\mathcal{Q})$ is the sign of the joint permutation of $\mathcal{P}$ and $\mathcal{Q}$, and $f^{\mathcal{P}(j)}_{b,\mu}=f^{js}_{b,\mu}(\beta_{j},E)$ and $f^{\mathcal{P}(j)}_{b,\mu}=f^{\pm s^{\pm}}_{b,\mu}(E)$ correspond to $J=js$ and $J=s^{\pm}$, respectively~(similar to $g_{b,\mu}^{\mathcal{Q}(j)}$).

Note that the terms corresponding to the zero solutions of $\beta$ in Eq.~(\ref{der}) are just finite product factors of $F(\beta_{I\in P},\beta_{J\in Q},E)$ for large $N$. The difference between the cases with and without the zero solutions of $\beta$ is just a finite correction of $F(\beta_{I\in P},\beta_{J\in Q},E)$. Hence, without loss of generality, we only consider the cases without the zero solutions of $\beta$ in the elaboration of the general theory hereafter. 
\end{widetext}
\section{The generalized Brillouin zones of general systems}
\label{gbz}
The GBZ is the key result of the non-Bloch band theory in many previous works~\cite{yao201801,yokomizo2019,yang202002}. In this appendix, we give a further elucidation on the GBZ for the general 1D non-Hermitian systems.

Utilizing the characteristic Eq.~(\ref{charac}) in the main text, we can obtain the forms of energy bands, $E^{\mu}(\beta),\mu=1,2,\ldots,q$, and $2m\equiv2qR$ nonzero solutions of $\beta$ for a fixed energy value $E$. We number the solutions of $\beta$ satisfying $|\beta_{1}|\leq\ldots\leq|\beta_{m}|\leq|\beta_{m+1}|\ldots\leq|\beta_{2m}|$. The continuous-band condition requires $|\beta_{m}|=|\beta_{m+1}|$~\cite{yao201801,yokomizo2019}, which shapes the GBZ. In general, different energy bands $E^{\mu}(\beta)$ correspond to different GBZs, the $\mu$-th sub-GBZs of total GBZ~\cite{yang202002}, which we elucidate as follows.

Motivated by the auxiliary generalized Brillouin zone~(aGBZ) in Ref.~\cite{yang202002}, we utilize the resultant of $f(E,\beta)$ and $f(E,\beta e^{i\theta})$ relative to $E$,
\begin{eqnarray}
G(\beta,\theta):=R_{E}[f(E,\beta),f(E,\beta e^{i\theta}),E],
\end{eqnarray} 
where $R_{E}$ represents the resultant relative to $E$~\cite{yang202002}, and $\theta$ takes all the values in the range $[0,2\pi]$. $G(\beta,\theta)=0$ gives all the solutions satisfying $|\beta_{p}(\theta)|=|\beta_{p+1}(\theta)|$, $p=1,2,\ldots,2m-1$, $\theta\in[0,2\pi]$, of which the subsets with and without $p=m$ are the values of $\beta$ on GBZs and other aGBZs, respectively. Based on these tricks, we elaborate on a general steps to outline the sub-GBZs of every energy bands.

Firstly, take all the values of $\theta\in[0,2\pi]$ and solve the equation $G(\beta,\theta)=0$, to obtain the solutions of $\beta$ as functions of $\theta$, satisfying $|\beta_{p}(\theta)|=|\beta_{p+1}(\theta)|,p=1,2,\ldots,2m-1$. These all values of $\beta(\theta)$ constitute the aGBZs in the complex plane of $\beta$. Secondly, substitute these solutions into each energy band $E^{\mu}(\beta)$, to obtain all the corresponding energy values of each band $E^{\mu}(\beta)$. Thirdly, substitute all the energy values of each band into the characteristic equation $f(E,\beta)=0$, to obtain the solutions of $\beta$ corresponding to each energy band. Finally, determine whether $|\beta_{p}(\theta)|=|\beta_{p+1}(\theta)|$ for all combinations of $\theta\in[0,2\pi]$ and energy bands $E^{\mu}(\beta)$, which are sub-aGBZs corresponding to each energy bands, respectively. We denote the sub-aGBZ of each energy band $E^{\mu}(\beta)$ as $\beta_{(p,p+1)}^{\mu}(\theta)$. The sub-GBZs correspond to the sub-aGBZs with $p=m$, which we  specially denote as $\beta_{GBZ}^{\mu}(\theta)$. Substituting these sub-aGBZs into each expression $E^{\mu}(\beta)$ of the energy band, we obtain the sub-AEBs, denoted as $E^{\mu}\big(\beta_{(p,p+1)}^{\mu}(\theta)\big)$. The physical CEBs, $E^{\mu}\big(\beta_{GBZ}^{\mu}(\theta)\big)$, correspond to the sub-AEBs with $p=m$. We emphasize that the multiple roots and zero solutions of $\beta$ in general cases do not influence the theory of GBZs nor aGBZs, because we only concentrate on the two equal-norm solutions, $\beta_{p}$ and $\beta_{p+1}$. Moreover, the points on the aGBZs, where the solutions of $\beta$ satisfy more than two equal norms, namely, $\ldots=|\beta_{p}|=|\beta_{p+1}|=|\beta_{p+2}|=\ldots$, are the points where two or more aGBZs intersect~\cite{conn2021}.

\begin{figure}
	\centering
	\subfigure[]{\includegraphics[width=0.19\textwidth]{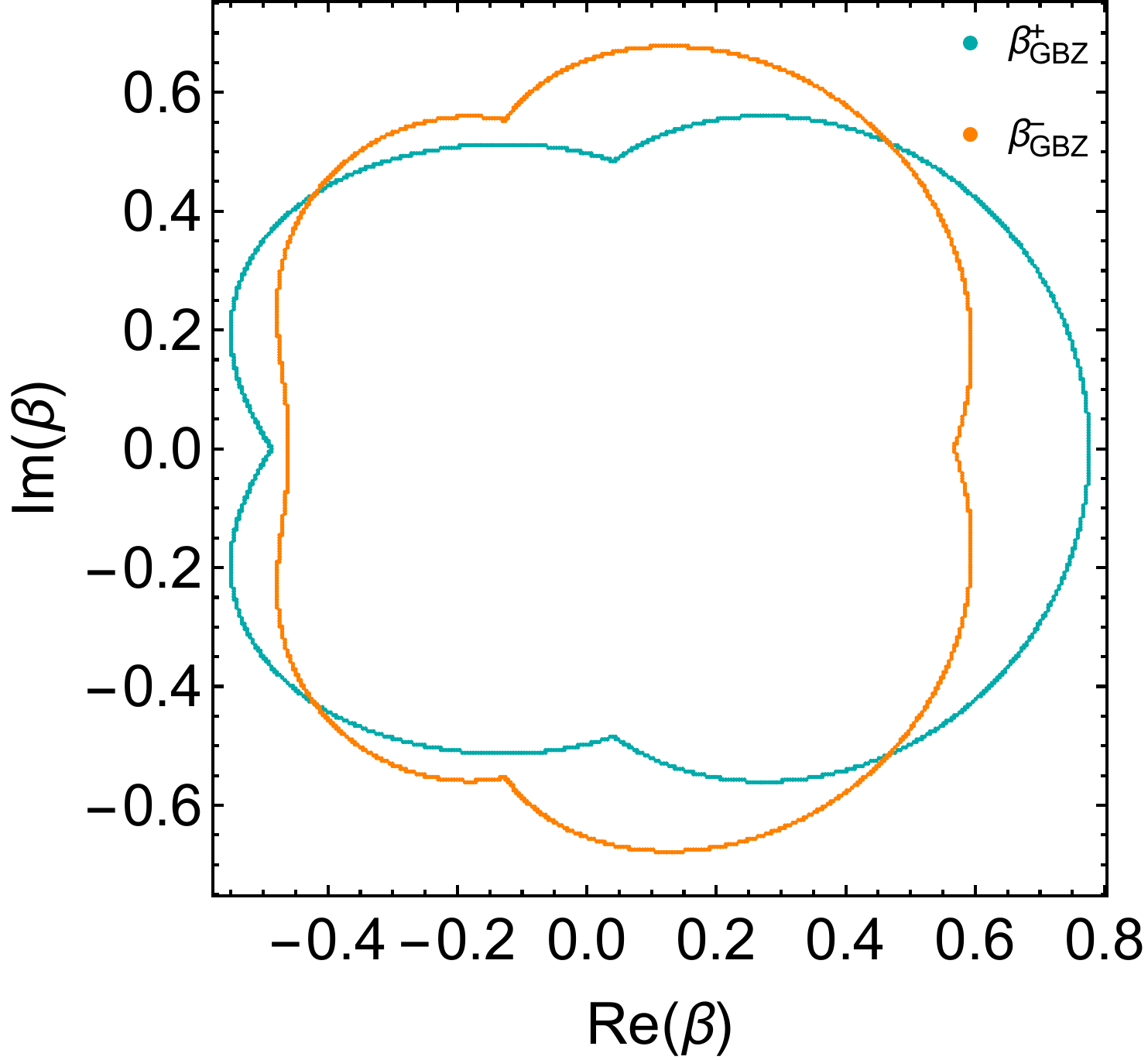}}
	\label{app1a}
	\subfigure[]{\includegraphics[width=0.26\textwidth]{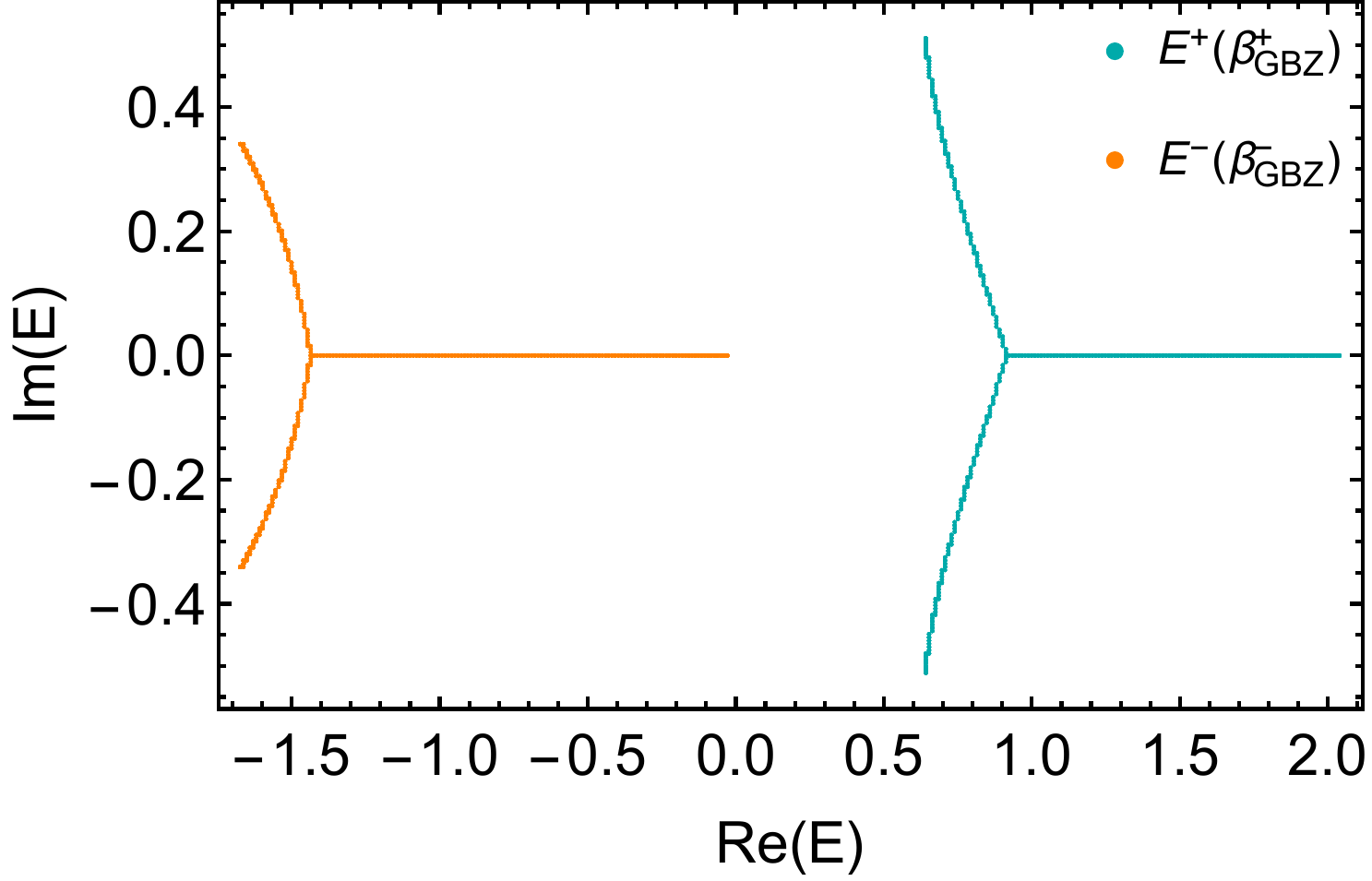}}
	\label{app1b}
	\caption{(a) The sub-GBZs, $\beta^{+}_{GBZ}$~(cyan) and  $\beta^{-}_{GBZ}$~(orange), of the non-Hermitian model Eq.~(\ref{modelapp}). (b) The sub-CEBs $E^{+}_{GBZ}(\beta^{+}_{GBZ})$~(cyan) and $E^{-}_{GBZ}(\beta^{-}_{GBZ})$~(orange).}
	\label{app}
\end{figure}

We apply the theory of GBZs to a two-band non-Hermitian model. The non-Bloch Hamiltonian is 
\begin{eqnarray}
\label{modelapp}
H(\beta)=\begin{bmatrix}
a\beta^{2}&b+\beta^{-1}\\
c+\beta&a\beta^{-2}
\end{bmatrix},
\end{eqnarray}
where $a=\frac{1}{5}$, $b=\frac{5}{3}$, and $c=\frac{1}{3}$. We obtain four nonzero solutions, $\beta_{i},i=1,2,3,4$, by the characteristic equation, $P(E,\beta)=0$ of this Hamiltonian. The two energy bands are
\begin{eqnarray}
E^{\pm}(\beta)=\frac{1}{2}\big[h(\beta)\pm\sqrt{h(\beta)^{2}-4\big(\frac{1}{25}-(\frac{1}{3}+\beta)(\frac{5}{3}+\beta^{-1})\big)}\big],\nonumber\\
\end{eqnarray}
where $h(\beta)=\frac{1}{5}(\beta^{2}+\beta^{-2})$. Following the steps in the above, we can obtain the two sub-GBZs with $|\beta_{2}|=|\beta_{3}|$ and sub-CEBs~(Fig.~\ref{app}). The other sub-aGBZs and sub-AEBs can also be obtain by following the above steps.

\section{Band degeneracy based on the theory of GBZs}
\label{appband}
Utilizing the theory of the GBZs, we study the energy band degeneracy of 1D non-Hermitian systems. Two sub-AEBs are degenerate at energy value $E_{0}$, only if
\begin{eqnarray}
\label{degene}
E^{\mu}\big(\beta_{(p,p+1)}^{\mu}\big)=E^{\nu}\big(\beta_{(p^{'},p^{'}+1)}^{\nu}\big)=E_{0},
\end{eqnarray}
for some points on the two corresponding sub-aGBZs $\beta_{(p,p+1)}^{\mu}$ and $\beta_{(p',p'+1)}^{\nu}$. Note that the case with $\mu=\nu$ and $p=p'$ is trivial, which is just the non-Hermitian generalization of the trivial case, $E^{\mu}(k)=E^{\mu}(k')$~($k,k'\in[0,2\pi]$), in momentum space of the Hermitian systems. We discuss the degeneracy of AEBs as the following cases: (i) When $p=p'+1$~($p=p'-1$) in Eq.~(\ref{degene}), the two sub-aGBZs, $\beta_{(p,p+1)}^{\mu}$ and $\beta_{(p-1,p)}^{\nu}$~($\beta_{(p,p+1)}^{\mu}$ and $\beta_{(p+1,p+2)}^{\nu}$), are degenerate at some points. In other word, they are the intersection points of these two sub-aGBZs, because the solutions of $f(E_{0},\beta)=0$ satisfies $|\beta_{p-1}|=|\beta_{p}|=|\beta_{p+1}|$~($|\beta_{p}|=|\beta_{p+1}|=|\beta_{p+2}|$). (ii) When $\mu\neq\nu$ and $p=p'$ in Eq.~(\ref{degene}), the two sub-aGBZs, $\beta_{(p,p+1)}^{\mu}$ and $\beta_{(p,p+1)}^{\nu}$, are degenerate at some points, which are the intersection points of these two sub-aGBZs. The solutions of $f(E_{0},\beta)=0$ satisfy $|\beta_{p}|=|\beta_{p+1}|$. (iii) When $p\neq p',p'\pm1$, the energy value $E_{0}$ is just a point, where the solutions of $\beta$ satisfy $|\beta_{p}|=|\beta_{p+1}|$ and  $|\beta_{p'}|=|\beta_{p'+1}|$. Moreover, $E_{0}$ corresponds to two different points on the two sub-aGBZs,  $\beta_{(p,p+1)}^{\mu}$ and $\beta_{(p',p'+1)}^{\nu}$, respectively.   

We only concentrate on the case with $\mu\neq\nu$ and $p=p'=m$ in the main text, namely, the degeneracy between different CEBs. The other cases are trivial or not physical. Since the values of $\beta$ on sub-GBZs correspond to the solutions of equation $f(E_{0},\beta)=0$ with $|\beta_{m}|=|\beta_{m+1}|$, two sub-CEBs are degenerate at one energy value, only when the two sub-GBZs of these two sub-CEBs are degenerate at some points. However, the inverse of this statement is not always true. For example, we consider the model Eq.~(\ref{modelapp}) given in the previous section. The two sub-GBZs, $\beta^{+}_{GBZ}$ and  $\beta^{-}_{GBZ}$, are degenerate at four points~[Fig.~\ref{app}(a)]. However, the two sub-CEBs $E^{+}_{GBZ}(\beta^{+}_{GBZ})$ and $E^{-}_{GBZ}(\beta^{-}_{GBZ})$ are not degenerate at any point~[Fig.~\ref{app}(b)], hence it is a gapped phase. 

\section{The generalized SSH model}
\label{detail1}
The boundary matrix of the generalized SSH model, with respect to the zero energy, is given by Eq.~(\ref{boundssh}) in the main text. 

When $-\lambda_{1}<t_{0}<\lambda_{1}$, the degenerate zero energies are IEBs. The norms of the four solutions~[Eq.~(\ref{solus})] are all less than $1$, thus the boundary matrix becomes
\begin{eqnarray}
\label{boundlimit}
B(0)=\begin{bmatrix}
t_{-1}^{+}&t_{-1}^{+}&0&0\\
0&0&t_{-1}^{-}&t_{-1}^{-}\\
0&0&0&0\\
0&0&0&0
\end{bmatrix},
\end{eqnarray}
in the thermodynamics limit. We obtain two analytic eigenstates $\alpha_{1}=(1,-1,0,0)^{T}$ and $\alpha_{2}=(0,0,1,-1)^{T}$, and the degenerate zero energies are not EPs. The two eigenstates in real space read
\begin{eqnarray}
\label{analytic}
&&\ket{\psi_{\alpha_{1}}}=\ket{\psi_{1}}-\ket{\psi_{2}},\nonumber\\
&&\ket{\psi_{\alpha_{2}}}=\ket{\psi_{3}}-\ket{\psi_{4}},
\end{eqnarray}
where 
\begin{eqnarray}
\ket{\psi_{i}}=\sum_{x=1}^{N}\beta_{i}^{x}\ket{u_{i}},i=1,2,3,4,
\end{eqnarray}
and
\begin{eqnarray}
&&\ket{u_{1}}=\ket{u_{2}}=(0,1)^{T},\nonumber\\
&&\ket{u_{3}}=\ket{u_{4}}=(1,0)^{T}.
\end{eqnarray}
We compare the distribution of the analytic eigenstate with the numerical eigenstate in the main text. Actually, the two numerical eigenstates are both consistent with $\ket{\psi_{\alpha_{2}}}$. That is because the two eigenstates of zero energy are given by
\begin{eqnarray}
\label{numerical}
&&\ket{\psi_{1}^{num}}=\ket{\psi_{\alpha_{2}}}+\eta\ket{\psi_{\alpha_{1}}},\nonumber\\
&&\ket{\psi_{2}^{num}}=\ket{\psi_{\alpha_{2}}}-\eta\ket{\psi_{\alpha_{1}}},
\end{eqnarray}
with $\eta\sim0$ in the numerical calculation. 

When $\lambda_{1}<t_{0}<\lambda_{2}$ and $-\lambda_{2}<t_{0}<-\lambda_{1}$, the degenerate zero energies are CEBs, not EPs. In the thermodynamics limit, the boundary matrix, the two analytic eigenstates, and the two numerical eigenstates are the same as Eqs.~(\ref{boundlimit}), (\ref{analytic}), and (\ref{numerical}), respectively. 

When $\lambda_{2}\leq t_{0}\leq\lambda_{3}$, $|\beta_{4}|^{2}<|\beta_{1}|^{2}=|\beta_{2}|^{2}<1\leq|\beta_{3}|^{2}$, and the boundary matrix becomes  
\begin{eqnarray}
\label{boundlimit1}
B(0)=\begin{bmatrix}
t_{-1}^{+}&t_{-1}^{+}&0&0\\
0&0&t_{-1}^{-}&t_{-1}^{-}\\
0&0&0&0\\
0&0&t_{1}^{-}\beta_{3}^{N}&0
\end{bmatrix},
\end{eqnarray}
in the thermodynamics limit. We obtain only one analytic eigenstate $\alpha_{1}=(1,-1,0,0)^{T}$~[$\ket{\psi_{\alpha_{1}}}$ in real space], thus the degenerate zero energies~(CEBs) are EPs~(defective eigenstates). In the numerical calculation, due to the finite $N$, we obtain two eigenenergies close enough to zero energy, and two numerical eigenstates are both equal to the analytic one, namely, $\ket{\psi_{\alpha_{1}}}$. Hence, the analytic result and numerical result are consistent with each other in the thermodynamics limit. When $-\lambda_{3}\leq t_{0}\leq-\lambda_{2}$, $|\beta_{3}|^{2}<|\beta_{1}|^{2}=|\beta_{2}|^{2}<1\leq|\beta_{4}|^{2}$, and we obtain the same result, namely, the presence of EPs. 

In the interval $-\lambda_{1}<t_{0}<0$, $|\beta_{3}|^{2}<|\beta_{4}|^{2}<|\beta_{2}|=|\beta_{1}|^{2}$, and we apply the theory of IEBs with finite $N$~[Sec.~\ref{iso}]. Motivated by the Ref.~\cite{lee2019}, we can approximately express the eigenvectors of the Hamiltonian
\begin{eqnarray}
H_{c}(\beta)=\begin{bmatrix}
0&(\beta-\beta_{1})(\beta-\beta_{2})/\beta\\
(\beta-\beta_{3})(\beta-\beta_{4})/\beta&0
\end{bmatrix},\nonumber\\
\end{eqnarray}
with respect to the energy $\Delta E$~(small enough) as
\begin{eqnarray}
&&\ket{u_{1}}=\big(\Delta E,\,\,\,(\beta_{1}-\beta_{3})(\beta_{1}-\beta_{4})\big)^{T},\nonumber\\
&&\ket{u_{2}}=\big(\Delta E,\,\,\,(\beta_{2}-\beta_{3})(\beta_{2}-\beta_{4})\big)^{T},\nonumber\\
&&\ket{u_{3}}=\big((\beta_{3}-\beta_{1})(\beta_{3}-\beta_{2}),\,\,\,\Delta E\big)^{T},\nonumber\\
&&\ket{u_{4}}=\big((\beta_{4}-\beta_{1})(\beta_{4}-\beta_{2}),\,\,\,\Delta E\big)^{T}.
\end{eqnarray}
The boundary matrix becomes
\begin{eqnarray}
B(\Delta E)=\begin{bmatrix}
t_{-1}^{+}\Delta_{1}&t_{-1}^{+}\Delta_{2}&t_{-1}^{+}\Delta E&t_{-1}^{+}\Delta E\\
t_{-1}^{-}\Delta E&t_{-1}^{-}\Delta E&t_{-1}^{-}\Delta_{3}&t_{-1}^{-}\Delta_{4}\\
t_{1}^{+}\Delta_{1}\beta_{1}^{N}&t_{1}^{+}\Delta_{2}\beta_{2}^{N}&t_{1}^{+}\Delta E\beta_{3}^{N}&t_{1}^{+}\Delta E\beta_{4}^{N}\\
t_{1}^{-}\Delta E\beta_{1}^{N}&t_{1}^{-}\Delta E\beta_{2}^{N}&t_{1}^{-}\Delta_{3}\beta_{3}^{N}&t_{1}^{-}\Delta_{4}\beta_{4}^{N}
\end{bmatrix},\nonumber\\
\end{eqnarray}
where 
\begin{eqnarray}
&&\Delta_{1}=(\beta_{1}-\beta_{3})(\beta_{1}-\beta_{4}),\nonumber\\
&&\Delta_{2}=(\beta_{2}-\beta_{3})(\beta_{2}-\beta_{4}),\nonumber\\
&&\Delta_{3}=(\beta_{3}-\beta_{1})(\beta_{3}-\beta_{2}),\nonumber\\
&&\Delta_{4}=(\beta_{4}-\beta_{1})(\beta_{4}-\beta_{2}).
\end{eqnarray}
From the determinant of the boundary matrix, we can read $F_{0}^{(0)}=F_{0}^{(1)}\sim0$, and the first two-order terms, $\mathcal{F}_{1}\big((\beta_{1}\beta_{4})^{N}-(\beta_{2}\beta_{4})^{N}\big)-\mathcal{F}_{2} (\beta_{1}\beta_{2})^{N}\Delta E^{2}$,
where 
\begin{eqnarray}
&&\mathcal{F}_{1}=t_{1}^{+}t_{1}^{-}t_{-1}^{+}t_{-1}^{-}\Delta_{1}\Delta_{2}\Delta_{3}\Delta_{4},\nonumber\\
&&\mathcal{F}_{2}=t_{1}^{+}t_{1}^{-}t_{-1}^{+}t_{-1}^{-}\big(\Delta_{1}\Delta_{3}-\Delta_{1}\Delta_{4}-\Delta_{2}\Delta_{3}+\Delta_{2}\Delta_{4}\big).\nonumber\\
\end{eqnarray}
Consequently, we obtain the exponential displacement of $\Delta E$ with the number of lattice sites $N$,
\begin{eqnarray}
\Delta E\sim\pm\sqrt{\frac{\mathcal{F}_{1}}{\mathcal{F}_{2}}\big[\big(\frac{\beta_{4}}{\beta_{2}}\big)^{N}-\big(\frac{\beta_{4}}{\beta_{1}}\big)^{N}\big]}.
\end{eqnarray}
Since $|\beta_{4}|<|\beta_{2}|=|\beta_{1}|$, $\Delta E\rightarrow0$ when $N\rightarrow\infty$, consistent with the result in the thermodynamics limit. In the interval $0<t_{0}<\lambda_{1}$, $|\beta_{4}|^{2}<|\beta_{3}|^{2}<|\beta_{2}|=|\beta_{1}|^{2}$, and we obtain the same result.

\section{The infernal points of 1D non-Hermitian systems}
\label{infernal}
\subsection{Formal theory}
Motivated by Ref.~\cite{denner2021}, we construct a formal theory applying to the infernal points. We start with a 1D non-Hermitian tight-binding model, with the hopping range $l$, and the internal degrees of freedom per unit cell $q$. In principle, we can always enlarge the unit cell, containing $l$ original unit cells. Thus, the Hamiltonian of this model, with the nearest-neighbor hopping, reads~($N$ new enlarged unit cells)
\begin{eqnarray}
H_{nn}=\begin{bmatrix}
m_{0}&H_{+}&0&\cdots&0\\
H_{-}&m_{0}&\ddots&\ddots&\vdots\\
0&\ddots&\ddots&\ddots&0\\
\vdots&\ddots&\ddots&m_{0}&H_{+}\\
0&\cdots&0&H_{-}&m_{0}\\
\end{bmatrix}_{N\times N},
\end{eqnarray}     
where $H_{nn}$ is a $N\times N$ block matrix and $m_{0}$, $H_{+}$ and $H_{-}$ are $ql \times ql$ matrices. The eigenenergies of $H_{nn}$ with $j$ enlarged unit cells are obtained by solving the characteristic equation $\det{[\tilde{H}(j)]}=0$, where 
\begin{eqnarray}
&&\tilde{H}(j)=H_{nn}-E\nonumber\\
&&\qquad\quad:=\begin{bmatrix}
H_{0}&H_{+}&0&\cdots&0\\
H_{-}&H_{0}&\ddots&\ddots&\vdots\\
0&\ddots&\ddots&\ddots&\vdots\\
\vdots&\ddots&\ddots&H_{0}&H_{+}\\
0&\cdots&0&H_{-}&H_{0}\\
\end{bmatrix}_{j\times j},
\end{eqnarray}
and $H_{0}=m_{0}-E$.
We define
\begin{eqnarray}
&&B^{(j)}=(0,\ldots,H_{+})^{T}_{j\times1},\nonumber\\
&&C^{(j)}=(0,\ldots,H_{-})_{1\times j}.\nonumber\\
\end{eqnarray} 
Rearrange $\tilde{H}(N)$ as a $2\times2$ block matrix,
\begin{eqnarray}
\tilde{H}(N)=\begin{bmatrix}
\tilde{H}(N-1)&B^{(N-1)}\\
C^{(N-1)}&H_{0}
\end{bmatrix},
\end{eqnarray}
and use Schur's determinant identity  
\begin{eqnarray}  
&&\det{[\tilde{H}(N)]}\nonumber\\
&&=\det{[H_{0}]}\det{[\tilde{H}(N-1)-B^{(N-1)}\cdot H_{0}^{-1}\cdot C^{(N-1)}]},\nonumber\\  
\end{eqnarray}  
if $H_{0}$ is invertible.
Note that
\begin{eqnarray}
&&B^{(N-1)}\cdot H_{0}^{-1}\cdot C^{(N-1)}\nonumber\\
&&=\begin{bmatrix}
0&\cdots&\cdots&\cdots&0\\
\vdots&\ddots&\ddots&\ddots&\vdots\\
\vdots&\ddots&\ddots&\ddots&\vdots\\
\vdots&\ddots&\ddots&0&0\\
0&\cdots&\cdots&0&H_{+}\cdot H_{0}^{-1}\cdot H_{-}
\end{bmatrix}_{(N-1)\times(N-1)},
\end{eqnarray}
thus,
\begin{eqnarray}
\label{block}
&&\tilde{H}(N-1)-B^{(N-1)}\cdot H_{0}^{-1}\cdot C^{(N-1)}\nonumber\\
&&=\begin{bmatrix}
\tilde{H}(N-2)&B^{(N-2)}\\
C^{(N-2)}&H_{0}-H_{+}\cdot H_{0}^{-1}\cdot H_{-}
\end{bmatrix},
\end{eqnarray}
By Schur's determinant identity, the determinant of Eq.~(\ref{block}) is 
\begin{eqnarray}  
&&\det{[H_{0}-H_{+}\cdot H_{0}^{-1}\cdot H_{-}]}\times\nonumber\\
&&\det{[\tilde{H}(N-2)-B^{(N-2)}\cdot (H_{0}-H_{+}\cdot H_{0}^{-1}\cdot H_{-})^{-1}\cdot C^{(N-2)}]},\nonumber\\
\end{eqnarray}  
if $H_{0}-H_{+}\cdot H_{0}^{-1}\cdot H_{-}$ is invertible.
By recursion, we can conclude that, 
\begin{widetext}
\begin{eqnarray}
\det{[\tilde{H}(j)-B^{(j)}\cdot (\Lambda^{(N-j)})^{-1}\cdot C^{(j)}]}=\det{[\Lambda^{(N-j+1)}]}
\times\det{[\tilde{H}(j-1)-B^{(j-1)}\cdot (\Lambda^{(N-j+1)})^{-1}\cdot C^{(j-1)}]},
\end{eqnarray}
\end{widetext}
where $j=2,\ldots,N-1$, 
\begin{eqnarray}
&&\Lambda^{(1)}=H_{0}-H_{+}\cdot H_{0}^{-1}\cdot H_{-},\nonumber\\
&&\Lambda^{(j)}=H_{0}-H_{+}\cdot(\Lambda^{(j-1)})^{-1}\cdot H_{-},
\end{eqnarray} 
and $\Lambda^{(N-j+1)}$ is invertible. Finally, we obtain
\begin{eqnarray}
\label{formalresu}
\det{[\tilde{H}(N)]}=\det{[H_{0}]}\prod_{j=1}^{N-1}\det{[\Lambda^{(j)}]}.
\end{eqnarray}
The above equation is valid in the region of $E$, where all of $H_{0}$ and $\Lambda^{(j)},j=1,2,\ldots,N-2$ are invertible. The full energy spectra of $H_{nn}$ are constituted by the solutions of $\det{[\Lambda^{(N-1)}]}=0$, and the solutions of $\det{[H_{0}]}=\det{[\Lambda^{(1)}]}=\ldots=\det{[\Lambda^{(N-2)}]}=0$, satisfying the characteristic equation. The formal theory is tedious for generic cases. However, if $\Lambda^{(j_{0})}=\Lambda^{(j_{0}+1)}$ for some special cases, then $\Lambda^{(j)}=\Lambda^{(j_{0})},j=j_{0}+1,j_{0}+2,\ldots,N-1$, and
\begin{eqnarray}
\det{[\tilde{H}(N)]}=\det{[H_{0}]}\prod_{j=1}^{j_{0}-1}\det{[\Lambda^{(j)}]}\big(\det{[\Lambda^{(j_{0})}]}\big)^{N-j_{0}}.\nonumber\\
\end{eqnarray}
When $N$ is large enough and $j_{0}\ll N$, the number of degenerate bands scales with $N$, thus the infernal point emerges.
Especially, when $j_{0}=1$, we obtain 
\begin{eqnarray}
\label{infernalesp}
\det{[\tilde{H}(N)]}=\det{[H_{0}]}\big(\det{[\Lambda^{(j_{0})}]}\big)^{N-1},
\end{eqnarray}
and the emergence of the infernal point.

\subsection{The non-Hermitian SSH model with infernal points}
We apply the formal theory to the non-Hermitian SSH model. The Hamiltonian of this model reads
\begin{eqnarray}
\hat{H}_{nssh}=\hat{C}^{\dagger}\cdot H_{nssh}\cdot\hat{C},
\end{eqnarray}
where $\hat{C}=(\hat{c}_{1,A},\hat{c}_{1,B},\ldots,\hat{c}_{N,A},\hat{c}_{N,B})^{T}$, with the creation operator $\hat{c}_{x,A}$ and $\hat{c}_{x,B}$ of the sub-lattices, and
\begin{eqnarray}
H_{nssh}=\begin{bmatrix}
0&t_{1}+\gamma&0&\cdots&\cdots&0\\
t_{1}-\gamma&0&t_{2}&\ddots&\ddots&\vdots\\
0&t_{2}&\ddots&\ddots&\ddots&\vdots\\
\vdots&\ddots&\ddots&\ddots&t_{2}&0\\
\vdots&\ddots&\ddots&t_{2}&0&t_{1}+\gamma\\
0&\cdots&\cdots&0&t_{1}-\gamma&0
\end{bmatrix}.
\end{eqnarray}
Using Eq.~(\ref{infernalesp}), we obtain 
\begin{eqnarray}
\det{[\tilde{H}_{nssh}(N)]}=E^{2}\times(E^{2}-t_{2}^{2})^{N-1},
\end{eqnarray}
when $t_{1}=\pm\gamma$. There are $\frac{1}{2}(N-2)$ energy values with $E=\pm t_{2}$ and two IEBs degenerate at zero energy, totally $2N$ energy values.

We solve the eigenstates for $t_{1}=\gamma$ analytically~(same for $t_{1}=-\gamma$). The Hamiltonian in real space reads
\begin{eqnarray}
H_{nssh}^{t_{1}=\gamma}=\begin{bmatrix}
0&2\gamma&0&\cdots&\cdots&0\\
0&0&t_{2}&\ddots&\ddots&\vdots\\
0&t_{2}&\ddots&\ddots&\ddots&\vdots\\
\vdots&\ddots&\ddots&\ddots&t_{2}&0\\
\vdots&\ddots&\ddots&t_{2}&0&2\gamma\\
0&\cdots&\cdots&0&0&0
\end{bmatrix},
\end{eqnarray}
and the Schr\"{o}dinger equation is $H_{nssh}^{t_{1}=\gamma}\psi=E\psi$. Assume the trial solution is 
\begin{eqnarray}
\psi=(\ldots,a^{(j-1)},b^{(j-1)},a^{(j)},b^{(j)},a^{(j+1)},b^{(j+1)},\ldots)^{T}.\nonumber\\
\end{eqnarray}
We obtain two bulk and two boundary equations under OBCs, which are
\begin{eqnarray}
&&\label{bulk1}
-Eb^{(j)}+t_{2}a^{(j+1)}=0,\\
&&\label{bulk2}
t_{2}b^{(j)}-Ea^{(j+1)}+2\gamma b^{(j+1)}=0,
\end{eqnarray} 
and
\begin{eqnarray}
&&\label{edge1}
-Ea^{(1)}+2\gamma b^{(1)}=0,\\
&&\label{edge2}
-Eb^{(N)}=0,
\end{eqnarray}
respectively.\\
(i) $E=0$:\\ Eq.~(\ref{edge1})$\Rightarrow$ $b^{(1)}=0$,\\ Eq.~(\ref{bulk2})$\Rightarrow$ $b^{(j)}=0$,\quad $j\geq1$,\\ 
Eq.~(\ref{bulk1})$\Rightarrow$$a^{(j+1)}=0$, $j\geq1$, only $a^{(1)}\neq0$, thus
\begin{eqnarray}
\psi_{0}=(1,0,\ldots,0)^{T}.
\end{eqnarray}
The topological edge states are defective, thus the energy $E=0$~(IEBs) is an EP.\\
(ii) $E=t_{2}$: \\
Eq.~(\ref{edge1})$\Rightarrow$ $a^{(1)}=\frac{2\gamma}{t_{2}}b^{(1)}$,\\
Eq.~(\ref{bulk1})$\Rightarrow$ $b^{(j)}=a^{(j+1)}$, $j\geq1$,\\
Eq.~(\ref{bulk2})$\Rightarrow$ $b^{(j+1)}=0$,\quad\, $j\geq1$, thus
\begin{eqnarray}
\psi_{t_{2}}=(\frac{2\gamma}{t_{2}},1,1,0,\ldots,0)^{T}.
\end{eqnarray}
(iii) $E=-t_{2}$: 
\begin{eqnarray}
\psi_{-t_{2}}=(\frac{2\gamma}{t_{2}},-1,1,0,\ldots,0)^{T}.
\end{eqnarray}
There are $N-1$ eigenvalues degenerate at $E=t_{2}$~($E=-t_{2}$), however, there is only one eigenstate. Therefore, the energy $E=t_{2}$~($E=-t_{2}$) is an infernal point.

Moreover, the topological edge states with respect to $E=0$ of this model are
\begin{eqnarray}
&&\ket{E}_{L}=\frac{1}{\mathcal{N}_{L}}\sum_{j=1}^{N}(-\frac{t_{1}-\gamma}{t_{2}})^{j-1}\ket{j,A},\nonumber\\
&&\ket{E}_{R}=\frac{1}{\mathcal{N}_{R}}\sum_{j=1}^{N}(-\frac{t_{2}}{t_{1}+\gamma})^{-N+j}\ket{j,B}.
\end{eqnarray}
Then, we can obtain the edge-state-subspace effective Hamiltonian
\begin{eqnarray}
&&H_{eff}=\big(\bra{E}_{L},\bra{E}_{R})^{T}\cdot \hat{H}_{nssh}\cdot(\ket{E}_{L},\ket{E}_{R}\big)\nonumber\\
&&\quad\quad\,\,\,=para\times 
\begin{bmatrix}
0&t_{1}+\gamma\\
t_{1}-\gamma&0
\end{bmatrix},
\end{eqnarray}
which is defective at $t_{1}=\pm\gamma$.

\subsection{The four-band model with infernal points}
We apply the formal theory to the four-band model~[Eq.~(\ref{fourband})]. Using Eq.~(\ref{infernalesp}), we obtain 
\begin{eqnarray}
\det{[\tilde{H}_{F}(N)]}=E^{4}\times(E^{2}-1)^{2(N-1)},
\end{eqnarray} 
when $t=\pm1$. There are $2(N-1)$ energy values with $E=\pm1$ and four IEBs degenerate at zero energy, totally $4N$ energy values.

We solve the eigenstates for $t=1$ analytically~(same for $t=-1$). Assume the trial solution is 
\begin{eqnarray}
\psi=(\ldots,a^{(j)},b^{(j)},c^{(j)},d^{(j)},\ldots)^{T}.
\end{eqnarray}
Using Schr\"{o}dinger equation $H_{F}^{t=1}\psi=E\psi$ under OBCs, we obtain the equations with respect to $E=0$,
\begin{eqnarray}
&&c^{(j-1)}+2b^{(j)}=0\label{b1},\\
&&d^{(j+1)}=0\label{b2},\\
&&2d^{(j)}+a^{(j+1)}=0\label{b3},\\
&&b^{(j-1)}=0\label{b4},
\end{eqnarray}
in the bulk,
\begin{eqnarray}
&&2b^{(1)}=0\label{l1},\\
&&d^{(2)}=0\label{l2},\\
&&2d^{(1)}+a^{(2)}=0\label{l3},
\end{eqnarray}
at the left boundary, and
\begin{eqnarray}
&&c^{(N-1)}+2b^{(N)}=0\label{r1},\\
&&2d^{(N)}=0\label{r2},\\
&&b^{(N-1)}=0\label{r3},
\end{eqnarray}
at the right boundary. By the above equations, we obtain $a^{(3)}=a^{(4)}=\ldots=a^{(N)}=0$,  $b^{(1)}=b^{(2)}=\ldots=b^{(N-1)}=0$, $c^{(1)}=c^{(2)}=\ldots=c^{(N-2)}=0$, and $d^{(2)}=d^{(3)}=\ldots=d^{(N)}=0$. We discuss the following cases:\\
(i) If $d^{(1)}=0$, Eq.~(\ref{l3}) gives $a^{(2)}=0$, thus, there is only one independent variable $a^{(1)}$ at the left boundary;\\
(ii) If $d^{(1)}\neq0$, Eq.~(\ref{l3}) gives $a^{(2)}=-2d^{(1)}$;\\
(iii) If $b^{(N)}=0$, Eq.~(\ref{r1}) gives $c^{(N-1)}=0$, thus there is only one independent variable $c^{(N)}$ at the right boundary;\\
(iv) If $b^{(N)}\neq0$, Eq.~(\ref{r1}) gives $c^{(N-1)}=-2b^{(N)}$.\\
According to the above, we obtain four independent edge states located at the left and right boundary by the above equations,
\begin{eqnarray}
&&\psi^{edge}_{1}=(1,0,0,0,\ldots,0,0,0,0)^{T},\nonumber\\
&&\psi^{edge}_{2}=(0,0,0,1,-2,0,0,0,\ldots,0,0,0,0)^{T},\nonumber\\
&&\psi^{edge}_{3}=(0,0,0,0,\ldots,0,0,1,0)^{T},\nonumber\\
&&\psi^{edge}_{4}=(0,0,0,0,\ldots,0,0,-2,0,0,1,0,0)^{T}.
\end{eqnarray}
Therefore, there are four eigenvalue, degenerate at $E=0$ with four eigenstates, which is not an EP. By the Schr\"{o}dinger equation under OBCs with respect to $E=1$, we obtain
\begin{eqnarray}
&&c^{(j-1)}+2b^{(j)}=a^{(j)},\label{bu1}\\
&&d^{(j+1)}=b^{(j)},\label{bu2}\\
&&2d^{(j)}+a^{(j+1)}=c^{(j)},\label{bu3}\\
&&b^{(j-1)}=d^{(j)}\label{bu4},
\end{eqnarray}
in the bulk,
\begin{eqnarray}
&&2b^{(1)}=a^{(1)}\label{le1},\\
&&d^{(2)}=b^{(1)}\label{le2},\\
&&2d^{(1)}+a^{(2)}=c^{(1)}\label{le3},\\
&&0=d^{(1)}\label{le4},
\end{eqnarray}
at the left boundary, and
\begin{eqnarray}
&&c^{(N-1)}+2b^{(N)}=a^{(N)}\label{ri1},\\
&&0=b^{(N)},\label{ri2}\\
&&2d^{(N)}=c^{(N)}\label{ri3},\\
&&b^{(N-1)}=d^{(N)}\label{ri4},
\end{eqnarray}
at the right boundary. From the above equations, we directly obtain $d^{(1)}=b^{(N)}=0$. Then, we discuss the following cases:\\ 
(i) If $a^{(1)}=0$, we obtain $b^{(1)}=b^{(2)}=\dots=b^{(N)}=0$, $d^{(1)}=d^{(2)}=\dots=d^{(N)}=0$, $a^{(j)}=c^{(j-1)}$, and $c^{(N)}=0$. Thus, there are $N-1$ independent eigenstates
\begin{eqnarray}
&&\psi_{j}=(\ldots,0,0,\mathop{\mathop{1,}_{\downarrow}}_{c^{(j)}} 0,\mathop{\mathop{1,}_{\downarrow}}_{a^{(j+1)}}0,0,0,\ldots)^{T},\nonumber\\
&&j=1,2,\ldots,N-1;
\end{eqnarray}
(ii) If $a^{(1)}=a\neq0$, we obtain
\begin{eqnarray}
&&a^{(4j-2)}=c^{(4j-3)},\nonumber\\
&&a^{(4j-1)}+a=c^{(4j-2)},\nonumber\\
&&a^{(4j)}=c^{(4j-1)},\nonumber\\
&&a^{(4j+1)}-a=c^{(4j)},\nonumber\\
&&b^{(4j-3)}=d^{(4j-2)}=\frac{a}{2},\nonumber\\
&&b^{(4j-2)}=d^{(4j-1)}=0,\nonumber\\
&&b^{(4j-1)}=d^{(4j)}=-\frac{a}{2},\nonumber\\
&&b^{(4j)}=d^{(4j+1)}=0.
\end{eqnarray}
Thus, we obtain one eigenstate in this case,
\begin{widetext}
\begin{eqnarray}
&&\tilde{\psi}_{N}=(a,\frac{a}{2},a^{(2)},0,\ldots,
\mathop{\mathop{a^{(4j-3)},}_{\downarrow}}_{a^{(4j-3)}}
\mathop{\mathop{\frac{a}{2},}_{\downarrow}}_{b^{(4j-3)}}
\mathop{\mathop{a^{(4j-2)},}_{\downarrow}}_{c^{(4j-3)}}
\mathop{\mathop{0,}_{\downarrow}}_{d^{(4j-3)}}
\mathop{\mathop{a^{(4j-2)},}_{\downarrow}}_{a^{(4j-2)}}
\mathop{\mathop{0,}_{\downarrow}}_{b^{(4j-2)}}
\mathop{\mathop{a^{(4j-1)}+a,}_{\downarrow}}_{c^{(4j-2)}}
\mathop{\mathop{\frac{a}{2},}_{\downarrow}}_{d^{(4j-2)}}\nonumber\\
&&\quad\quad\quad\mathop{\mathop{a^{(4j-1)},}_{\downarrow}}_{a^{(4j-1)}}
\mathop{\mathop{-\frac{a}{2},}_{\downarrow}}_{b^{(4j-1)}}
\mathop{\mathop{a^{(4j)},}_{\downarrow}}_{c^{(4j-1)}}
\mathop{\mathop{0,}_{\downarrow}}_{d^{(4j-1)}}
\mathop{\mathop{a^{(4j)},}_{\downarrow}}_{a^{(4j)}}
\mathop{\mathop{0,}_{\downarrow}}_{b^{(4j)}}
\mathop{\mathop{a^{(4j+1)}-a,}_{\downarrow}}_{c^{(4j)}} 
\mathop{\mathop{-\frac{a}{2},}_{\downarrow}}_{d^{(4j)}} 
\ldots)^{T}.
\end{eqnarray}
\end{widetext}
The independent eigenstate is given by
\begin{widetext}
\begin{eqnarray}
&&\psi_{N}=\tilde{\psi}_{N}-\sum_{j=1}^{N-1}a^{(j+1)}\psi_{j}\nonumber\\
&&\quad\,\,\,\,=(a,\frac{a}{2},0,0,\ldots,0,
\mathop{\mathop{\frac{a}{2},}_{\downarrow}}_{b^{(4j-3)}}
0,0,0,0,
\mathop{\mathop{a,}_{\downarrow}}_{c^{(4j-2)}} \mathop{\mathop{\frac{a}{2},}_{\downarrow}}_{d^{(4j-2)}} 
0, \mathop{\mathop{-\frac{a}{2},}_{\downarrow}}_{b^{(4j-1)}}
0,0,0,0,\mathop{\mathop{-a,}_{\downarrow}}_{c^{(4j)}}
\mathop{\mathop{-\frac{a}{2},}_{\downarrow}}_{d^{(4j)}}
\ldots)^{T}.
\end{eqnarray}
\end{widetext}
According to the above, there are $N$ eigenstates with respect to $E=1$, and it is the same as $E=-1$. Therefore, there are $2N-2$ eigenvalues degenerate at $E=1$~($E=-1$) with only $N$ eigenstates, which is an infernal point.

\bibliography{reference}

\begin{thebibliography}{76}%
\makeatletter
\providecommand \@ifxundefined [1]{%
 \@ifx{#1\undefined}
}%
\providecommand \@ifnum [1]{%
 \ifnum #1\expandafter \@firstoftwo
 \else \expandafter \@secondoftwo
 \fi
}%
\providecommand \@ifx [1]{%
 \ifx #1\expandafter \@firstoftwo
 \else \expandafter \@secondoftwo
 \fi
}%
\providecommand \natexlab [1]{#1}%
\providecommand \enquote  [1]{``#1''}%
\providecommand \bibnamefont  [1]{#1}%
\providecommand \bibfnamefont [1]{#1}%
\providecommand \citenamefont [1]{#1}%
\providecommand \href@noop [0]{\@secondoftwo}%
\providecommand \href [0]{\begingroup \@sanitize@url \@href}%
\providecommand \@href[1]{\@@startlink{#1}\@@href}%
\providecommand \@@href[1]{\endgroup#1\@@endlink}%
\providecommand \@sanitize@url [0]{\catcode `\\12\catcode `\$12\catcode
  `\&12\catcode `\#12\catcode `\^12\catcode `\_12\catcode `\%12\relax}%
\providecommand \@@startlink[1]{}%
\providecommand \@@endlink[0]{}%
\providecommand \url  [0]{\begingroup\@sanitize@url \@url }%
\providecommand \@url [1]{\endgroup\@href {#1}{\urlprefix }}%
\providecommand \urlprefix  [0]{URL }%
\providecommand \Eprint [0]{\href }%
\providecommand \doibase [0]{http://dx.doi.org/}%
\providecommand \selectlanguage [0]{\@gobble}%
\providecommand \bibinfo  [0]{\@secondoftwo}%
\providecommand \bibfield  [0]{\@secondoftwo}%
\providecommand \translation [1]{[#1]}%
\providecommand \BibitemOpen [0]{}%
\providecommand \bibitemStop [0]{}%
\providecommand \bibitemNoStop [0]{.\EOS\space}%
\providecommand \EOS [0]{\spacefactor3000\relax}%
\providecommand \BibitemShut  [1]{\csname bibitem#1\endcsname}%
\let\auto@bib@innerbib\@empty
%</preamble>
\bibitem [{\citenamefont {Bergholtz}\ \emph {et~al.}(2021)\citenamefont
  {Bergholtz}, \citenamefont {Budich},\ and\ \citenamefont
  {Kunst}}]{bergholtzrev}%
  \BibitemOpen
  \bibfield  {author} {\bibinfo {author} {\bibfnamefont {Emil~J.}\ \bibnamefont
  {Bergholtz}}, \bibinfo {author} {\bibfnamefont {Jan~Carl}\ \bibnamefont
  {Budich}}, \ and\ \bibinfo {author} {\bibfnamefont {Flore~K.}\ \bibnamefont
  {Kunst}},\ }\bibfield  {title} {\enquote {\bibinfo {title} {Exceptional
  topology of non-hermitian systems},}\ }\href {\doibase
  10.1103/RevModPhys.93.015005} {\bibfield  {journal} {\bibinfo  {journal}
  {Rev. Mod. Phys.}\ }\textbf {\bibinfo {volume} {93}},\ \bibinfo {pages}
  {015005} (\bibinfo {year} {2021})}\BibitemShut {NoStop}%
\bibitem [{\citenamefont {Lee}(2016)}]{lee2016}%
  \BibitemOpen
  \bibfield  {author} {\bibinfo {author} {\bibfnamefont {Tony~E.}\ \bibnamefont
  {Lee}},\ }\bibfield  {title} {\enquote {\bibinfo {title} {Anomalous edge
  state in a non-hermitian lattice},}\ }\href {\doibase
  10.1103/PhysRevLett.116.133903} {\bibfield  {journal} {\bibinfo  {journal}
  {Phys. Rev. Lett.}\ }\textbf {\bibinfo {volume} {116}},\ \bibinfo {pages}
  {133903} (\bibinfo {year} {2016})}\BibitemShut {NoStop}%
\bibitem [{\citenamefont {Leykam}\ \emph {et~al.}(2017)\citenamefont {Leykam},
  \citenamefont {Bliokh}, \citenamefont {Huang}, \citenamefont {Chong},\ and\
  \citenamefont {Nori}}]{leykam2018}%
  \BibitemOpen
  \bibfield  {author} {\bibinfo {author} {\bibfnamefont {Daniel}\ \bibnamefont
  {Leykam}}, \bibinfo {author} {\bibfnamefont {Konstantin~Y.}\ \bibnamefont
  {Bliokh}}, \bibinfo {author} {\bibfnamefont {Chunli}\ \bibnamefont {Huang}},
  \bibinfo {author} {\bibfnamefont {Y.~D.}\ \bibnamefont {Chong}}, \ and\
  \bibinfo {author} {\bibfnamefont {Franco}\ \bibnamefont {Nori}},\ }\bibfield
  {title} {\enquote {\bibinfo {title} {Edge modes, degeneracies, and
  topological numbers in non-hermitian systems},}\ }\href {\doibase
  10.1103/PhysRevLett.118.040401} {\bibfield  {journal} {\bibinfo  {journal}
  {Phys. Rev. Lett.}\ }\textbf {\bibinfo {volume} {118}},\ \bibinfo {pages}
  {040401} (\bibinfo {year} {2017})}\BibitemShut {NoStop}%
\bibitem [{\citenamefont {Shen}\ \emph {et~al.}(2018)\citenamefont {Shen},
  \citenamefont {Zhen},\ and\ \citenamefont {Fu}}]{shen2018}%
  \BibitemOpen
  \bibfield  {author} {\bibinfo {author} {\bibfnamefont {Huitao}\ \bibnamefont
  {Shen}}, \bibinfo {author} {\bibfnamefont {Bo}~\bibnamefont {Zhen}}, \ and\
  \bibinfo {author} {\bibfnamefont {Liang}\ \bibnamefont {Fu}},\ }\bibfield
  {title} {\enquote {\bibinfo {title} {Topological band theory for
  non-hermitian hamiltonians},}\ }\href {\doibase
  10.1103/PhysRevLett.120.146402} {\bibfield  {journal} {\bibinfo  {journal}
  {Phys. Rev. Lett.}\ }\textbf {\bibinfo {volume} {120}},\ \bibinfo {pages}
  {146402} (\bibinfo {year} {2018})}\BibitemShut {NoStop}%
\bibitem [{\citenamefont {Kunst}\ \emph {et~al.}(2018)\citenamefont {Kunst},
  \citenamefont {Edvardsson}, \citenamefont {Budich},\ and\ \citenamefont
  {Bergholtz}}]{kunst2018}%
  \BibitemOpen
  \bibfield  {author} {\bibinfo {author} {\bibfnamefont {Flore~K.}\
  \bibnamefont {Kunst}}, \bibinfo {author} {\bibfnamefont {Elisabet}\
  \bibnamefont {Edvardsson}}, \bibinfo {author} {\bibfnamefont {Jan~Carl}\
  \bibnamefont {Budich}}, \ and\ \bibinfo {author} {\bibfnamefont {Emil~J.}\
  \bibnamefont {Bergholtz}},\ }\bibfield  {title} {\enquote {\bibinfo {title}
  {Biorthogonal bulk-boundary correspondence in non-hermitian systems},}\
  }\href {\doibase 10.1103/PhysRevLett.121.026808} {\bibfield  {journal}
  {\bibinfo  {journal} {Phys. Rev. Lett.}\ }\textbf {\bibinfo {volume} {121}},\
  \bibinfo {pages} {026808} (\bibinfo {year} {2018})}\BibitemShut {NoStop}%
\bibitem [{\citenamefont {Yao}\ and\ \citenamefont {Wang}(2018)}]{yao201801}%
  \BibitemOpen
  \bibfield  {author} {\bibinfo {author} {\bibfnamefont {Shunyu}\ \bibnamefont
  {Yao}}\ and\ \bibinfo {author} {\bibfnamefont {Zhong}\ \bibnamefont {Wang}},\
  }\bibfield  {title} {\enquote {\bibinfo {title} {Edge states and topological
  invariants of non-hermitian systems},}\ }\href {\doibase
  10.1103/PhysRevLett.121.086803} {\bibfield  {journal} {\bibinfo  {journal}
  {Phys. Rev. Lett.}\ }\textbf {\bibinfo {volume} {121}},\ \bibinfo {pages}
  {086803} (\bibinfo {year} {2018})}\BibitemShut {NoStop}%
\bibitem [{\citenamefont {Yao}\ \emph {et~al.}(2018)\citenamefont {Yao},
  \citenamefont {Song},\ and\ \citenamefont {Wang}}]{yao201802}%
  \BibitemOpen
  \bibfield  {author} {\bibinfo {author} {\bibfnamefont {Shunyu}\ \bibnamefont
  {Yao}}, \bibinfo {author} {\bibfnamefont {Fei}\ \bibnamefont {Song}}, \ and\
  \bibinfo {author} {\bibfnamefont {Zhong}\ \bibnamefont {Wang}},\ }\bibfield
  {title} {\enquote {\bibinfo {title} {Non-hermitian chern bands},}\ }\href
  {\doibase 10.1103/PhysRevLett.121.136802} {\bibfield  {journal} {\bibinfo
  {journal} {Phys. Rev. Lett.}\ }\textbf {\bibinfo {volume} {121}},\ \bibinfo
  {pages} {136802} (\bibinfo {year} {2018})}\BibitemShut {NoStop}%
\bibitem [{\citenamefont {Esaki}\ \emph {et~al.}(2011)\citenamefont {Esaki},
  \citenamefont {Sato}, \citenamefont {Hasebe},\ and\ \citenamefont
  {Kohmoto}}]{sato2011}%
  \BibitemOpen
  \bibfield  {author} {\bibinfo {author} {\bibfnamefont {Kenta}\ \bibnamefont
  {Esaki}}, \bibinfo {author} {\bibfnamefont {Masatoshi}\ \bibnamefont {Sato}},
  \bibinfo {author} {\bibfnamefont {Kazuki}\ \bibnamefont {Hasebe}}, \ and\
  \bibinfo {author} {\bibfnamefont {Mahito}\ \bibnamefont {Kohmoto}},\
  }\bibfield  {title} {\enquote {\bibinfo {title} {Edge states and topological
  phases in non-hermitian systems},}\ }\href {\doibase
  10.1103/PhysRevB.84.205128} {\bibfield  {journal} {\bibinfo  {journal} {Phys.
  Rev. B}\ }\textbf {\bibinfo {volume} {84}},\ \bibinfo {pages} {205128}
  (\bibinfo {year} {2011})}\BibitemShut {NoStop}%
\bibitem [{\citenamefont {Kawabata}\ \emph {et~al.}(2018)\citenamefont
  {Kawabata}, \citenamefont {Shiozaki},\ and\ \citenamefont
  {Ueda}}]{kawabata2018}%
  \BibitemOpen
  \bibfield  {author} {\bibinfo {author} {\bibfnamefont {Kohei}\ \bibnamefont
  {Kawabata}}, \bibinfo {author} {\bibfnamefont {Ken}\ \bibnamefont
  {Shiozaki}}, \ and\ \bibinfo {author} {\bibfnamefont {Masahito}\ \bibnamefont
  {Ueda}},\ }\bibfield  {title} {\enquote {\bibinfo {title} {Anomalous helical
  edge states in a non-hermitian chern insulator},}\ }\href {\doibase
  10.1103/PhysRevB.98.165148} {\bibfield  {journal} {\bibinfo  {journal} {Phys.
  Rev. B}\ }\textbf {\bibinfo {volume} {98}},\ \bibinfo {pages} {165148}
  (\bibinfo {year} {2018})}\BibitemShut {NoStop}%
\bibitem [{\citenamefont {Xiong}(2018)}]{xiong2018}%
  \BibitemOpen
  \bibfield  {author} {\bibinfo {author} {\bibfnamefont {Ye}~\bibnamefont
  {Xiong}},\ }\bibfield  {title} {\enquote {\bibinfo {title} {Why does bulk
  boundary correspondence fail in some non-hermitian topological models},}\
  }\href {\doibase 10.1088/2399-6528/aab64a} {\bibfield  {journal} {\bibinfo
  {journal} {Journal of Physics Communications}\ }\textbf {\bibinfo {volume}
  {2}},\ \bibinfo {pages} {035043} (\bibinfo {year} {2018})}\BibitemShut
  {NoStop}%
\bibitem [{\citenamefont {Gong}\ \emph {et~al.}(2018)\citenamefont {Gong},
  \citenamefont {Ashida}, \citenamefont {Kawabata}, \citenamefont {Takasan},
  \citenamefont {Higashikawa},\ and\ \citenamefont {Ueda}}]{gong2018}%
  \BibitemOpen
  \bibfield  {author} {\bibinfo {author} {\bibfnamefont {Zongping}\
  \bibnamefont {Gong}}, \bibinfo {author} {\bibfnamefont {Yuto}\ \bibnamefont
  {Ashida}}, \bibinfo {author} {\bibfnamefont {Kohei}\ \bibnamefont
  {Kawabata}}, \bibinfo {author} {\bibfnamefont {Kazuaki}\ \bibnamefont
  {Takasan}}, \bibinfo {author} {\bibfnamefont {Sho}\ \bibnamefont
  {Higashikawa}}, \ and\ \bibinfo {author} {\bibfnamefont {Masahito}\
  \bibnamefont {Ueda}},\ }\bibfield  {title} {\enquote {\bibinfo {title}
  {Topological phases of non-hermitian systems},}\ }\href {\doibase
  10.1103/PhysRevX.8.031079} {\bibfield  {journal} {\bibinfo  {journal} {Phys.
  Rev. X}\ }\textbf {\bibinfo {volume} {8}},\ \bibinfo {pages} {031079}
  (\bibinfo {year} {2018})}\BibitemShut {NoStop}%
\bibitem [{\citenamefont {Longhi}(2019)}]{longhi2019}%
  \BibitemOpen
  \bibfield  {author} {\bibinfo {author} {\bibfnamefont {S.}~\bibnamefont
  {Longhi}},\ }\bibfield  {title} {\enquote {\bibinfo {title} {Topological
  phase transition in non-hermitian quasicrystals},}\ }\href {\doibase
  10.1103/PhysRevLett.122.237601} {\bibfield  {journal} {\bibinfo  {journal}
  {Phys. Rev. Lett.}\ }\textbf {\bibinfo {volume} {122}},\ \bibinfo {pages}
  {237601} (\bibinfo {year} {2019})}\BibitemShut {NoStop}%
\bibitem [{\citenamefont {Yokomizo}\ and\ \citenamefont
  {Murakami}(2019)}]{yokomizo2019}%
  \BibitemOpen
  \bibfield  {author} {\bibinfo {author} {\bibfnamefont {Kazuki}\ \bibnamefont
  {Yokomizo}}\ and\ \bibinfo {author} {\bibfnamefont {Shuichi}\ \bibnamefont
  {Murakami}},\ }\bibfield  {title} {\enquote {\bibinfo {title} {Non-bloch band
  theory of non-hermitian systems},}\ }\href {\doibase
  10.1103/PhysRevLett.123.066404} {\bibfield  {journal} {\bibinfo  {journal}
  {Phys. Rev. Lett.}\ }\textbf {\bibinfo {volume} {123}},\ \bibinfo {pages}
  {066404} (\bibinfo {year} {2019})}\BibitemShut {NoStop}%
\bibitem [{\citenamefont {Song}\ \emph {et~al.}(2019)\citenamefont {Song},
  \citenamefont {Yao},\ and\ \citenamefont {Wang}}]{song2019}%
  \BibitemOpen
  \bibfield  {author} {\bibinfo {author} {\bibfnamefont {Fei}\ \bibnamefont
  {Song}}, \bibinfo {author} {\bibfnamefont {Shunyu}\ \bibnamefont {Yao}}, \
  and\ \bibinfo {author} {\bibfnamefont {Zhong}\ \bibnamefont {Wang}},\
  }\bibfield  {title} {\enquote {\bibinfo {title} {Non-hermitian skin effect
  and chiral damping in open quantum systems},}\ }\href {\doibase
  10.1103/PhysRevLett.123.170401} {\bibfield  {journal} {\bibinfo  {journal}
  {Phys. Rev. Lett.}\ }\textbf {\bibinfo {volume} {123}},\ \bibinfo {pages}
  {170401} (\bibinfo {year} {2019})}\BibitemShut {NoStop}%
\bibitem [{\citenamefont {Imura}\ and\ \citenamefont
  {Takane}(2019)}]{imura2019}%
  \BibitemOpen
  \bibfield  {author} {\bibinfo {author} {\bibfnamefont {Ken-Ichiro}\
  \bibnamefont {Imura}}\ and\ \bibinfo {author} {\bibfnamefont {Yositake}\
  \bibnamefont {Takane}},\ }\bibfield  {title} {\enquote {\bibinfo {title}
  {Generalized bulk-edge correspondence for non-hermitian topological
  systems},}\ }\href {\doibase 10.1103/PhysRevB.100.165430} {\bibfield
  {journal} {\bibinfo  {journal} {Phys. Rev. B}\ }\textbf {\bibinfo {volume}
  {100}},\ \bibinfo {pages} {165430} (\bibinfo {year} {2019})}\BibitemShut
  {NoStop}%
\bibitem [{\citenamefont {Edvardsson}\ \emph {et~al.}(2019)\citenamefont
  {Edvardsson}, \citenamefont {Kunst},\ and\ \citenamefont
  {Bergholtz}}]{edvardsson2019}%
  \BibitemOpen
  \bibfield  {author} {\bibinfo {author} {\bibfnamefont {Elisabet}\
  \bibnamefont {Edvardsson}}, \bibinfo {author} {\bibfnamefont {Flore~K.}\
  \bibnamefont {Kunst}}, \ and\ \bibinfo {author} {\bibfnamefont {Emil~J.}\
  \bibnamefont {Bergholtz}},\ }\bibfield  {title} {\enquote {\bibinfo {title}
  {Non-hermitian extensions of higher-order topological phases and their
  biorthogonal bulk-boundary correspondence},}\ }\href {\doibase
  10.1103/PhysRevB.99.081302} {\bibfield  {journal} {\bibinfo  {journal} {Phys.
  Rev. B}\ }\textbf {\bibinfo {volume} {99}},\ \bibinfo {pages} {081302}
  (\bibinfo {year} {2019})}\BibitemShut {NoStop}%
\bibitem [{\citenamefont {Ezawa}(2019)}]{ezawa2019}%
  \BibitemOpen
  \bibfield  {author} {\bibinfo {author} {\bibfnamefont {Motohiko}\
  \bibnamefont {Ezawa}},\ }\bibfield  {title} {\enquote {\bibinfo {title}
  {Non-hermitian boundary and interface states in nonreciprocal higher-order
  topological metals and electrical circuits},}\ }\href {\doibase
  10.1103/PhysRevB.99.121411} {\bibfield  {journal} {\bibinfo  {journal} {Phys.
  Rev. B}\ }\textbf {\bibinfo {volume} {99}},\ \bibinfo {pages} {121411}
  (\bibinfo {year} {2019})}\BibitemShut {NoStop}%
\bibitem [{\citenamefont {Liu}\ \emph {et~al.}(2019)\citenamefont {Liu},
  \citenamefont {Zhang}, \citenamefont {Ai}, \citenamefont {Gong},
  \citenamefont {Kawabata}, \citenamefont {Ueda},\ and\ \citenamefont
  {Nori}}]{kawabata2019second}%
  \BibitemOpen
  \bibfield  {author} {\bibinfo {author} {\bibfnamefont {Tao}\ \bibnamefont
  {Liu}}, \bibinfo {author} {\bibfnamefont {Yu-Ran}\ \bibnamefont {Zhang}},
  \bibinfo {author} {\bibfnamefont {Qing}\ \bibnamefont {Ai}}, \bibinfo
  {author} {\bibfnamefont {Zongping}\ \bibnamefont {Gong}}, \bibinfo {author}
  {\bibfnamefont {Kohei}\ \bibnamefont {Kawabata}}, \bibinfo {author}
  {\bibfnamefont {Masahito}\ \bibnamefont {Ueda}}, \ and\ \bibinfo {author}
  {\bibfnamefont {Franco}\ \bibnamefont {Nori}},\ }\bibfield  {title} {\enquote
  {\bibinfo {title} {Second-order topological phases in non-hermitian
  systems},}\ }\href {\doibase 10.1103/PhysRevLett.122.076801} {\bibfield
  {journal} {\bibinfo  {journal} {Phys. Rev. Lett.}\ }\textbf {\bibinfo
  {volume} {122}},\ \bibinfo {pages} {076801} (\bibinfo {year}
  {2019})}\BibitemShut {NoStop}%
\bibitem [{\citenamefont {Lee}\ and\ \citenamefont {Thomale}(2019)}]{lee2019}%
  \BibitemOpen
  \bibfield  {author} {\bibinfo {author} {\bibfnamefont {Ching~Hua}\
  \bibnamefont {Lee}}\ and\ \bibinfo {author} {\bibfnamefont {Ronny}\
  \bibnamefont {Thomale}},\ }\bibfield  {title} {\enquote {\bibinfo {title}
  {Anatomy of skin modes and topology in non-hermitian systems},}\ }\href
  {\doibase 10.1103/PhysRevB.99.201103} {\bibfield  {journal} {\bibinfo
  {journal} {Phys. Rev. B}\ }\textbf {\bibinfo {volume} {99}},\ \bibinfo
  {pages} {201103} (\bibinfo {year} {2019})}\BibitemShut {NoStop}%
\bibitem [{\citenamefont {Okuma}\ and\ \citenamefont {Sato}(2019)}]{okuma2019}%
  \BibitemOpen
  \bibfield  {author} {\bibinfo {author} {\bibfnamefont {Nobuyuki}\
  \bibnamefont {Okuma}}\ and\ \bibinfo {author} {\bibfnamefont {Masatoshi}\
  \bibnamefont {Sato}},\ }\bibfield  {title} {\enquote {\bibinfo {title}
  {Topological phase transition driven by infinitesimal instability: Majorana
  fermions in non-hermitian spintronics},}\ }\href {\doibase
  10.1103/PhysRevLett.123.097701} {\bibfield  {journal} {\bibinfo  {journal}
  {Phys. Rev. Lett.}\ }\textbf {\bibinfo {volume} {123}},\ \bibinfo {pages}
  {097701} (\bibinfo {year} {2019})}\BibitemShut {NoStop}%
\bibitem [{\citenamefont {Wang}\ \emph {et~al.}(2020)\citenamefont {Wang},
  \citenamefont {Guo},\ and\ \citenamefont {Kou}}]{su2020}%
  \BibitemOpen
  \bibfield  {author} {\bibinfo {author} {\bibfnamefont {Xiao-Ran}\
  \bibnamefont {Wang}}, \bibinfo {author} {\bibfnamefont {Cui-Xian}\
  \bibnamefont {Guo}}, \ and\ \bibinfo {author} {\bibfnamefont {Su-Peng}\
  \bibnamefont {Kou}},\ }\bibfield  {title} {\enquote {\bibinfo {title}
  {Defective edge states and number-anomalous bulk-boundary correspondence in
  non-hermitian topological systems},}\ }\href {\doibase
  10.1103/PhysRevB.101.121116} {\bibfield  {journal} {\bibinfo  {journal}
  {Phys. Rev. B}\ }\textbf {\bibinfo {volume} {101}},\ \bibinfo {pages}
  {121116} (\bibinfo {year} {2020})}\BibitemShut {NoStop}%
\bibitem [{\citenamefont {Borgnia}\ \emph {et~al.}(2020)\citenamefont
  {Borgnia}, \citenamefont {Kruchkov},\ and\ \citenamefont
  {Slager}}]{slager2020}%
  \BibitemOpen
  \bibfield  {author} {\bibinfo {author} {\bibfnamefont {Dan~S.}\ \bibnamefont
  {Borgnia}}, \bibinfo {author} {\bibfnamefont {Alex~Jura}\ \bibnamefont
  {Kruchkov}}, \ and\ \bibinfo {author} {\bibfnamefont {Robert-Jan}\
  \bibnamefont {Slager}},\ }\bibfield  {title} {\enquote {\bibinfo {title}
  {Non-hermitian boundary modes and topology},}\ }\href {\doibase
  10.1103/PhysRevLett.124.056802} {\bibfield  {journal} {\bibinfo  {journal}
  {Phys. Rev. Lett.}\ }\textbf {\bibinfo {volume} {124}},\ \bibinfo {pages}
  {056802} (\bibinfo {year} {2020})}\BibitemShut {NoStop}%
\bibitem [{\citenamefont {Yang}\ \emph
  {et~al.}(2020{\natexlab{a}})\citenamefont {Yang}, \citenamefont {Chiu},
  \citenamefont {Fang},\ and\ \citenamefont {Hu}}]{jones2020}%
  \BibitemOpen
  \bibfield  {author} {\bibinfo {author} {\bibfnamefont {Zhesen}\ \bibnamefont
  {Yang}}, \bibinfo {author} {\bibfnamefont {Ching-Kai}\ \bibnamefont {Chiu}},
  \bibinfo {author} {\bibfnamefont {Chen}\ \bibnamefont {Fang}}, \ and\
  \bibinfo {author} {\bibfnamefont {Jiangping}\ \bibnamefont {Hu}},\ }\bibfield
   {title} {\enquote {\bibinfo {title} {Jones polynomial and knot transitions
  in hermitian and non-hermitian topological semimetals},}\ }\href {\doibase
  10.1103/PhysRevLett.124.186402} {\bibfield  {journal} {\bibinfo  {journal}
  {Phys. Rev. Lett.}\ }\textbf {\bibinfo {volume} {124}},\ \bibinfo {pages}
  {186402} (\bibinfo {year} {2020}{\natexlab{a}})}\BibitemShut {NoStop}%
\bibitem [{\citenamefont {Okuma}\ \emph {et~al.}(2020)\citenamefont {Okuma},
  \citenamefont {Kawabata}, \citenamefont {Shiozaki},\ and\ \citenamefont
  {Sato}}]{origin2020}%
  \BibitemOpen
  \bibfield  {author} {\bibinfo {author} {\bibfnamefont {Nobuyuki}\
  \bibnamefont {Okuma}}, \bibinfo {author} {\bibfnamefont {Kohei}\ \bibnamefont
  {Kawabata}}, \bibinfo {author} {\bibfnamefont {Ken}\ \bibnamefont
  {Shiozaki}}, \ and\ \bibinfo {author} {\bibfnamefont {Masatoshi}\
  \bibnamefont {Sato}},\ }\bibfield  {title} {\enquote {\bibinfo {title}
  {Topological origin of non-hermitian skin effects},}\ }\href {\doibase
  10.1103/PhysRevLett.124.086801} {\bibfield  {journal} {\bibinfo  {journal}
  {Phys. Rev. Lett.}\ }\textbf {\bibinfo {volume} {124}},\ \bibinfo {pages}
  {086801} (\bibinfo {year} {2020})}\BibitemShut {NoStop}%
\bibitem [{\citenamefont {Kawabata}\ \emph
  {et~al.}(2019{\natexlab{a}})\citenamefont {Kawabata}, \citenamefont
  {Shiozaki}, \citenamefont {Ueda},\ and\ \citenamefont {Sato}}]{kawabataprx}%
  \BibitemOpen
  \bibfield  {author} {\bibinfo {author} {\bibfnamefont {Kohei}\ \bibnamefont
  {Kawabata}}, \bibinfo {author} {\bibfnamefont {Ken}\ \bibnamefont
  {Shiozaki}}, \bibinfo {author} {\bibfnamefont {Masahito}\ \bibnamefont
  {Ueda}}, \ and\ \bibinfo {author} {\bibfnamefont {Masatoshi}\ \bibnamefont
  {Sato}},\ }\bibfield  {title} {\enquote {\bibinfo {title} {Symmetry and
  topology in non-hermitian physics},}\ }\href {\doibase
  10.1103/PhysRevX.9.041015} {\bibfield  {journal} {\bibinfo  {journal} {Phys.
  Rev. X}\ }\textbf {\bibinfo {volume} {9}},\ \bibinfo {pages} {041015}
  (\bibinfo {year} {2019}{\natexlab{a}})}\BibitemShut {NoStop}%
\bibitem [{\citenamefont {Kawabata}\ \emph
  {et~al.}(2020{\natexlab{a}})\citenamefont {Kawabata}, \citenamefont {Okuma},\
  and\ \citenamefont {Sato}}]{kawabata20}%
  \BibitemOpen
  \bibfield  {author} {\bibinfo {author} {\bibfnamefont {Kohei}\ \bibnamefont
  {Kawabata}}, \bibinfo {author} {\bibfnamefont {Nobuyuki}\ \bibnamefont
  {Okuma}}, \ and\ \bibinfo {author} {\bibfnamefont {Masatoshi}\ \bibnamefont
  {Sato}},\ }\bibfield  {title} {\enquote {\bibinfo {title} {Non-bloch band
  theory of non-hermitian hamiltonians in the symplectic class},}\ }\href
  {\doibase 10.1103/PhysRevB.101.195147} {\bibfield  {journal} {\bibinfo
  {journal} {Phys. Rev. B}\ }\textbf {\bibinfo {volume} {101}},\ \bibinfo
  {pages} {195147} (\bibinfo {year} {2020}{\natexlab{a}})}\BibitemShut
  {NoStop}%
\bibitem [{\citenamefont {Okugawa}\ \emph {et~al.}(2020)\citenamefont
  {Okugawa}, \citenamefont {Takahashi},\ and\ \citenamefont
  {Yokomizo}}]{okugawa2020}%
  \BibitemOpen
  \bibfield  {author} {\bibinfo {author} {\bibfnamefont {Ryo}\ \bibnamefont
  {Okugawa}}, \bibinfo {author} {\bibfnamefont {Ryo}\ \bibnamefont
  {Takahashi}}, \ and\ \bibinfo {author} {\bibfnamefont {Kazuki}\ \bibnamefont
  {Yokomizo}},\ }\bibfield  {title} {\enquote {\bibinfo {title} {Second-order
  topological non-hermitian skin effects},}\ }\href {\doibase
  10.1103/PhysRevB.102.241202} {\bibfield  {journal} {\bibinfo  {journal}
  {Phys. Rev. B}\ }\textbf {\bibinfo {volume} {102}},\ \bibinfo {pages}
  {241202} (\bibinfo {year} {2020})}\BibitemShut {NoStop}%
\bibitem [{\citenamefont {Kawabata}\ \emph
  {et~al.}(2020{\natexlab{b}})\citenamefont {Kawabata}, \citenamefont {Sato},\
  and\ \citenamefont {Shiozaki}}]{kawabatahigher}%
  \BibitemOpen
  \bibfield  {author} {\bibinfo {author} {\bibfnamefont {Kohei}\ \bibnamefont
  {Kawabata}}, \bibinfo {author} {\bibfnamefont {Masatoshi}\ \bibnamefont
  {Sato}}, \ and\ \bibinfo {author} {\bibfnamefont {Ken}\ \bibnamefont
  {Shiozaki}},\ }\bibfield  {title} {\enquote {\bibinfo {title} {Higher-order
  non-hermitian skin effect},}\ }\href {\doibase 10.1103/PhysRevB.102.205118}
  {\bibfield  {journal} {\bibinfo  {journal} {Phys. Rev. B}\ }\textbf {\bibinfo
  {volume} {102}},\ \bibinfo {pages} {205118} (\bibinfo {year}
  {2020}{\natexlab{b}})}\BibitemShut {NoStop}%
\bibitem [{\citenamefont {Zhang}\ \emph
  {et~al.}(2020{\natexlab{a}})\citenamefont {Zhang}, \citenamefont {Yang},\
  and\ \citenamefont {Fang}}]{yang202001}%
  \BibitemOpen
  \bibfield  {author} {\bibinfo {author} {\bibfnamefont {Kai}\ \bibnamefont
  {Zhang}}, \bibinfo {author} {\bibfnamefont {Zhesen}\ \bibnamefont {Yang}}, \
  and\ \bibinfo {author} {\bibfnamefont {Chen}\ \bibnamefont {Fang}},\
  }\bibfield  {title} {\enquote {\bibinfo {title} {Correspondence between
  winding numbers and skin modes in non-hermitian systems},}\ }\href {\doibase
  10.1103/PhysRevLett.125.126402} {\bibfield  {journal} {\bibinfo  {journal}
  {Phys. Rev. Lett.}\ }\textbf {\bibinfo {volume} {125}},\ \bibinfo {pages}
  {126402} (\bibinfo {year} {2020}{\natexlab{a}})}\BibitemShut {NoStop}%
\bibitem [{\citenamefont {Yang}\ \emph
  {et~al.}(2020{\natexlab{b}})\citenamefont {Yang}, \citenamefont {Zhang},
  \citenamefont {Fang},\ and\ \citenamefont {Hu}}]{yang202002}%
  \BibitemOpen
  \bibfield  {author} {\bibinfo {author} {\bibfnamefont {Zhesen}\ \bibnamefont
  {Yang}}, \bibinfo {author} {\bibfnamefont {Kai}\ \bibnamefont {Zhang}},
  \bibinfo {author} {\bibfnamefont {Chen}\ \bibnamefont {Fang}}, \ and\
  \bibinfo {author} {\bibfnamefont {Jiangping}\ \bibnamefont {Hu}},\ }\bibfield
   {title} {\enquote {\bibinfo {title} {Non-hermitian bulk-boundary
  correspondence and auxiliary generalized brillouin zone theory},}\ }\href
  {\doibase 10.1103/PhysRevLett.125.226402} {\bibfield  {journal} {\bibinfo
  {journal} {Phys. Rev. Lett.}\ }\textbf {\bibinfo {volume} {125}},\ \bibinfo
  {pages} {226402} (\bibinfo {year} {2020}{\natexlab{b}})}\BibitemShut
  {NoStop}%
\bibitem [{\citenamefont {Hu}\ and\ \citenamefont {Zhao}(2021)}]{knot2021}%
  \BibitemOpen
  \bibfield  {author} {\bibinfo {author} {\bibfnamefont {Haiping}\ \bibnamefont
  {Hu}}\ and\ \bibinfo {author} {\bibfnamefont {Erhai}\ \bibnamefont {Zhao}},\
  }\bibfield  {title} {\enquote {\bibinfo {title} {Knots and non-hermitian
  bloch bands},}\ }\href {\doibase 10.1103/PhysRevLett.126.010401} {\bibfield
  {journal} {\bibinfo  {journal} {Phys. Rev. Lett.}\ }\textbf {\bibinfo
  {volume} {126}},\ \bibinfo {pages} {010401} (\bibinfo {year}
  {2021})}\BibitemShut {NoStop}%
\bibitem [{\citenamefont {Fu}\ \emph {et~al.}(2021)\citenamefont {Fu},
  \citenamefont {Hu},\ and\ \citenamefont {Wan}}]{fu2021}%
  \BibitemOpen
  \bibfield  {author} {\bibinfo {author} {\bibfnamefont {Yongxu}\ \bibnamefont
  {Fu}}, \bibinfo {author} {\bibfnamefont {Jihan}\ \bibnamefont {Hu}}, \ and\
  \bibinfo {author} {\bibfnamefont {Shaolong}\ \bibnamefont {Wan}},\ }\bibfield
   {title} {\enquote {\bibinfo {title} {Non-hermitian second-order skin and
  topological modes},}\ }\href {\doibase 10.1103/PhysRevB.103.045420}
  {\bibfield  {journal} {\bibinfo  {journal} {Phys. Rev. B}\ }\textbf {\bibinfo
  {volume} {103}},\ \bibinfo {pages} {045420} (\bibinfo {year}
  {2021})}\BibitemShut {NoStop}%
\bibitem [{\citenamefont {Xie}\ \emph {et~al.}(2021)\citenamefont {Xie},
  \citenamefont {Wu}, \citenamefont {Zhang}, \citenamefont {Jin},\ and\
  \citenamefont {Song}}]{xie2021}%
  \BibitemOpen
  \bibfield  {author} {\bibinfo {author} {\bibfnamefont {L.~C.}\ \bibnamefont
  {Xie}}, \bibinfo {author} {\bibfnamefont {H.~C.}\ \bibnamefont {Wu}},
  \bibinfo {author} {\bibfnamefont {X.~Z.}\ \bibnamefont {Zhang}}, \bibinfo
  {author} {\bibfnamefont {L.}~\bibnamefont {Jin}}, \ and\ \bibinfo {author}
  {\bibfnamefont {Z.}~\bibnamefont {Song}},\ }\bibfield  {title} {\enquote
  {\bibinfo {title} {Two-dimensional anisotropic non-hermitian lieb lattice},}\
  }\href {\doibase 10.1103/PhysRevB.104.125406} {\bibfield  {journal} {\bibinfo
   {journal} {Phys. Rev. B}\ }\textbf {\bibinfo {volume} {104}},\ \bibinfo
  {pages} {125406} (\bibinfo {year} {2021})}\BibitemShut {NoStop}%
\bibitem [{\citenamefont {Longhi}(2021)}]{longhi2021}%
  \BibitemOpen
  \bibfield  {author} {\bibinfo {author} {\bibfnamefont {Stefano}\ \bibnamefont
  {Longhi}},\ }\bibfield  {title} {\enquote {\bibinfo {title} {Non-hermitian
  skin effect beyond the tight-binding models},}\ }\href {\doibase
  10.1103/PhysRevB.104.125109} {\bibfield  {journal} {\bibinfo  {journal}
  {Phys. Rev. B}\ }\textbf {\bibinfo {volume} {104}},\ \bibinfo {pages}
  {125109} (\bibinfo {year} {2021})}\BibitemShut {NoStop}%
\bibitem [{\citenamefont {Li}\ and\ \citenamefont {Wan}(2021)}]{li2021}%
  \BibitemOpen
  \bibfield  {author} {\bibinfo {author} {\bibfnamefont {Haoshu}\ \bibnamefont
  {Li}}\ and\ \bibinfo {author} {\bibfnamefont {Shaolong}\ \bibnamefont
  {Wan}},\ }\href@noop {} {\enquote {\bibinfo {title} {Exact formulas of the
  end-to-end green's functions in non-hermitian systems},}\ } (\bibinfo {year}
  {2021}),\ \Eprint {http://arxiv.org/abs/2109.03045} {arXiv:2109.03045
  [cond-mat.mes-hall]} \BibitemShut {NoStop}%
\bibitem [{\citenamefont {Heiss}(2012)}]{heiss2012}%
  \BibitemOpen
  \bibfield  {author} {\bibinfo {author} {\bibfnamefont {W~D}\ \bibnamefont
  {Heiss}},\ }\bibfield  {title} {\enquote {\bibinfo {title} {The physics of
  exceptional points},}\ }\href {\doibase 10.1088/1751-8113/45/44/444016}
  {\bibfield  {journal} {\bibinfo  {journal} {Journal of Physics A:
  Mathematical and Theoretical}\ }\textbf {\bibinfo {volume} {45}},\ \bibinfo
  {pages} {444016} (\bibinfo {year} {2012})}\BibitemShut {NoStop}%
\bibitem [{\citenamefont {Yin}\ \emph {et~al.}(2018)\citenamefont {Yin},
  \citenamefont {Jiang}, \citenamefont {Li}, \citenamefont {L\"u},\ and\
  \citenamefont {Chen}}]{yin2018}%
  \BibitemOpen
  \bibfield  {author} {\bibinfo {author} {\bibfnamefont {Chuanhao}\
  \bibnamefont {Yin}}, \bibinfo {author} {\bibfnamefont {Hui}\ \bibnamefont
  {Jiang}}, \bibinfo {author} {\bibfnamefont {Linhu}\ \bibnamefont {Li}},
  \bibinfo {author} {\bibfnamefont {Rong}\ \bibnamefont {L\"u}}, \ and\
  \bibinfo {author} {\bibfnamefont {Shu}\ \bibnamefont {Chen}},\ }\bibfield
  {title} {\enquote {\bibinfo {title} {Geometrical meaning of winding number
  and its characterization of topological phases in one-dimensional chiral
  non-hermitian systems},}\ }\href {\doibase 10.1103/PhysRevA.97.052115}
  {\bibfield  {journal} {\bibinfo  {journal} {Phys. Rev. A}\ }\textbf {\bibinfo
  {volume} {97}},\ \bibinfo {pages} {052115} (\bibinfo {year}
  {2018})}\BibitemShut {NoStop}%
\bibitem [{\citenamefont {Martinez~Alvarez}\ \emph {et~al.}(2018)\citenamefont
  {Martinez~Alvarez}, \citenamefont {Barrios~Vargas},\ and\ \citenamefont
  {Foa~Torres}}]{alvarez2018}%
  \BibitemOpen
  \bibfield  {author} {\bibinfo {author} {\bibfnamefont {V.~M.}\ \bibnamefont
  {Martinez~Alvarez}}, \bibinfo {author} {\bibfnamefont {J.~E.}\ \bibnamefont
  {Barrios~Vargas}}, \ and\ \bibinfo {author} {\bibfnamefont {L.~E.~F.}\
  \bibnamefont {Foa~Torres}},\ }\bibfield  {title} {\enquote {\bibinfo {title}
  {Non-hermitian robust edge states in one dimension: Anomalous localization
  and eigenspace condensation at exceptional points},}\ }\href {\doibase
  10.1103/PhysRevB.97.121401} {\bibfield  {journal} {\bibinfo  {journal} {Phys.
  Rev. B}\ }\textbf {\bibinfo {volume} {97}},\ \bibinfo {pages} {121401}
  (\bibinfo {year} {2018})}\BibitemShut {NoStop}%
\bibitem [{\citenamefont {Kawabata}\ \emph
  {et~al.}(2019{\natexlab{b}})\citenamefont {Kawabata}, \citenamefont
  {Bessho},\ and\ \citenamefont {Sato}}]{kawabata2019}%
  \BibitemOpen
  \bibfield  {author} {\bibinfo {author} {\bibfnamefont {Kohei}\ \bibnamefont
  {Kawabata}}, \bibinfo {author} {\bibfnamefont {Takumi}\ \bibnamefont
  {Bessho}}, \ and\ \bibinfo {author} {\bibfnamefont {Masatoshi}\ \bibnamefont
  {Sato}},\ }\bibfield  {title} {\enquote {\bibinfo {title} {Classification of
  exceptional points and non-hermitian topological semimetals},}\ }\href
  {\doibase 10.1103/PhysRevLett.123.066405} {\bibfield  {journal} {\bibinfo
  {journal} {Phys. Rev. Lett.}\ }\textbf {\bibinfo {volume} {123}},\ \bibinfo
  {pages} {066405} (\bibinfo {year} {2019}{\natexlab{b}})}\BibitemShut
  {NoStop}%
\bibitem [{\citenamefont {Okugawa}\ and\ \citenamefont
  {Yokoyama}(2019)}]{okugawa2019}%
  \BibitemOpen
  \bibfield  {author} {\bibinfo {author} {\bibfnamefont {Ryo}\ \bibnamefont
  {Okugawa}}\ and\ \bibinfo {author} {\bibfnamefont {Takehito}\ \bibnamefont
  {Yokoyama}},\ }\bibfield  {title} {\enquote {\bibinfo {title} {Topological
  exceptional surfaces in non-hermitian systems with parity-time and
  parity-particle-hole symmetries},}\ }\href {\doibase
  10.1103/PhysRevB.99.041202} {\bibfield  {journal} {\bibinfo  {journal} {Phys.
  Rev. B}\ }\textbf {\bibinfo {volume} {99}},\ \bibinfo {pages} {041202}
  (\bibinfo {year} {2019})}\BibitemShut {NoStop}%
\bibitem [{\citenamefont {Budich}\ \emph {et~al.}(2019)\citenamefont {Budich},
  \citenamefont {Carlstr\"om}, \citenamefont {Kunst},\ and\ \citenamefont
  {Bergholtz}}]{bergholtz201901}%
  \BibitemOpen
  \bibfield  {author} {\bibinfo {author} {\bibfnamefont {Jan~Carl}\
  \bibnamefont {Budich}}, \bibinfo {author} {\bibfnamefont {Johan}\
  \bibnamefont {Carlstr\"om}}, \bibinfo {author} {\bibfnamefont {Flore~K.}\
  \bibnamefont {Kunst}}, \ and\ \bibinfo {author} {\bibfnamefont {Emil~J.}\
  \bibnamefont {Bergholtz}},\ }\bibfield  {title} {\enquote {\bibinfo {title}
  {Symmetry-protected nodal phases in non-hermitian systems},}\ }\href
  {\doibase 10.1103/PhysRevB.99.041406} {\bibfield  {journal} {\bibinfo
  {journal} {Phys. Rev. B}\ }\textbf {\bibinfo {volume} {99}},\ \bibinfo
  {pages} {041406} (\bibinfo {year} {2019})}\BibitemShut {NoStop}%
\bibitem [{\citenamefont {Carlstr\"om}\ \emph {et~al.}(2019)\citenamefont
  {Carlstr\"om}, \citenamefont {St\aa{}lhammar}, \citenamefont {Budich},\ and\
  \citenamefont {Bergholtz}}]{bergholtz201902}%
  \BibitemOpen
  \bibfield  {author} {\bibinfo {author} {\bibfnamefont {Johan}\ \bibnamefont
  {Carlstr\"om}}, \bibinfo {author} {\bibfnamefont {Marcus}\ \bibnamefont
  {St\aa{}lhammar}}, \bibinfo {author} {\bibfnamefont {Jan~Carl}\ \bibnamefont
  {Budich}}, \ and\ \bibinfo {author} {\bibfnamefont {Emil~J.}\ \bibnamefont
  {Bergholtz}},\ }\bibfield  {title} {\enquote {\bibinfo {title} {Knotted
  non-hermitian metals},}\ }\href {\doibase 10.1103/PhysRevB.99.161115}
  {\bibfield  {journal} {\bibinfo  {journal} {Phys. Rev. B}\ }\textbf {\bibinfo
  {volume} {99}},\ \bibinfo {pages} {161115} (\bibinfo {year}
  {2019})}\BibitemShut {NoStop}%
\bibitem [{\citenamefont {Yang}\ and\ \citenamefont {Hu}(2019)}]{yang2019}%
  \BibitemOpen
  \bibfield  {author} {\bibinfo {author} {\bibfnamefont {Zhesen}\ \bibnamefont
  {Yang}}\ and\ \bibinfo {author} {\bibfnamefont {Jiangping}\ \bibnamefont
  {Hu}},\ }\bibfield  {title} {\enquote {\bibinfo {title} {Non-hermitian
  hopf-link exceptional line semimetals},}\ }\href {\doibase
  10.1103/PhysRevB.99.081102} {\bibfield  {journal} {\bibinfo  {journal} {Phys.
  Rev. B}\ }\textbf {\bibinfo {volume} {99}},\ \bibinfo {pages} {081102}
  (\bibinfo {year} {2019})}\BibitemShut {NoStop}%
\bibitem [{\citenamefont {Li}\ \emph {et~al.}(2019)\citenamefont {Li},
  \citenamefont {Lee},\ and\ \citenamefont {Gong}}]{li2019}%
  \BibitemOpen
  \bibfield  {author} {\bibinfo {author} {\bibfnamefont {Linhu}\ \bibnamefont
  {Li}}, \bibinfo {author} {\bibfnamefont {Ching~Hua}\ \bibnamefont {Lee}}, \
  and\ \bibinfo {author} {\bibfnamefont {Jiangbin}\ \bibnamefont {Gong}},\
  }\bibfield  {title} {\enquote {\bibinfo {title} {Geometric characterization
  of non-hermitian topological systems through the singularity ring in
  pseudospin vector space},}\ }\href {\doibase 10.1103/PhysRevB.100.075403}
  {\bibfield  {journal} {\bibinfo  {journal} {Phys. Rev. B}\ }\textbf {\bibinfo
  {volume} {100}},\ \bibinfo {pages} {075403} (\bibinfo {year}
  {2019})}\BibitemShut {NoStop}%
\bibitem [{\citenamefont {Zhang}\ \emph
  {et~al.}(2020{\natexlab{b}})\citenamefont {Zhang}, \citenamefont {Yang},\
  and\ \citenamefont {Hu}}]{zhang2020}%
  \BibitemOpen
  \bibfield  {author} {\bibinfo {author} {\bibfnamefont {Zhicheng}\
  \bibnamefont {Zhang}}, \bibinfo {author} {\bibfnamefont {Zhesen}\
  \bibnamefont {Yang}}, \ and\ \bibinfo {author} {\bibfnamefont {Jiangping}\
  \bibnamefont {Hu}},\ }\bibfield  {title} {\enquote {\bibinfo {title}
  {Bulk-boundary correspondence in non-hermitian hopf-link exceptional line
  semimetals},}\ }\href {\doibase 10.1103/PhysRevB.102.045412} {\bibfield
  {journal} {\bibinfo  {journal} {Phys. Rev. B}\ }\textbf {\bibinfo {volume}
  {102}},\ \bibinfo {pages} {045412} (\bibinfo {year}
  {2020}{\natexlab{b}})}\BibitemShut {NoStop}%
\bibitem [{\citenamefont {Rui}\ \emph {et~al.}(2019)\citenamefont {Rui},
  \citenamefont {Zhao},\ and\ \citenamefont {Schnyder}}]{rui2019}%
  \BibitemOpen
  \bibfield  {author} {\bibinfo {author} {\bibfnamefont {W.~B.}\ \bibnamefont
  {Rui}}, \bibinfo {author} {\bibfnamefont {Y.~X.}\ \bibnamefont {Zhao}}, \
  and\ \bibinfo {author} {\bibfnamefont {Andreas~P.}\ \bibnamefont
  {Schnyder}},\ }\bibfield  {title} {\enquote {\bibinfo {title} {Topology and
  exceptional points of massive dirac models with generic non-hermitian
  perturbations},}\ }\href {\doibase 10.1103/PhysRevB.99.241110} {\bibfield
  {journal} {\bibinfo  {journal} {Phys. Rev. B}\ }\textbf {\bibinfo {volume}
  {99}},\ \bibinfo {pages} {241110} (\bibinfo {year} {2019})}\BibitemShut
  {NoStop}%
\bibitem [{\citenamefont {Xue}\ \emph {et~al.}(2020)\citenamefont {Xue},
  \citenamefont {Wang}, \citenamefont {Zhang},\ and\ \citenamefont
  {Chong}}]{xue2020}%
  \BibitemOpen
  \bibfield  {author} {\bibinfo {author} {\bibfnamefont {Haoran}\ \bibnamefont
  {Xue}}, \bibinfo {author} {\bibfnamefont {Qiang}\ \bibnamefont {Wang}},
  \bibinfo {author} {\bibfnamefont {Baile}\ \bibnamefont {Zhang}}, \ and\
  \bibinfo {author} {\bibfnamefont {Y.~D.}\ \bibnamefont {Chong}},\ }\bibfield
  {title} {\enquote {\bibinfo {title} {Non-hermitian dirac cones},}\ }\href
  {\doibase 10.1103/PhysRevLett.124.236403} {\bibfield  {journal} {\bibinfo
  {journal} {Phys. Rev. Lett.}\ }\textbf {\bibinfo {volume} {124}},\ \bibinfo
  {pages} {236403} (\bibinfo {year} {2020})}\BibitemShut {NoStop}%
\bibitem [{\citenamefont {Yokomizo}\ and\ \citenamefont
  {Murakami}(2020)}]{yokomizo2020}%
  \BibitemOpen
  \bibfield  {author} {\bibinfo {author} {\bibfnamefont {Kazuki}\ \bibnamefont
  {Yokomizo}}\ and\ \bibinfo {author} {\bibfnamefont {Shuichi}\ \bibnamefont
  {Murakami}},\ }\bibfield  {title} {\enquote {\bibinfo {title} {Topological
  semimetal phase with exceptional points in one-dimensional non-hermitian
  systems},}\ }\href {\doibase 10.1103/PhysRevResearch.2.043045} {\bibfield
  {journal} {\bibinfo  {journal} {Phys. Rev. Research}\ }\textbf {\bibinfo
  {volume} {2}},\ \bibinfo {pages} {043045} (\bibinfo {year}
  {2020})}\BibitemShut {NoStop}%
\bibitem [{\citenamefont {Zhong}\ \emph {et~al.}(2020)\citenamefont {Zhong},
  \citenamefont {Kou}, \citenamefont {\"Ozdemir},\ and\ \citenamefont
  {El-Ganainy}}]{zhong2020}%
  \BibitemOpen
  \bibfield  {author} {\bibinfo {author} {\bibfnamefont {Q.}~\bibnamefont
  {Zhong}}, \bibinfo {author} {\bibfnamefont {J.}~\bibnamefont {Kou}}, \bibinfo
  {author} {\bibfnamefont {\ifmmode \mbox{\c{S}}\else \c{S}\fi{}.~K.}\
  \bibnamefont {\"Ozdemir}}, \ and\ \bibinfo {author} {\bibfnamefont
  {R.}~\bibnamefont {El-Ganainy}},\ }\bibfield  {title} {\enquote {\bibinfo
  {title} {Hierarchical construction of higher-order exceptional points},}\
  }\href {\doibase 10.1103/PhysRevLett.125.203602} {\bibfield  {journal}
  {\bibinfo  {journal} {Phys. Rev. Lett.}\ }\textbf {\bibinfo {volume} {125}},\
  \bibinfo {pages} {203602} (\bibinfo {year} {2020})}\BibitemShut {NoStop}%
\bibitem [{\citenamefont {Crippa}\ \emph {et~al.}(2021)\citenamefont {Crippa},
  \citenamefont {Budich},\ and\ \citenamefont {Sangiovanni}}]{crippa2021}%
  \BibitemOpen
  \bibfield  {author} {\bibinfo {author} {\bibfnamefont {L.}~\bibnamefont
  {Crippa}}, \bibinfo {author} {\bibfnamefont {J.~C.}\ \bibnamefont {Budich}},
  \ and\ \bibinfo {author} {\bibfnamefont {G.}~\bibnamefont {Sangiovanni}},\
  }\bibfield  {title} {\enquote {\bibinfo {title} {Fourth-order exceptional
  points in correlated quantum many-body systems},}\ }\href {\doibase
  10.1103/PhysRevB.104.L121109} {\bibfield  {journal} {\bibinfo  {journal}
  {Phys. Rev. B}\ }\textbf {\bibinfo {volume} {104}},\ \bibinfo {pages}
  {L121109} (\bibinfo {year} {2021})}\BibitemShut {NoStop}%
\bibitem [{\citenamefont {Yang}\ \emph {et~al.}(2021)\citenamefont {Yang},
  \citenamefont {Schnyder}, \citenamefont {Hu},\ and\ \citenamefont
  {Chiu}}]{yang2021}%
  \BibitemOpen
  \bibfield  {author} {\bibinfo {author} {\bibfnamefont {Zhesen}\ \bibnamefont
  {Yang}}, \bibinfo {author} {\bibfnamefont {A.~P.}\ \bibnamefont {Schnyder}},
  \bibinfo {author} {\bibfnamefont {Jiangping}\ \bibnamefont {Hu}}, \ and\
  \bibinfo {author} {\bibfnamefont {Ching-Kai}\ \bibnamefont {Chiu}},\
  }\bibfield  {title} {\enquote {\bibinfo {title} {Fermion doubling theorems in
  two-dimensional non-hermitian systems for fermi points and exceptional
  points},}\ }\href {\doibase 10.1103/PhysRevLett.126.086401} {\bibfield
  {journal} {\bibinfo  {journal} {Phys. Rev. Lett.}\ }\textbf {\bibinfo
  {volume} {126}},\ \bibinfo {pages} {086401} (\bibinfo {year}
  {2021})}\BibitemShut {NoStop}%
\bibitem [{\citenamefont {Denner}\ \emph {et~al.}(2021)\citenamefont {Denner},
  \citenamefont {Skurativska}, \citenamefont {Schindler}, \citenamefont
  {Fischer}, \citenamefont {Thomale}, \citenamefont {Bzdu{\v{s}}ek},\ and\
  \citenamefont {Neupert}}]{denner2021}%
  \BibitemOpen
  \bibfield  {author} {\bibinfo {author} {\bibfnamefont {M.~Michael}\
  \bibnamefont {Denner}}, \bibinfo {author} {\bibfnamefont {Anastasiia}\
  \bibnamefont {Skurativska}}, \bibinfo {author} {\bibfnamefont {Frank}\
  \bibnamefont {Schindler}}, \bibinfo {author} {\bibfnamefont {Mark~H.}\
  \bibnamefont {Fischer}}, \bibinfo {author} {\bibfnamefont {Ronny}\
  \bibnamefont {Thomale}}, \bibinfo {author} {\bibfnamefont {Tom{\'a}{\v{s}}}\
  \bibnamefont {Bzdu{\v{s}}ek}}, \ and\ \bibinfo {author} {\bibfnamefont
  {Titus}\ \bibnamefont {Neupert}},\ }\bibfield  {title} {\enquote {\bibinfo
  {title} {Exceptional topological insulators},}\ }\href {\doibase
  10.1038/s41467-021-25947-z} {\bibfield  {journal} {\bibinfo  {journal}
  {Nature Communications}\ }\textbf {\bibinfo {volume} {12}},\ \bibinfo {pages}
  {5681} (\bibinfo {year} {2021})}\BibitemShut {NoStop}%
\bibitem [{\citenamefont {Zhang}\ \emph {et~al.}(2021)\citenamefont {Zhang},
  \citenamefont {Li}, \citenamefont {Liu}, \citenamefont {Tai}, \citenamefont
  {Thomale},\ and\ \citenamefont {Lee}}]{tidal}%
  \BibitemOpen
  \bibfield  {author} {\bibinfo {author} {\bibfnamefont {Xiao}\ \bibnamefont
  {Zhang}}, \bibinfo {author} {\bibfnamefont {Guangjie}\ \bibnamefont {Li}},
  \bibinfo {author} {\bibfnamefont {Yuhan}\ \bibnamefont {Liu}}, \bibinfo
  {author} {\bibfnamefont {Tommy}\ \bibnamefont {Tai}}, \bibinfo {author}
  {\bibfnamefont {Ronny}\ \bibnamefont {Thomale}}, \ and\ \bibinfo {author}
  {\bibfnamefont {Ching~Hua}\ \bibnamefont {Lee}},\ }\bibfield  {title}
  {\enquote {\bibinfo {title} {Tidal surface states as fingerprints of
  non-hermitian nodal knot metals},}\ }\href {\doibase
  10.1038/s42005-021-00535-1} {\bibfield  {journal} {\bibinfo  {journal}
  {Communications Physics}\ }\textbf {\bibinfo {volume} {4}},\ \bibinfo {pages}
  {47} (\bibinfo {year} {2021})}\BibitemShut {NoStop}%
\bibitem [{\citenamefont {Mandal}\ and\ \citenamefont
  {Bergholtz}(2021)}]{mandal2021}%
  \BibitemOpen
  \bibfield  {author} {\bibinfo {author} {\bibfnamefont {Ipsita}\ \bibnamefont
  {Mandal}}\ and\ \bibinfo {author} {\bibfnamefont {Emil~J.}\ \bibnamefont
  {Bergholtz}},\ }\bibfield  {title} {\enquote {\bibinfo {title} {Symmetry and
  higher-order exceptional points},}\ }\href {\doibase
  10.1103/PhysRevLett.127.186601} {\bibfield  {journal} {\bibinfo  {journal}
  {Phys. Rev. Lett.}\ }\textbf {\bibinfo {volume} {127}},\ \bibinfo {pages}
  {186601} (\bibinfo {year} {2021})}\BibitemShut {NoStop}%
\bibitem [{\citenamefont {Delplace}\ \emph {et~al.}(2021)\citenamefont
  {Delplace}, \citenamefont {Yoshida},\ and\ \citenamefont
  {Hatsugai}}]{delplace2021}%
  \BibitemOpen
  \bibfield  {author} {\bibinfo {author} {\bibfnamefont {Pierre}\ \bibnamefont
  {Delplace}}, \bibinfo {author} {\bibfnamefont {Tsuneya}\ \bibnamefont
  {Yoshida}}, \ and\ \bibinfo {author} {\bibfnamefont {Yasuhiro}\ \bibnamefont
  {Hatsugai}},\ }\bibfield  {title} {\enquote {\bibinfo {title}
  {Symmetry-protected multifold exceptional points and their topological
  characterization},}\ }\href {\doibase 10.1103/PhysRevLett.127.186602}
  {\bibfield  {journal} {\bibinfo  {journal} {Phys. Rev. Lett.}\ }\textbf
  {\bibinfo {volume} {127}},\ \bibinfo {pages} {186602} (\bibinfo {year}
  {2021})}\BibitemShut {NoStop}%
\bibitem [{\citenamefont {Ghorashi}\ \emph
  {et~al.}(2021{\natexlab{a}})\citenamefont {Ghorashi}, \citenamefont {Li},
  \citenamefont {Sato},\ and\ \citenamefont {Hughes}}]{ghorashi202101}%
  \BibitemOpen
  \bibfield  {author} {\bibinfo {author} {\bibfnamefont {Sayed Ali~Akbar}\
  \bibnamefont {Ghorashi}}, \bibinfo {author} {\bibfnamefont {Tianhe}\
  \bibnamefont {Li}}, \bibinfo {author} {\bibfnamefont {Masatoshi}\
  \bibnamefont {Sato}}, \ and\ \bibinfo {author} {\bibfnamefont {Taylor~L.}\
  \bibnamefont {Hughes}},\ }\bibfield  {title} {\enquote {\bibinfo {title}
  {Non-hermitian higher-order dirac semimetals},}\ }\href {\doibase
  10.1103/PhysRevB.104.L161116} {\bibfield  {journal} {\bibinfo  {journal}
  {Phys. Rev. B}\ }\textbf {\bibinfo {volume} {104}},\ \bibinfo {pages}
  {L161116} (\bibinfo {year} {2021}{\natexlab{a}})}\BibitemShut {NoStop}%
\bibitem [{\citenamefont {Ghorashi}\ \emph
  {et~al.}(2021{\natexlab{b}})\citenamefont {Ghorashi}, \citenamefont {Li},\
  and\ \citenamefont {Sato}}]{ghorashi202102}%
  \BibitemOpen
  \bibfield  {author} {\bibinfo {author} {\bibfnamefont {Sayed Ali~Akbar}\
  \bibnamefont {Ghorashi}}, \bibinfo {author} {\bibfnamefont {Tianhe}\
  \bibnamefont {Li}}, \ and\ \bibinfo {author} {\bibfnamefont {Masatoshi}\
  \bibnamefont {Sato}},\ }\bibfield  {title} {\enquote {\bibinfo {title}
  {Non-hermitian higher-order weyl semimetals},}\ }\href {\doibase
  10.1103/PhysRevB.104.L161117} {\bibfield  {journal} {\bibinfo  {journal}
  {Phys. Rev. B}\ }\textbf {\bibinfo {volume} {104}},\ \bibinfo {pages}
  {L161117} (\bibinfo {year} {2021}{\natexlab{b}})}\BibitemShut {NoStop}%
\bibitem [{\citenamefont {Liu}\ \emph {et~al.}(2021)\citenamefont {Liu},
  \citenamefont {He}, \citenamefont {Yang},\ and\ \citenamefont
  {Nori}}]{liu2021}%
  \BibitemOpen
  \bibfield  {author} {\bibinfo {author} {\bibfnamefont {Tao}\ \bibnamefont
  {Liu}}, \bibinfo {author} {\bibfnamefont {James~Jun}\ \bibnamefont {He}},
  \bibinfo {author} {\bibfnamefont {Zhongmin}\ \bibnamefont {Yang}}, \ and\
  \bibinfo {author} {\bibfnamefont {Franco}\ \bibnamefont {Nori}},\ }\bibfield
  {title} {\enquote {\bibinfo {title} {Higher-order weyl-exceptional-ring
  semimetals},}\ }\href {\doibase 10.1103/PhysRevLett.127.196801} {\bibfield
  {journal} {\bibinfo  {journal} {Phys. Rev. Lett.}\ }\textbf {\bibinfo
  {volume} {127}},\ \bibinfo {pages} {196801} (\bibinfo {year}
  {2021})}\BibitemShut {NoStop}%
\bibitem [{\citenamefont {St\aa{}lhammar}\ and\ \citenamefont
  {Bergholtz}(2021)}]{bergholtz2021}%
  \BibitemOpen
  \bibfield  {author} {\bibinfo {author} {\bibfnamefont {Marcus}\ \bibnamefont
  {St\aa{}lhammar}}\ and\ \bibinfo {author} {\bibfnamefont {Emil~J.}\
  \bibnamefont {Bergholtz}},\ }\bibfield  {title} {\enquote {\bibinfo {title}
  {Classification of exceptional nodal topologies protected by $\mathcal{PT}$
  symmetry},}\ }\href {\doibase 10.1103/PhysRevB.104.L201104} {\bibfield
  {journal} {\bibinfo  {journal} {Phys. Rev. B}\ }\textbf {\bibinfo {volume}
  {104}},\ \bibinfo {pages} {L201104} (\bibinfo {year} {2021})}\BibitemShut
  {NoStop}%
\bibitem [{\citenamefont {Le}\ \emph {et~al.}(2021)\citenamefont {Le},
  \citenamefont {Yang}, \citenamefont {Cui}, \citenamefont {Schnyder},\ and\
  \citenamefont {Chiu}}]{le2021generalized}%
  \BibitemOpen
  \bibfield  {author} {\bibinfo {author} {\bibfnamefont {Congcong}\
  \bibnamefont {Le}}, \bibinfo {author} {\bibfnamefont {Zhesen}\ \bibnamefont
  {Yang}}, \bibinfo {author} {\bibfnamefont {Fan}\ \bibnamefont {Cui}},
  \bibinfo {author} {\bibfnamefont {A.~P.}\ \bibnamefont {Schnyder}}, \ and\
  \bibinfo {author} {\bibfnamefont {Ching-Kai}\ \bibnamefont {Chiu}},\
  }\href@noop {} {\enquote {\bibinfo {title} {The generalized nielsen-ninomiya
  theorem for the 17 wallpaper: Classification of 2d nodal superconductors,
  dirac semimetals, and non-hermitian nodal systems},}\ } (\bibinfo {year}
  {2021}),\ \Eprint {http://arxiv.org/abs/2108.04534} {arXiv:2108.04534
  [cond-mat.mes-hall]} \BibitemShut {NoStop}%
\bibitem [{\citenamefont {Wang}\ \emph {et~al.}(2021)\citenamefont {Wang},
  \citenamefont {Yu}, \citenamefont {Yang},\ and\ \citenamefont
  {Kou}}]{wang2021}%
  \BibitemOpen
  \bibfield  {author} {\bibinfo {author} {\bibfnamefont {X.~R.}\ \bibnamefont
  {Wang}}, \bibinfo {author} {\bibfnamefont {X.~J.}\ \bibnamefont {Yu}},
  \bibinfo {author} {\bibfnamefont {F.}~\bibnamefont {Yang}}, \ and\ \bibinfo
  {author} {\bibfnamefont {S.~P.}\ \bibnamefont {Kou}},\ }\href@noop {}
  {\enquote {\bibinfo {title} {Physical exceptional points without degeneracy
  of energy levels},}\ } (\bibinfo {year} {2021}),\ \Eprint
  {http://arxiv.org/abs/2109.05980} {arXiv:2109.05980 [quant-ph]} \BibitemShut
  {NoStop}%
\bibitem [{Note1()}]{Note1}%
  \BibitemOpen
  \bibinfo {note} {We emphasize that we take OBC along one specific direction
  to obtain the open-boundary spectra of non-Hermitian systems in this
  paper.}\BibitemShut {Stop}%
\bibitem [{Note2()}]{Note2}%
  \BibitemOpen
  \bibinfo {note} {Without loss of generality, we take the same hopping range
  $R$ in the left and right directions, since the general theory of 1D
  non-Hermitian systems is not fundamentally affected in the situation with
  different hopping ranges.}\BibitemShut {Stop}%
\bibitem [{\citenamefont {Alase}\ \emph {et~al.}(2017)\citenamefont {Alase},
  \citenamefont {Cobanera}, \citenamefont {Ortiz},\ and\ \citenamefont
  {Viola}}]{alase2017}%
  \BibitemOpen
  \bibfield  {author} {\bibinfo {author} {\bibfnamefont {Abhijeet}\
  \bibnamefont {Alase}}, \bibinfo {author} {\bibfnamefont {Emilio}\
  \bibnamefont {Cobanera}}, \bibinfo {author} {\bibfnamefont {Gerardo}\
  \bibnamefont {Ortiz}}, \ and\ \bibinfo {author} {\bibfnamefont {Lorenza}\
  \bibnamefont {Viola}},\ }\bibfield  {title} {\enquote {\bibinfo {title}
  {Generalization of bloch's theorem for arbitrary boundary conditions:
  Theory},}\ }\href {\doibase 10.1103/PhysRevB.96.195133} {\bibfield  {journal}
  {\bibinfo  {journal} {Phys. Rev. B}\ }\textbf {\bibinfo {volume} {96}},\
  \bibinfo {pages} {195133} (\bibinfo {year} {2017})}\BibitemShut {NoStop}%
\bibitem [{Note3()}]{Note3}%
  \BibitemOpen
  \bibinfo {note} {We take the first three-order terms in the expansion of
  $\protect \qopname \relax m{det}{[B(E)]}$, to discuss the asymptotic behavior
  of the IEBs. If one of the first three-order terms vanishes, we need to
  consider the fourth-order term, and it is the same as the subsequent
  orders.}\BibitemShut {Stop}%
\bibitem [{Note4()}]{Note4}%
  \BibitemOpen
  \bibinfo {note} {In this paper, we concentrate on the general theory of 1D
  non-Hermitian systems without conjugated time-reversal
  symmetry~(TRS$^{\dagger }$). The continuous bands condition of systems in
  symplectic class~(with TRS$^{\dagger }$) is given by $|\beta _{m-1}|=|\beta
  _{m}|$ or $|\beta _{m+1}|=|\beta _{m+2}|$, where $m$ is an even integer~\cite
  {kawabata20}. The general theory of 1D non-Hermitian systems is not
  fundamentally affected in symplectic class.}\BibitemShut {Stop}%
\bibitem [{\citenamefont {Liu}\ \emph {et~al.}(2017)\citenamefont {Liu},
  \citenamefont {Vafa},\ and\ \citenamefont {Xu}}]{knot1}%
  \BibitemOpen
  \bibfield  {author} {\bibinfo {author} {\bibfnamefont {Chunxiao}\
  \bibnamefont {Liu}}, \bibinfo {author} {\bibfnamefont {Farzan}\ \bibnamefont
  {Vafa}}, \ and\ \bibinfo {author} {\bibfnamefont {Cenke}\ \bibnamefont
  {Xu}},\ }\bibfield  {title} {\enquote {\bibinfo {title} {Symmetry-protected
  topological hopf insulator and its generalizations},}\ }\href {\doibase
  10.1103/PhysRevB.95.161116} {\bibfield  {journal} {\bibinfo  {journal} {Phys.
  Rev. B}\ }\textbf {\bibinfo {volume} {95}},\ \bibinfo {pages} {161116}
  (\bibinfo {year} {2017})}\BibitemShut {NoStop}%
\bibitem [{\citenamefont {Kennedy}(2016)}]{knot2}%
  \BibitemOpen
  \bibfield  {author} {\bibinfo {author} {\bibfnamefont {Ricardo}\ \bibnamefont
  {Kennedy}},\ }\bibfield  {title} {\enquote {\bibinfo {title} {Topological
  hopf-chern insulators and the hopf superconductor},}\ }\href {\doibase
  10.1103/PhysRevB.94.035137} {\bibfield  {journal} {\bibinfo  {journal} {Phys.
  Rev. B}\ }\textbf {\bibinfo {volume} {94}},\ \bibinfo {pages} {035137}
  (\bibinfo {year} {2016})}\BibitemShut {NoStop}%
\bibitem [{\citenamefont {Chen}\ \emph {et~al.}(2017)\citenamefont {Chen},
  \citenamefont {Lu},\ and\ \citenamefont {Hou}}]{knot3}%
  \BibitemOpen
  \bibfield  {author} {\bibinfo {author} {\bibfnamefont {Wei}\ \bibnamefont
  {Chen}}, \bibinfo {author} {\bibfnamefont {Hai-Zhou}\ \bibnamefont {Lu}}, \
  and\ \bibinfo {author} {\bibfnamefont {Jing-Min}\ \bibnamefont {Hou}},\
  }\bibfield  {title} {\enquote {\bibinfo {title} {Topological semimetals with
  a double-helix nodal link},}\ }\href {\doibase 10.1103/PhysRevB.96.041102}
  {\bibfield  {journal} {\bibinfo  {journal} {Phys. Rev. B}\ }\textbf {\bibinfo
  {volume} {96}},\ \bibinfo {pages} {041102} (\bibinfo {year}
  {2017})}\BibitemShut {NoStop}%
\bibitem [{\citenamefont {Yan}\ \emph {et~al.}(2017)\citenamefont {Yan},
  \citenamefont {Bi}, \citenamefont {Shen}, \citenamefont {Lu}, \citenamefont
  {Zhang},\ and\ \citenamefont {Wang}}]{knot4}%
  \BibitemOpen
  \bibfield  {author} {\bibinfo {author} {\bibfnamefont {Zhongbo}\ \bibnamefont
  {Yan}}, \bibinfo {author} {\bibfnamefont {Ren}\ \bibnamefont {Bi}}, \bibinfo
  {author} {\bibfnamefont {Huitao}\ \bibnamefont {Shen}}, \bibinfo {author}
  {\bibfnamefont {Ling}\ \bibnamefont {Lu}}, \bibinfo {author} {\bibfnamefont
  {Shou-Cheng}\ \bibnamefont {Zhang}}, \ and\ \bibinfo {author} {\bibfnamefont
  {Zhong}\ \bibnamefont {Wang}},\ }\bibfield  {title} {\enquote {\bibinfo
  {title} {Nodal-link semimetals},}\ }\href {\doibase
  10.1103/PhysRevB.96.041103} {\bibfield  {journal} {\bibinfo  {journal} {Phys.
  Rev. B}\ }\textbf {\bibinfo {volume} {96}},\ \bibinfo {pages} {041103}
  (\bibinfo {year} {2017})}\BibitemShut {NoStop}%
\bibitem [{\citenamefont {Bi}\ \emph {et~al.}(2017)\citenamefont {Bi},
  \citenamefont {Yan}, \citenamefont {Lu},\ and\ \citenamefont {Wang}}]{knot5}%
  \BibitemOpen
  \bibfield  {author} {\bibinfo {author} {\bibfnamefont {Ren}\ \bibnamefont
  {Bi}}, \bibinfo {author} {\bibfnamefont {Zhongbo}\ \bibnamefont {Yan}},
  \bibinfo {author} {\bibfnamefont {Ling}\ \bibnamefont {Lu}}, \ and\ \bibinfo
  {author} {\bibfnamefont {Zhong}\ \bibnamefont {Wang}},\ }\bibfield  {title}
  {\enquote {\bibinfo {title} {Nodal-knot semimetals},}\ }\href {\doibase
  10.1103/PhysRevB.96.201305} {\bibfield  {journal} {\bibinfo  {journal} {Phys.
  Rev. B}\ }\textbf {\bibinfo {volume} {96}},\ \bibinfo {pages} {201305}
  (\bibinfo {year} {2017})}\BibitemShut {NoStop}%
\bibitem [{\citenamefont {Chang}\ and\ \citenamefont {Yee}(2017)}]{knot6}%
  \BibitemOpen
  \bibfield  {author} {\bibinfo {author} {\bibfnamefont {Po-Yao}\ \bibnamefont
  {Chang}}\ and\ \bibinfo {author} {\bibfnamefont {Chuck-Hou}\ \bibnamefont
  {Yee}},\ }\bibfield  {title} {\enquote {\bibinfo {title} {Weyl-link
  semimetals},}\ }\href {\doibase 10.1103/PhysRevB.96.081114} {\bibfield
  {journal} {\bibinfo  {journal} {Phys. Rev. B}\ }\textbf {\bibinfo {volume}
  {96}},\ \bibinfo {pages} {081114} (\bibinfo {year} {2017})}\BibitemShut
  {NoStop}%
\bibitem [{\citenamefont {Ezawa}(2017)}]{knot7}%
  \BibitemOpen
  \bibfield  {author} {\bibinfo {author} {\bibfnamefont {Motohiko}\
  \bibnamefont {Ezawa}},\ }\bibfield  {title} {\enquote {\bibinfo {title}
  {Topological semimetals carrying arbitrary hopf numbers: Fermi surface
  topologies of a hopf link, solomon's knot, trefoil knot, and other linked
  nodal varieties},}\ }\href {\doibase 10.1103/PhysRevB.96.041202} {\bibfield
  {journal} {\bibinfo  {journal} {Phys. Rev. B}\ }\textbf {\bibinfo {volume}
  {96}},\ \bibinfo {pages} {041202} (\bibinfo {year} {2017})}\BibitemShut
  {NoStop}%
\bibitem [{\citenamefont {Zhou}\ \emph {et~al.}(2018)\citenamefont {Zhou},
  \citenamefont {Xiong}, \citenamefont {Wan},\ and\ \citenamefont
  {An}}]{knot8}%
  \BibitemOpen
  \bibfield  {author} {\bibinfo {author} {\bibfnamefont {Yao}\ \bibnamefont
  {Zhou}}, \bibinfo {author} {\bibfnamefont {Feng}\ \bibnamefont {Xiong}},
  \bibinfo {author} {\bibfnamefont {Xiangang}\ \bibnamefont {Wan}}, \ and\
  \bibinfo {author} {\bibfnamefont {Jin}\ \bibnamefont {An}},\ }\bibfield
  {title} {\enquote {\bibinfo {title} {Hopf-link topological nodal-loop
  semimetals},}\ }\href {\doibase 10.1103/PhysRevB.97.155140} {\bibfield
  {journal} {\bibinfo  {journal} {Phys. Rev. B}\ }\textbf {\bibinfo {volume}
  {97}},\ \bibinfo {pages} {155140} (\bibinfo {year} {2018})}\BibitemShut
  {NoStop}%
\bibitem [{\citenamefont {Stålhammar}\ \emph {et~al.}(2019)\citenamefont
  {Stålhammar}, \citenamefont {Rødland}, \citenamefont {Arone}, \citenamefont
  {Budich},\ and\ \citenamefont {Bergholtz}}]{bergholtzhyper}%
  \BibitemOpen
  \bibfield  {author} {\bibinfo {author} {\bibfnamefont {Marcus}\ \bibnamefont
  {Stålhammar}}, \bibinfo {author} {\bibfnamefont {Lukas}\ \bibnamefont
  {Rødland}}, \bibinfo {author} {\bibfnamefont {Gregory}\ \bibnamefont
  {Arone}}, \bibinfo {author} {\bibfnamefont {Jan~Carl}\ \bibnamefont
  {Budich}}, \ and\ \bibinfo {author} {\bibfnamefont {Emil~J.}\ \bibnamefont
  {Bergholtz}},\ }\bibfield  {title} {\enquote {\bibinfo {title} {{Hyperbolic
  Nodal Band Structures and Knot Invariants}},}\ }\href {\doibase
  10.21468/SciPostPhys.7.2.019} {\bibfield  {journal} {\bibinfo  {journal}
  {SciPost Phys.}\ }\textbf {\bibinfo {volume} {7}},\ \bibinfo {pages} {19}
  (\bibinfo {year} {2019})}\BibitemShut {NoStop}%
\bibitem [{\citenamefont {Wu}\ \emph {et~al.}(2022)\citenamefont {Wu},
  \citenamefont {Xie}, \citenamefont {Zhou},\ and\ \citenamefont
  {An}}]{conn2021}%
  \BibitemOpen
  \bibfield  {author} {\bibinfo {author} {\bibfnamefont {Deguang}\ \bibnamefont
  {Wu}}, \bibinfo {author} {\bibfnamefont {Jiao}\ \bibnamefont {Xie}}, \bibinfo
  {author} {\bibfnamefont {Yao}\ \bibnamefont {Zhou}}, \ and\ \bibinfo {author}
  {\bibfnamefont {Jin}\ \bibnamefont {An}},\ }\bibfield  {title} {\enquote
  {\bibinfo {title} {Connections between the open-boundary spectrum and the
  generalized brillouin zone in non-hermitian systems},}\ }\href {\doibase
  10.1103/PhysRevB.105.045422} {\bibfield  {journal} {\bibinfo  {journal}
  {Phys. Rev. B}\ }\textbf {\bibinfo {volume} {105}},\ \bibinfo {pages}
  {045422} (\bibinfo {year} {2022})}\BibitemShut {NoStop}%
\end{thebibliography}%
\end{document}